%% file: YB-review.tex
\documentclass[a4paper]{iopart}

\usepackage[utf8]{inputenc}
\usepackage[T1]{fontenc}

\usepackage{sakamotostyle}
\usepackage{mathptmx}

\usepackage[perpage,symbol*,marginal]{footmisc}

\usepackage{Definitions}
\usepackage{lipsum}
\usepackage{hyperref}
\usepackage{slashed}
\usepackage{bbm}

\DeclareMathOperator{\STrace}{STr} %
\DeclareDocumentCommand\STr{}{\opbraces{\STrace}}

\acrodef{yb}[\textsc{yb}]{Yang--Baxter}
\acrodef{nc}[\textsc{nc}]{non-commutativity}
\acrodef{sym}[\textsc{sym}]{super Yang--Mills}
\acrodef{himr}[\textsc{himr}  background]{Hashimoto--Itzhaki--Maldacena--Russo background~\cite{hep-th/9907166,hep-th/9908134}}
\acrodef{cybe}[\textsc{cybe}]{classical Yang--Baxter equation}
\acrodef{kl}[\textsc{kl}]{Kosmann--Lie}
\acrodef{sw}[\textsc{sw}]{Seiberg--Witten}
\acrodef{susy}[\textsc{susy}]{supersymmetry}
\acrodef{wz}[\textsc{wz}]{Wess--Zumino}
\acrodef{wznw}[\textsc{wznw}]{Wess--Zumino--Novikov--Witten}
\acrodef{eom}[\textsc{eom}]{equations of motion}
\acrodef{dof}[\textsc{dof}]{degrees of freedom}
\acrodef{gs}[\textsc{gs}]{Green--Schwarz}
\acrodef{gse}[\textsc{gse}]{generalized supergravity equations}
\acrodef{mcybe}[m\textsc{cybe}]{modified classical Yang--Baxter equation}
\acrodef{dft}[\textsc{dft}]{double field theory}

\numberwithin{equation}{section}

\begin{document}
\topical[Yang-Baxter deformations and generalized supergravity]{Yang-Baxter deformations and generalized supergravity \qquad
-- A short summary --}
\author{Domenico Orlando$^{1,2}$, Susanne Reffert$^2$, Jun-ichi Sakamoto$^{3,4}$, 
\\Yuta Sekiguchi$^{2,5}$, Kentaroh Yoshida$^6$}
\address{$^1$ INFN, sezione di Torino and Arnold--Regge Center. Via Pietro Giuria 1, 10125 Torino, Italy}
\address{$^2$ Institute for Theoretical Physics, 
Albert Einstein Center for Fundamental Physics, 
University of Bern, Sidlerstrasse 5, CH-3012 Bern, 
Switzerland}
\address{$^3$ Department of Physics and Center for Theoretical Sciences,
National Taiwan University, Taipei 10617, Taiwan}
\address{$^4$ Osaka City University Advanced Mathematical Institute (OCAMI),
3-3-138, Sugimoto, Sumiyoshi-ku, Osaka, 558-8585, Japan}
\address{$^5$ Dipartimento di Fisica, Universit\`a di Torino, Via Pietro Giuria 1, 10125 Torino, Italy}
\address{$^6$ Department of Physics, Kyoto University, 
Kyoto 606-8502, Japan}
\ead{sreffert@itp.unibe.ch}

\begin{abstract}
Integrable deformations of type IIB superstring theory on AdS$_5\times$S$^5$ have played an important role over the last years. The Yang--Baxter deformation is a systematic way of generating such integrable deformations. Since its introduction, this topic has seen important conceptual progress and has among others led to the intriguing discovery generalized supergravity, a new low-energy effective theory. 

This review endeavors to not only introduce the historical development of the Yang--Baxter deformation, but also its relation to generalized supergravity, non-geometric backgrounds, non-abelian T-duality and preserved Killing spinors. We supplement the general treatment with a wealth of explicit examples.
\end{abstract}

\submitto{\jpa}
\maketitle

\input{introduction}
\input{YB}
\input{Killing}

\input{conclusions}
\appendix

\input{appendix}

\bibliography{references}{}
\bibliographystyle{iopart-num}

\end{document}

%% file: introduction.tex
\section{Introduction}

The integrability of type IIB superstring on the  AdS$_{5}\times$S$^5$ 
background\footnote{This system is often abbreviated as the AdS$_{5}\times$S$^5$ superstring.}~\cite{Bena:2003wd}  
has led to many important advances as it allows the application of highly developed techniques from integrable systems to a variety of string-theoretic problems (see~\cite{Beisert:2010jr} for a comprehensive review of this subject).
Finding deformations of AdS$_{5}\times$S$^5$ which retain the property of integrability allows us to further expand the reach of such techniques to more and more string backgrounds. While at first, integrable deformations have been found on a case-by-case basis, finding systematic ways of generating integrable backgrounds has become more and more important.

\medskip

The \ac{yb} deformation~\cite{Klimcik:2002zj,Klimcik:2008eq,Delduc:2013fga,Matsumoto:2015jja} is such a systematic way of performing integrable deformations of two-dimensional principal chiral models and symmetric coset sigma models, and can also be applied to the AdS$_{5}\times$S$^5$ superstring~\cite{Delduc:2013qra,Kawaguchi:2014qwa}.
The deformations can be labeled by classical $r$-matrices satisfying the classical \textsc{yb} equation.
It is possible to derive the associated deformed string backgrounds by performing a supercoset construction~\cite{Arutyunov:2013ega,Arutyunov:2015qva,Kyono:2016jqy}.
In recent years, our understanding of these integrable deformations has undergone a lot of progress.
This review endeavors to put YB deformations into the context of these recent insights, highlighting their connections to generalized supergravity~\cite{Arutyunov:2015mqj,Wulff:2016tju} as well as understanding them as string duality transformations~\cite{Matsumoto:2014nra,Matsumoto:2014gwa,Matsumoto:2015uja,Osten:2016dvf,Orlando:2016qqu,Hoare:2016wsk,Borsato:2016pas,Sakamoto:2017cpu,Sakamoto:2018krs} and relating certain subclasses to non-geometric fluxes~\cite{Fernandez-Melgarejo:2017oyu}.
At the same time, we give a comprehensive list of explicit examples of the various types of 
YB-deformed backgrounds which have over the years appeared in the literature.

\bigskip
We start out in Section~\ref{sec:AdS5xS5} with the very basics by introducing the AdS$_{5}\times$S$^5$ superstring~\cite{Metsaev:1998it} and its most important properties such as e.g. its classical integrability~\cite{Bena:2003wd}.
After this, we develop the YB deformations of the AdS$_{5}\times$S$^5$ superstring~\cite{Delduc:2013qra,Kawaguchi:2014qwa} in Section~\ref{sec:YB-sigma}.
After giving the action on the YB-deformed AdS$_{5}\times$S$^5$ superstring (Sec.~\ref{subsec:YB-sigma-action}), we show its classical integrability (Sec.~\ref{sec:YBd_classInt}) and $\kappa$--symmetry (Sec.~\ref{sec:kappa-YB})~\cite{Delduc:2013qra,Kawaguchi:2014qwa} 
and discuss YB-deformed backgrounds from the Green--Schwarz action~\cite{Arutyunov:2013ega,Arutyunov:2015qva,Kyono:2016jqy} (Sec.~\ref{subsec:YBfromGS}).
Next, we introduce an important recent development, namely generalized supergravity~\cite{Arutyunov:2015mqj,Wulff:2016tju}  
(Sec.~\ref{sec:GSE-YBsec}). Whether a given r-matrix encoding a YB deformation gives rise to a supergravity solution or a solution of generalized supergravity depends on whether or not it satisfies the unimodularity condition~\cite{Borsato:2016ose}, discussed in Section~\ref{subsec:unimod_cond}. In Section~\ref{subsec-class-r}, we classify r-matrices according to whether they are Abelian or non-Abelian and their rank. In Section~\ref{sec:stringDualityTrans}, we show that YB deformations can be beautifully reinterpreted in the framework of string duality transformations~\cite{Sakamoto:2017cpu,Sakamoto:2018krs}.
To do so, we include a brief review of double field theory~\cite{Hull:2009mi,Hohm:2010jy,Hohm:2010pp,Hull:2009zb,Siegel:1993bj,Siegel:1993th,Siegel:1993xq}. 
After having developed the general theory, we introduce a number of examples of homogeneous YB-deformed $\AdS{5} \times \rmS^5$ backgrounds in Section~\ref{sec:ExampleYBAdS5}. 
Section~\ref{sec:YB-T-fold} focuses on YB deformations of Minkowski~\cite{Matsumoto:2015ypa,Borowiec:2015wua,Kyono:2015zeu,Pachol:2015mfa,Fernandez-Melgarejo:2017oyu} and $\AdS5\times \rmS^5$ backgrounds. We show that the deformed backgrounds we considered here 
are $T$-folds~\cite{Fernandez-Melgarejo:2017oyu}, a particular class of non-geometric backgrounds, again providing a number of explicit examples. 
In Section~\ref{sec:KillingYB}, we finally turn to the interplay between integrability and preserved supersymmetries of a deformed background and present a formula for determining the preserved Killing spinors~\cite{Orlando:2018kms,Orlando:2018qaq}. 
In~\ref{app:conventions}, we present our conventions and in~\ref{app:psu-algebra} we collect useful formulas on the $\alg{psu}(2,2|4)$ algebra. 

%% file: YB.tex
\section{The $\AdS{5}\times $S$^5$ superstring}\label{sec:AdS5xS5}

In this section, we will briefly review some basic facts on the $\AdS{5}\times $S$^5$ superstring. 
For a comprehensive review, see \cite{Arutyunov:2009ga}. 

\subsection{Metsaev--Tseytlin action}

The dynamics of the superstring on the $\AdS{5}\times $S$^5$ background is described by the supercoset
\begin{align}
 \frac{PSU(2,2|4)}{SO(1,4)\times SO(5)}\,.
\end{align}
The corresponding action in the \ac{gs} formulation has been written down by Metsaev and Tseytlin~\cite{Metsaev:1998it} and has the form
\begin{equation}
  S=-\frac{T}{2}\int \dd{\tau} \dd{\sigma} \Pg_-^{\WSa\WSb}  \STr[A_{\alpha}  d_-(A_{\beta})] ,
\label{AdS5S5-action}
\end{equation}
where $T \equiv R^2/2\pi\alpha'$ is the effective string tension and $R$ is the radius of $\AdS{5}$ and $\rmS^5$.
The $\Pg_{\pm}^{\WSa\WSb}$ are linear combinations of the metric on the world-sheet $\gamma^{\alpha\beta}$
and the anti-symmetric tensor $\epsilon^{\alpha\beta}$\,,
\begin{align}
 \Pg_{\pm}^{\WSa\WSb}\equiv\,\frac{\gga^{\WSa\WSb}\pm \varepsilon^{\WSa\WSb}}{2}\,.
\end{align}
We work in  conformal gauge $\gamma^{\alpha\beta}={\rm diag}(-1,1)$ and normalize the epsilon tensor as $\varepsilon^{ \tau \sigma} = 1/\sqrt{-\gamma}$\,.
$A$ is the left-invariant $1$-form for an element $g$ of $SU(2,2|4)$ defined by
\begin{align}
 A&=g^{-1}\,\rmd g \,,& g&\in SU(2,2|4)\,.
\end{align}
It satisfies the Maurer--Cartan equation
\begin{align}
 \rmd A + A\wedge A = 0\,. 
\label{eq:MC}
\end{align}
Given the projection operators $P^{(i)}\,(i=0,1,2,3)$ on each $\mathbb{Z}_4$-graded subspaces of $\alg{g}\equiv\alg{su}(2,2|4)$, the projection operators $d_\pm$ are defined as the linear combination of the $P^{(i)}$
\begin{align}
 d_{\pm} \equiv \mp P^{(1)}+2\,P^{(2)}\pm P^{(3)} \,,
\label{eq:dpm}
\end{align}
satisfying the relation
\begin{align}
 \STr[ X\,d_\pm(Y) ] = \STr[ d_\mp (X)\, Y ] \,.
\label{eq:dpm-transpose}
\end{align}
If we expand the left-invariant $1$-form $A$ as
\begin{align}
 A &= A^{(0)} +A^{(1)} +A^{(2)} + A^{(3)}\,, & \text{where } A^{(i)} &=P^{(i)}(A) \,,
\end{align}
the action (\ref{AdS5S5-action}) can be rewritten as
\begin{align}
  S=\frac{T}{2}\int \STr(A^{(2)}\wedge *_{\gga} A^{(2)} - A^{(1)}\wedge A^{(3)} )\,,
\end{align}
which is the sum of the kinetic and the \ac{wz} term.
The ratio of the coefficients of the two terms is determined by the requirement that the action be $\kappa$-symmetric.%

\subsection{Classical integrability}

From a practical point of view, the property of integrability is of paramount importance, as it allows the application of powerful computational techniques.

We will see that the \ac{eom} of the action (\ref{AdS5S5-action}) and the flatness condition (\ref{eq:MC}) can be combined into a flatness condition for the Lax pair with a parameter $u$\,.
From the Lax pair, we can construct infinitely many conserved charges via the monodromy matrix.
In this sense, the $\AdS{5}\times \text{S}^5$ superstring is classically integrable 
as shown by Bena, Polchinski and Roiban \cite{Bena:2003wd}.
In this subsection, we show the classical integrability of the $\AdS{5}\times \text{S}^5$ superstring by constructing the Lax pair explicitly.

\paragraph{\ac{eom}.}

Let us start with the \ac{eom} of the action (\ref{AdS5S5-action}). They are given by
\begin{align}
\cE=\cD_\alpha d_-(A^\alpha_{(-)})+\cD_\alpha d_+(A_{(+)}^\alpha)
+[A_{(+)\alpha}\,,d_-(A_{(-)}^\alpha)]+[A_{(-)\alpha}\,,d_+(A_{(+)}^\alpha)]=0\,,
\label{uEOM}
\end{align}
where $\cD_{\alpha}$ is the covariant derivative associated with $\gamma^{\alpha\beta}$ and worldsheet vectors marked with $(\pm)$ are have been acted on with the projection operator $\Pg_{\pm}^{\WSa\WSb}$, 
\begin{align}
A_{(\pm)}^\alpha=P^{\alpha\beta}_\pm A_\beta\,.
\end{align}
The flatness condition (\ref{eq:MC}) of $A$ can be rewritten as
\begin{equation}
  \begin{aligned}
    \Z&=\frac{1}{2}\sqrt{-\gamma}\,\varepsilon^{\alpha\beta}(\partial_\alpha A_\beta-\partial_\beta A_\alpha+[A_\alpha\,,A_\beta]) \\
    &=\cD_\alpha A_{(+)}^\alpha-\cD_\alpha A_{(-)}^\alpha+[A_{(-)\alpha}\,,A_{(+)}^\alpha]=0\,.
  \end{aligned}
\label{uflat}
\end{equation}
For later convenience, we will decompose the \ac{eom} (\ref{uEOM}) and the flatness condition (\ref{uflat}) on each of the $\mathbb{Z}_4$-graded components.
The bosonic parts are
\begin{align}
\mathsf{B}_1&:=\Z^{(0)}
=\cD_\alpha A_{(+)}^{\alpha(0)}-\cD_\alpha A_{(-)}^{\alpha(0)}
+[A_{(-)\alpha}^{(0)}\,,A_{(+)}^{\alpha(0)}]+[A_{(-)\alpha}^{(2)}\,,A_{(+)}^{\alpha(2)}]\no\\
&\qquad\qquad\qquad+[A_{(-)\alpha}^{(1)}\,,A_{(+)}^{\alpha(3)}]+[A_{(-)\alpha}^{(3)}\,,A_{(+)}^{\alpha(1)}]=0\,,\\
\mathsf{B}_2&:=\frac{1}{4}(\cE^{(2)}+2\,\Z^{(2)})
=\cD_\alpha A_{(+)}^{\alpha(2)}+[A_{(-)\alpha}^{(0)}\,,A_{(+)}^{\alpha(2)}]+[A_{(-)\alpha}^{(3)}\,,A_{(+)}^{\alpha(3)}]=0\,,\\ 
\mathsf{B}_3&:=\frac{1}{4}(\cE^{(2)}-2\,\Z^{(2)})
=\cD_\alpha A_{(-)}^{\alpha(2)}-[A_{(-)\alpha}^{(2)}\,,A_{(+)}^{\alpha(0)}]-[A_{(-)\alpha}^{(1)}\,,A_{(+)}^{\alpha(1)}]=0\,,
\end{align}
and the fermionic parts are
\begin{align}
\mathsf{F}_1&:=\frac{1}{4}(3\Z^{(1)}-\cE^{(1)})\no\\
&=\cD_\alpha A_{(+)}^{\alpha(1)}-\cD_\alpha A_{(-)}^{\alpha(1)}+[A_{(-)\alpha}^{(0)}\,,A_{(+)}^{\alpha(1)}]+[A_{(-)\alpha}^{(1)}\,,A_+^{\alpha(0)}]+[A_{-\alpha}^{(2)}\,,A_+^{\alpha(3)}]=0\,,\\ 
\mathsf{F}_2&:=\frac{1}{4}(3\Z^{(3)}+\cE^{(3)})\no\\
&=\cD_\alpha A_{(+)}^{\alpha(3)}-\cD_\alpha A_{(-)}^{\alpha(3)}+[A_{(-)\alpha}^{(0)}\,,A_{(+)}^{\alpha(3)}]+[A_{(-)\alpha}^{(1)}\,,A_+^{\alpha(2)}]+[A_{(-)\alpha}^{(3)}\,,A_{(+)}^{\alpha(0)}]=0\,,\\
\mathsf{F}_3&:=\frac{1}{4}(\cE^{(1)}+\Z^{(1)})
=[A_{(-)\alpha}^{(3)}\,,A_{(+)}^{\alpha(2)}]=0\,,\\ 
\mathsf{F}_4&:=\frac{1}{4}(-\cE^{(3)}+\Z^{(3)})
=[A_{(-)\alpha}^{(2)}\,,A_{(+)}^{\alpha(1)}]=0 \,.
\end{align}

\paragraph{Construction of the Lax pair.}

The Lax pair of the $\AdS{5}\times \text{S}^5$ superstring is given by (see~\cite{Bena:2003wd})
\begin{align}
\cL_\alpha\equiv M_{(-)\alpha}+L_{(+)\alpha}\,,
\label{uLax}
\end{align}
where $M^\alpha_{(-)}$ and $L^\alpha_{(+)}$ are
\begin{align}
M^\alpha_{(-)}&=A^{\alpha(0)}_{(-)}+u\,A_{(-)}^{\alpha(1)}+u^{2}A_{(-)}^{\alpha(2)}+u^{-1} A_{(-)}^{\alpha(3)}\,, \\
L^\alpha_{(+)}&=A^{\alpha(0)}_{(+)}+u\,A_{(+)}^{\alpha(1)}+u^{-2}A_{(+)}^{\alpha(2)}+u^{-1} A_{(+)}^{\alpha(3)}\,.
\end{align}
Here $u$ is the spectral parameter.
The flatness condition of the Lax pair (\ref{uLax}) 
\begin{align}
\frac{1}{2}\epsilon^{\alpha\beta}
(\partial_\alpha \cL_\beta-\partial_\beta \cL_\alpha+[\cL_\alpha\,,\cL_\beta])
=0\,,
\label{flat-super}
\end{align}
is equivalent to the \ac{eom} (\ref{uEOM}) and the flatness condition (\ref{uflat}).
To see this observe that the condition can be rewritten as a sum over the \ac{eom}
\begin{equation}
  \begin{aligned}
\frac{1}{2}\epsilon^{\alpha\beta}
(\partial_\alpha \cL_\beta-\partial_\beta \cL_\alpha+[\cL_\alpha\,,\cL_\beta])
 ={}& u^0\,\mathsf{B}_1+u^{-2}\,\mathsf{B}_2-u^2\,\mathsf{B}_3 \\
&+u\,\mathsf{F}_1+u^{-1} \mathsf{F}_2+u^{-3}\,\mathsf{F}_3+u^3\,\mathsf{F}_4 = 0\,.
\end{aligned}
\end{equation}
Therefore, the Lax pair (\ref{uLax}) is on-shell a flat current.
We can then define the monodromy matrix
\begin{align}
T(u)=\mathsf{P} \exp(\int_{C} \dd{\sigma} \cL_\sigma(u) )\,,
\end{align}
where $\mathsf{P}$ denotes the equal-time path ordering in terms of $\sigma$, and
$C$ is a closed path on the worldsheet.
The flatness condition (\ref{flat-super}) implies that the integral is invariant under deformations of the contour, which in turn means that  $T(u)$ does not depend on $\tau$.
Therefore we obtain infinitely many conserved charges as the coefficients of an expansion in \(u\) of the monodromy matrix.

\subsection{$\kappa$-symmetry}\label{sec:u-kappa}

The action (\ref{AdS5S5-action}) of the $\AdS{5}\times \text{S}^5$ superstring is invariant under $\kappa$-symmetry~\cite{Metsaev:1998it}, as we will briefly show in this subsection.

The action of $\kappa$-symmetry is realized as a combination of the variation of the group element $g$ and the world-sheet metric $\gamma^{\alpha\beta}$,
\begin{align}
g^{-1}\delta_{\kappa}g&=
P^{\alpha\beta}_-\{\gQ^1\kappa_{1\alpha},A_{\beta}^{(2)}\}
+P^{\alpha\beta}_+\{\gQ^2\kappa_{2\alpha},A_{\beta}^{(2)}\}
\,, \label{ukappa1} \\
\delta_\kappa(\sqrt{-\ga} \ga^{\alpha\beta})&=
\frac{1}{4}\sqrt{-\ga} \STr[ \Upsilon\Bigl([\gQ^1 \kappa^\alpha_{1(+)},A_{+(+)}^{(1)\beta}]
+[\gQ^2\kappa^\alpha_{2(-)},A_{-(-)}^{(3)\beta}]\Bigr)+(\alpha\leftrightarrow \beta)]\,,
\label{ukappa2}
\end{align}
where  $\kappa_{I\alpha}\,(I=1,2)$ are local fermionic parameters and we have defined $\Upsilon\:={\rm diag}\,(\Id_4, -\Id_4)$\,.
We first decompose the variation of the action into two parts,
\begin{align}
\delta_{\kappa}S=\delta_gS+\delta_{\gamma}S\,.
\end{align}
The variation $\delta_g S$ coming from the group element $g$ is given by
\begin{align}
\delta_g S&=\frac{T}{2}\int \rmd^2\sigma\,\sqrt{-\ga}\,\str(g^{-1}\delta\,g\,\cE)\,,
\end{align}
where $\cE$ is the \ac{eom} (\ref{uEOM}).
If we take $g^{-1}\delta\,g=\epsilon=\epsilon^{(1)}+\epsilon^{(3)}$ with 
\begin{align}
\epsilon^{(1)}&=P^{\alpha\beta}_-\{\gQ^1\kappa_{1\alpha},A_{\beta}^{(2)}\}\,, &
\epsilon^{(3)}&=P^{\alpha\beta}_+\{\gQ^2\kappa_{2\alpha},A_{\beta}^{(2)}\}\,,
\label{kappaan}
\end{align}
$\delta_g S$ can be rewritten as
\begin{align}
\delta_g S&=\frac{T}{2}\int \rmd^2\sigma\,
\sqrt{-\ga}\,\str\left[\epsilon^{(1)}\left(\cE^{(3)}-\cZ^{(3)}\right)+\epsilon^{(3)}\left(\cE^{(1)}+\cZ^{(1)}\right)\right]\no \\
&=-2T\int \rmd^2\sigma\,\sqrt{-\ga}\,\str\left(\epsilon^{(1)}[A_{-\alpha}^{(2)},A_+^{\alpha(1)}]+\epsilon^{(3)}[A_{+\alpha}^{(2)},A_-^{\alpha(3)}]\right)\,,
\label{eq:deltag-S-ukappa}
\end{align}
where we have used 
\begin{align}
\cE^{(1)}+\cZ^{(1)}&=-4[A_{+\alpha}^{(2)}\,,A_-^{\alpha(3)}]\,, &
\cE^{(3)}-\cZ^{(3)}&=-4[A_{-\alpha}^{(2)}\,,A_+^{\alpha(1)}]\,.
\end{align}
By using the expression\,(\ref{kappaan}), we can rewrite each of the terms in (\ref{eq:deltag-S-ukappa}) as
\begin{align}
\begin{split}
\str\left(\epsilon^{(1)}[A_{-\alpha}^{(2)},A_+^{\alpha(1)}]\right)
&=\str\left(A_{-\alpha}^{(2)}A_{-\beta}^{(2)}[A_{+\alpha}^{(1)}\,,\gQ^1\kappa_{+1}^\beta]\right)\,,\\
\str\left(\epsilon^{(3)}[A_{+\alpha}^{(2)},A_-^{\alpha(3)}]\right)
&=\str\left(A_{+\alpha}^{(2)}A_{+\beta}^{(2)}[A_{-\alpha}^{(3)}\,,\gQ^2\kappa_{-2}^\beta]\right)\,.
\end{split}
\label{eq:kappa-para-str}
\end{align}
An arbitrary grade-$2$ traceless element $A^{(2)}$ of $\mathfrak{su}(2,2|4)$ fulfills the relation
\begin{align}
A^{(2)}_{\alpha\pm}A^{(2)}_{\beta\pm}
=\frac{1}{8}\Upsilon\,\str(A^{(2)}_{\alpha\pm}A^{(2)}_{\beta\pm})+c_{\alpha\beta} Z\,,
\label{symformula}
\end{align}
where $Z$ is the central charge of $\mathfrak{su}(2,2|4)$ and $c_{\alpha\beta}$ is a symmetric function in $\alpha$ and $\beta$\,.
By using the expressions (\ref{eq:kappa-para-str}), (\ref{symformula}),
$\delta_g S$ becomes
\begin{align}
\delta_g S&=\frac{T}{4}\int \rmd^2\sigma\,
\sqrt{-\ga}\,\biggl[\str\left(A^{(2)}_{\alpha-}A^{(2)}_{\beta-}\right)
\str\left([\gQ^1\kappa_{+1}^{\beta}\,,A_+^{\alpha(1)}]\right)\nonumber \\
&\qquad\qquad\qquad+\str\left(A^{(2)}_{\alpha+}A^{(2)}_{\beta+}\right)
\str\left([\gQ^2\kappa_{-2}^{\beta}\,,A_-^{\alpha(3)}]\right)\biggr]\,.
\label{ukappag}
\end{align}
Next we study the variation $\delta_{\gamma}S$ coming from the world-sheet metric $\ga^{\alpha\beta}$\,.
From (\ref{ukappa2}), $\delta_{\gamma}S$ is
\begin{align}
\delta_\ga S&=-\frac{T}{4}\int \rmd^2\sigma\,
\sqrt{-\ga}\,\str(A_\alpha^{(2)} A_\beta^{(2)})
\str\left[\Upsilon\left([\gQ^1\kappa_{+1}^{\beta}\,,A_+^{(1)\alpha}]
+[\gQ^2\kappa_{-2}^{\beta}\,,A_-^{(3)\alpha}]\right)\right]\no \\
&=-\frac{T}{4}\int \rmd^2\sigma\,
\sqrt{-\ga}\,\biggl[\str\left(A^{(2)}_{\alpha-}A^{(2)}_{\beta-}\right)
\str\left([\gQ^1\kappa_{+1}^{\beta}\,,A_+^{\alpha(1)}]\right)\no \\
&\qquad\qquad\qquad\qquad
+\str\left(A^{(2)}_{\alpha+}A^{(2)}_{\beta+}\right)\str\left([\gQ^2\kappa_{-2}^{\beta}\,,A_-^{\alpha(3)}]\right)\biggr]\,.
\label{ukappaga}
\end{align}
Here, we used the relation
\begin{align}
A_{\alpha\pm}B^\alpha=A_{\alpha\pm}B_\mp^\alpha\,,
\label{eq:AB-relation}
\end{align}
where $A_\alpha\,,B_\alpha$ are arbitrary vectors.
This variation manifestly cancels out $\delta_g S$,
\begin{align}
\delta_\kappa S=\delta_gS+\delta_{\gamma}S=0\,.
\end{align}
We see that the action (\ref{AdS5S5-action}) is invariant under the $\kappa$-symmetry transformations (\ref{ukappa1}), (\ref{ukappa2}).

\subsection{The $\AdS{5} \times \rmS^5$ background from the GS action}
\label{subsec:AdS5S5-from-GS}

Next, let us explain how to read off the $\AdS{5} \times \rmS^5$ background from the action (\ref{AdS5S5-action}).

\subsubsection{The canonical form of the GS action.}

To read off the target space background from the action of the $\AdS{5} \times \rmS^5$ superstring,
we need to first introduce the canonical form of the \ac{gs} action.

The canonical form of the type II \ac{gs} superstring action at second order in $\theta$~\cite{Cvetic:1999zs} is given by
\begin{align}
  \begin{split}
    S=-\dlT\int \dd[2]{\sigma} {}& \Big[ \Pg_{-}^{\WSa\WSb}\, (\CG_{mn}+B_{mn})\, \partial_{\WSa}X^m\,\partial_{\WSb}X^n\\
    & +\ii\, \bigl(\Pg_{+}^{\WSa\WSb} \,\partial_{\WSa} X^m\, \brTheta_{1}\, \Gamma_m\, D_{+\WSb}\Theta_{1} 
    + \Pg_{-}^{\WSa\WSb} \,\partial_{\WSa} X^m\, \brTheta_{2}\, \Gamma_m\, D_{-\WSb}\Theta_{2} \bigr)\\
    &-\frac{\ii}{8}\, \Pg_{+}^{\WSa\WSb}\, \brTheta_{1}\, \Gamma_m\, \bisF \, \Gamma_n\, \Theta_{2} \, \partial_{\WSa} X^m\, \partial_{\WSb} X^n +\cO(\theta^4) \Big]\,,
  \end{split}
      \label{eq:GS-action-canonical-YBsec}
\end{align}
where $\Gamma_a$\,($\Gamma_m=e_m{}^a\Gamma_a$) are the $32\times 32$ gamma matrices.
The differential operators $D_{\pm \WSa}$ are defined by
\begin{align}
 D_{\pm \WSa}& \equiv \partial_{\WSa} + \frac{1}{4}\, \partial_{\WSa} X^m\, \omega_{\pm m}{}^{\Loa\Lob}\, \Gamma_{\Loa\Lob}\,,\\
\omega_{\pm m\Loa\Lob}&\equiv \omega_{m\Loa\Lob} \pm \frac{1}{2}\, e_m{}^{\Loc}\,H_{\Loc\Loa\Lob} \,,
\end{align}
where $\omega^{\Loa\Lob}=\omega_m{}^{\Loa\Lob}\,\rmd X^m$ is the spin connection on the target space.
$\Theta_I$ and $\brTheta_I$ are the $32$-component Majorana spinor and its conjugate (see~\ref{app:psu-algebra} for details).

\medskip

The metric $g_{mn}$ and the $B$-field $B_{mn}$ of the target space can be read off from the first line in (\ref{eq:GS-action-canonical-YBsec}).
We can read off the dilaton $\Phi$ and the R-R field strengths $\hat{F}_{a_1\dots a_n}$
from the R-R bispinor $\bisF$ which is defined by
\begin{align}
\bisF=\sum_{p}\frac{1}{p!}e^{\Phi}\,\hat{F}_{a_1\dots a_p}\Gamma^{a_1\dots a_p}
\,.
\label{eq:bi-spinor}
\end{align}
If the Lagrangian (\ref{eq:GS-action-canonical-YBsec}) describes type IIB superstring,
the summation runs over $p=1,3,5,7,9$.
Each R--R field strength satisfies
\begin{align}
 \hat{F}_p = (-1)^{\frac{p(p-1)}{2}}\, * \hat{F}_{10-p} \,,
\end{align}
where the Hodge star $*$ is defined in~\ref{app:conventions}.
By comparing the action (\ref{AdS5S5-action}) expanded in terms of $\theta$ to the canonical form (\ref{eq:GS-action-canonical-YBsec}),
we can obtain the explicit expression for the $\AdS{5} \times \rmS^5$ superstring.

\subsubsection{Group parametrization.}

To derive the $\AdS{5} \times \rmS^5$ background from the \ac{gs} action,
we introduce a coordinate system via a parametrization of the group element $g$\,.
We first decompose the group element into bosonic and fermionic parts,
\begin{align}
 g =g_{\bos}\cdot g_{\fer}\in SU(2,2|4)\,.
\label{bf-decomposition}
\end{align}
We parametrize the bosonic part $g_{\bos}$ as
\begin{align}
\begin{split}
 &g_{\bos} = g_{\AdS5}\cdot g_{\rmS^5}\,,\\
 &g_{\AdS5} \equiv \exp(x^\mu\,P_\mu)\cdot \exp(\ln(z) D)\,,
\\
 &g_{\rmS^5} \equiv \exp(\phi_1\, h_1+\phi_2\,h_2+\phi_3\,h_3)\cdot \exp(\xi\,\gJ_{56})\cdot \exp(r\,\gP_5)\,.
\end{split}
\label{eq:group-parameterization}
\end{align}
Here, $P_{\mu}$ ($\mu=0,\dotsc,3$) and $D$ are the translation and dilatation generators of the conformal algebra $\alg{so}(2,4)$.
We define the Cartan generators of the $\alg{so}(6)$ algebra $h_{i}$, $(i=1,2,3)$  by
\begin{align}
 h_1 &\equiv \gJ_{57}\,, &
 h_2 &\equiv \gJ_{68}\,, &
 h_3 &\equiv \gP_9\,. 
\end{align}
We parameterize the fermionic part $g_{\fer}$ as
\begin{align}
 g_{\fer} &= \exp(\gQ^I\, \theta_I) \,, &
 \gQ^I\,\theta_I &=(\gQ^I)^{\check{\SPa}\hat{\SPa}}\,\theta_{I\check{\SPa}\hat{\SPa}} \,,
\label{eq:group-parameterization-fermi}
\end{align}
where the supercharges $(\gQ^I)^{\check{\SPa}\hat{\SPa}}$ $(I=1,2)$ are labeled by two indices $(\check{\SPa}\,,\hat{\SPa}=1,\dotsc, 4)$ and $\theta_{I\check{\SPa}\hat{\SPa}}\,(I=1,2)$ are $16$-component Majorana--Weyl fermions.
 A matrix representation of the above generators of $\mathfrak{su}(2,2|4)$ is given in~\ref{app:psu-algebra}.

\subsubsection{Expansion of the left-invariant current.}

Next, we will expand the left-invariant current $A$ to second order in the spacetime fermions $\theta$,
\begin{equation}
  A =A_{(0)}+A_{(1)}+A_{(2)}+\cO(\theta^3)\,.
\end{equation}
Now, since we chose the parametrization (\ref{bf-decomposition}), (\ref{eq:group-parameterization}), (\ref{eq:group-parameterization-fermi}) for $g$,
the left-invariant current $A$ can be expanded as
\begin{equation}
  \label{eq:A-AdS5xS5}
  \begin{aligned}
 A &= g_{\fer}^{-1}\,A_{(0)}\,g_{\fer} + \gQ^I\,\rmd\theta_I \\
 &= A_{(0)} + [A_{(0)},\,\gQ^I\,\theta_I] + \frac{1}{2}\,\bigl[[A_{(0)},\,\gQ^I\,\theta_I],\,\gQ^J\,\theta_J\bigr] + \gQ^I\,\rmd\theta_I + \cO(\theta^3)\,,
\end{aligned}
\end{equation}
where $A_{(p)}$ is defined as $\cO(\theta^p)$ of the left-invariant current $A$,
and $A_{(0)}$ is given by
\begin{align}
 A_{(0)}&\equiv g_{\bos}^{-1}\,\rmd g_{\bos} = \Bigl(e_m{}^{\Loa}\,\gP_{\Loa} - \frac{1}{2}\, \omega_m{}^{\Loa\Lob}\,\gJ_{\Loa\Lob}\Bigr)\,\rmd X^m \,.
\end{align}
The vielbein $e^{\Loa}=e_{m}{}^{\Loa}\,\rmd X^m$ has the form
\begin{align}
 e^{\Loa} &= \biggl(\frac{\rmd x^0}{z},\frac{\rmd x^1}{z} ,\frac{\rmd x^2}{z},\frac{\rmd x^3}{z}, \frac{\rmd z}{z},
\rmd r, \sin r\,\rmd \xi, \sin r\,\cos\xi\,\rmd\phi_1,\sin r\,\sin\xi\,\rmd\phi_2,\cos r\,\rmd\phi_3\biggr)\,,
\end{align}
and $\omega^{\Loa\Lob}=\omega_m{}^{\Loa\Lob}\,\rmd X^m$ is the associated spin connection.

Moreover, by using the commutation relations of $\mathfrak{su}(2,2|4)$ (see~\ref{app:conventions} for our conventions), each commutator in (\ref{eq:A-AdS5xS5}) can be evaluated as
\begin{equation}
 [A_{(0)},\,\gQ^I\,\theta_I] = \gQ^I\,\Bigl(\frac{1}{4}\,\delta^{IJ}\,\omega^{\Loa\Lob} \,\gamma_{\Loa\Lob} 
  + \frac{\ii}{2}\,\epsilon^{IJ}\, e^{\Loa}\, \hat{\gamma}_{\Loa} \Bigr)\,\theta_J \label{A0Q}\,,
\end{equation}
\begin{equation}
  \begin{aligned}
 \bigl[[A_{(0)},\,\gQ^I\,\theta_I],\,\gQ^J\,\theta_J\bigr] 
 ={}& \ii\,\brtheta_I\,\hat{\gamma}^{\Loa}\,\Bigl(\frac{1}{4}\,\delta^{IJ}\,\omega^{\Loc\Lod}\, \gamma_{\Loc\Lod} 
  + \frac{\ii}{2}\,\epsilon^{IJ}\, e^{\Lob}\, \hat{\gamma}_{\Lob} \Bigr)\,\theta_J\,\gP_{\Loa} 
\\
 &+ \frac{1}{4}\,\epsilon^{IK}\, \brtheta_I\, \gamma^{\Loc\Lod}\,\Bigl(\frac{1}{4}\,\delta^{KJ}\,\omega^{\Loa\Lob}\, \gamma_{\Loa\Lob} 
  + \frac{\ii}{2}\,\epsilon^{KJ}\, e^{\Loa}\, \hat{\gamma}_{\Loa} \Bigr)\,\theta_J\,\rmd X^m\, R_{\Loc\Lod}{}^{\Loe\Lof}\,\gJ_{\Loe\Lof} \\
 & + \text{(irrelevant terms proportional to the central charge $Z$)}\,.
\end{aligned}
\label{A0QQ}
\end{equation}
Here, $R_{\Loa\Lob\Loc\Lod}$ is the Riemann tensor in the tangent space of the $\AdS{5} \times \rmS^5$ background.
For the derivation of (\ref{A0QQ}),
we have used $\delta^{IJ}\,\brtheta_I\,\hat{\gamma}^{\Loa}\,\rmd \theta_J = 0$ and
$\epsilon^{IJ}\,\brtheta_I\,\gamma^{\Loa\Lob}\,\rmd \theta_J = 0$\,.

From the above calculations, the left-invariant current $A$ up to second order in $\theta$ becomes
\begin{equation}
  \begin{aligned}
 A ={}& \Bigl(e^{\Loa}+\frac{\ii}{2}\,\brtheta_I\,\hat{\gamma}^{\Loa}\,D^{IJ}\theta_J\Bigr)\,\gP_{\Loa} 
-\frac{1}{2}\,\Bigl(\omega^{\Loa\Lob}-\frac{1}{4}\,\epsilon^{IK}\,\brtheta_I\,\gamma^{\Loc\Lod}\,R_{\Loc\Lod}{}^{\Loa\Lob}\,D^{KJ}\theta_J\Bigr)\,\gJ_{\Loa\Lob} \\
 & +\gQ^I\,D^{IJ}\theta_J + \cO(\theta^3)\,,
\end{aligned}
\label{eq:A-ex}
\end{equation}
where we defined the differential operator
\begin{align}
 D^{IJ} &\equiv \delta^{IJ}\,\Bigl(\rmd + \frac{1}{4}\,\omega^{\Loa\Lob}\,\gamma_{\Loa\Lob}\Bigr)+\frac{\ii}{2}\,\epsilon^{IJ}\,e^{\Loa}\,\hat{\gamma}_{\Loa}\,.
\end{align}
In particular, $A_{(1)}$\,, $A_{(2)}$ are given by
\begin{align}
 A_{(1)} &= \gQ^I\,D^{IJ}\theta_J\,, \\
A_{(2)}&=\frac{\ii}{2}\,\brtheta_I\,\hat{\gamma}^{\Loa}\,D^{IJ}\theta_J\,\gP_{\Loa} 
+\frac{1}{8}\,\epsilon^{IK}\,\brtheta_I\,\gamma^{\Loc\Lod}\,R_{\Loc\Lod}{}^{\Loa\Lob}\,D^{KJ}\theta_J\,\gJ_{\Loa\Lob}\,.
\end{align}

\subsubsection{Evaluation of the bi-linear current part.}

Using the expansion \eqref{eq:A-ex} we obtain
\begin{align}
 \frac{1}{2}\,\str\bigl[\,A_{\WSa}\,d_-(A_{\WSb})\,\bigr]
 &= \eta_{\Loa\Lob}\,e_{\WSa}{}^{\Loa}\,e_{\WSb}{}^{\Lob} 
  + \ii\, \bigl[\,e_{\WSb}{}^{\Loa}\, (\brtheta_1\,\hat{\gamma}_{\Loa}\, \partial_{\WSa} \theta_1) 
  + e_{\WSa}{}^{\Loa}\, (\brtheta_2\,\hat{\gamma}_{\Loa}\, \partial_{\WSb} \theta_2) \,\bigr]
\no\\
 &\quad + \frac{\ii}{4}\,\Bigl[e_{\WSb}{}^{\Lob}\, e_{\WSa}{}^{\Loa}\,\omega_{\Loa}{}^{\Loc\Lod} \,(\brtheta_1\,\hat{\gamma}_{\Lob}\,\gamma_{\Loc\Lod} \,\theta_1)
 + e_{\WSa}{}^{\Loa}\, e_{\WSb}{}^{\Lob}\, \omega_{\Lob}{}^{\Loc\Lod} \, (\brtheta_2\,\hat{\gamma}_{\Loa}\,\gamma_{\Loc\Lod}\,\theta_2) \Bigr]
\no\\
 &\quad - e_{\WSb}{}^{\Loa} \,e_{\WSa}{}^{\Lob}\, \brtheta_1\,\hat{\gamma}_{\Loa}\, \hat{\gamma}_{\Lob}\,\theta_2+\cO(\theta^3)\,,
\end{align}
where $e_{\WSa}{}^{\Loa}\equiv e_m{}^{\Loa}\,\partial_{\WSa}X^m$\,. 
Further using \eqref{eq:lift-32-AdS5-1}, \eqref{eq:lift-32-AdS5-2}, and \eqref{eq:lift-32-AdS5-3}, we obtain
\begin{align}
 \frac{1}{2} \STr[ A_{\WSa} d_-(A_{\WSb})]
 &= \CG_{mn}\,\partial_{\WSa} X^m\,\partial_{\WSb} X^n 
  + \ii\, \bigl[\,e_{\WSb}{}^{\Loa}\, \brTheta_1\,\Gamma_{\Loa}\, \partial_{\WSa} \Theta_1 
  + e_{\WSa}{}^{\Loa}\, \brTheta_2\,\Gamma_{\Loa}\, \partial_{\WSb} \Theta_2 \,\bigr]
\no\\
 &\quad + \frac{\ii}{4}\,\Bigl[e_{\WSb}{}^{\Lob}\, e_{\WSa}{}^{\Loa}\,\omega_{\Loa}{}^{\Loc\Lod} \, \brTheta_1\,\Gamma_{\Lob}\,\Gamma_{\Loc\Lod} \,\Theta_1 
 + e_{\WSa}{}^{\Loa}\, e_{\WSb}{}^{\Lob}\, \omega_{\Lob}{}^{\Loc\Lod} \, \brTheta_2\,\Gamma_{\Loa}\,\Gamma_{\Loc\Lod}\,\Theta_2 \Bigr]
\no\\
 &\quad - \frac{\ii}{8}\,e_{\WSb}{}^{\Loa} \,e_{\WSa}{}^{\Lob}\, \brTheta_I\,\Gamma_{\Loa}\,\bisF_5\,\Gamma_{\Lob}\,\Theta_J
+\cO(\Theta^3)\,,
\label{STrAA}
\end{align}
where $\bisF_5$ is a bispinor
\begin{align}
 \bisF_5 \equiv\frac{1}{5!}\,e^{\phi}\hat{F}_{\Loa_1\cdots \Loa_5}\,\Gamma^{\Loa_1\cdots \Loa_5} =4\,\bigl(\Gamma^{01234}+\Gamma^{56789}\bigr)\,. \label{eq:F5-bispinor}
\end{align}
This describes the R--R $5$-form field strength in the tangent space of the $\AdS{5} \times \rmS^5$ background.

\medskip

The expression (\ref{STrAA}) implies that the action \eqref{AdS5S5-action} takes the canonical form (\ref{eq:GS-action-canonical-YBsec}) of the \ac{gs} action.
Therefore, the target space of the action \eqref{AdS5S5-action} is
the familiar $\AdS{5} \times \rmS^5$ background with the RR $5$-form
\begin{align}
 \rmd s^2 &=\rmd s_{\AdS{5}}^2+\rmd s_{\rmS^5}^2\,, \\
\Exp{\Phi}\hat{F}_5 &= 4\,\bigl(\omega_{\AdS5}+\omega_{\rmS^5}\bigr) \,.
\label{eq:uF5}
\end{align}
Under the parametrization (\ref{eq:group-parameterization}) of $g_{b}$, the metrics of $\AdS5$ and $\rmS^5$ are
\begin{align}
\label{eq:AdS5S5-metric}
\rmd s_{\AdS{5}}^2&=\frac{-(\rmd x^0)^2+(\rmd x^1)^2+(\rmd x^2)^2+(\rmd x^3)^2 + \rmd z^2}{z^2}\,,\\
\rmd s_{\rmS^5}^2&=\rmd r^2 + \sin^2 r\, \rmd\xi^2 + \cos^2\xi\,\sin^2 r\, \rmd\phi_1^2 + \sin^2r\,\sin^2\xi\, \rmd\phi_2^2 + \cos^2r\, \rmd\phi_3^2\,,
\end{align}
and the volume forms $\omega_{\AdS5}$, $\omega_{\rmS^5}$ of $\AdS5$ and $\rmS^5$ are given by
\begin{align}
 &\omega_{\AdS5} \equiv - \frac{\rmd x^0\wedge\rmd x^1\wedge \rmd x^2\wedge\rmd x^3\wedge\rmd z}{z^5}\,, 
\\
 &\omega_{\rmS^5} \equiv \sin^3r \cos r \sin\xi \cos\xi\,\rmd r\wedge \rmd\xi\wedge\rmd \phi_1\wedge \rmd\phi_2\wedge\rmd \phi_3\qquad (\omega_{\AdS{5}} = *_{10}\omega_{\rmS^5}) \,. 
\end{align}
In the following discussion, we set the dilaton to zero, $\Phi=0$\,.

\subsection{Killing vectors}

For later use, we calculate the Killing vectors $\hat{T}_i\equiv\hat{T}_i^m\,\partial_m$ associated to the bosonic symmetries $T_i$ of the $\AdS{5}$ background. 
The Killing vectors on this background can be expressed as (see Appendix C in~\cite{Sakamoto:2018krs} for more details)
\begin{align}
 \hat{T}_{i}=\hat{T}_{i}{}^{m}\,\partial_{m} 
 =\bigl[\Ad_{g_{\bos}^{-1}}\bigr]_{i}{}^{\Loa}\, e_{\Loa}{}^{m}\, \partial_{m}
 = \str\bigl(g_{\bos}^{-1}\,T_{i}\,g_{\bos}\,\gP_{\Loa}\bigr)\,e^{\Loa m}\, \partial_{m}\,,
\label{eq:Killing-Formula}
\end{align}
where we introduced the notation $g\,T_i\,g^{-1}\equiv [\Ad_{g}]_{i}{}^{j}\,T_j$\,. 
Using our parametrization~\eqref{eq:group-parameterization}, the Killing vectors on the $\AdS{5}$ background are given by
\begin{equation}
  \begin{aligned}
 \hat{P}_\mu &\equiv \str\bigl(g_{\bos}^{-1}\,P_{\mu}\,g_{\bos}\, \gP_{\Loa} \bigr)\,e^{\Loa m}\,\partial_m = \partial_\mu \,,
\\
 \hat{K}_\mu &\equiv \str\bigl(g_{\bos}^{-1}\,K_{\mu}\,g_{\bos}\, \gP_{\Loa} \bigr)\,e^{\Loa m}\,\partial_m = \bigl(x^\nu\,x_\nu +z^2\bigr)\,\partial_\mu - 2\,x_\mu\,(x^\nu\,\partial_\nu+z\,\partial_z)\,,
\\
 \hat{M}_{\mu\nu} &\equiv \str\bigl(g_{\bos}^{-1}\,M_{\mu\nu}\,g_{\bos}\, \gP_{\Loa} \bigr)\,e^{\Loa m}\,\partial_m = x_\mu\,\partial_\nu -x_\nu\,\partial_\mu \,,
\\
 \hat{D} &\equiv \str\bigl(g_{\bos}^{-1}\,D\,g_{\bos}\, \gP_{\Loa} \bigr)\,e^{\Loa m}\,\partial_m = x^\mu\,\partial_\mu + z\,\partial_z \,.
\end{aligned}
\end{equation}
The Lie brackets of these vector fields satisfy the same commutation relations~\eqref{eq:so(2-4)-algebra} as the conformal algebra $\alg{so}(2,4)$ (with negative sign, $[\hat{T}_i,\,\hat{T}_j]=-f_{ij}{}^k\,\hat{T}_k$):
\begin{equation}
  \begin{aligned}
 [\hat{P}_\mu,\, \hat{K}_\nu] &= -2\,\bigl(\eta_{\mu\nu}\, \hat{D} - \hat{M}_{\mu\nu}\bigr)\,, \\
 [\hat{D},\, \hat{P}_{\mu}] &= -\hat{P}_\mu\,, \\
  [\hat{D},\,\hat{K}_\mu] &= \hat{K}_\mu\,,
\\
 [\hat{M}_{\mu\nu},\, \hat{P}_\rho] &= -\eta_{\mu\rho}\, \hat{P}_\nu +\eta_{\nu\rho}\, \hat{P}_\mu \,, \\
 [\hat{M}_{\mu\nu},\, \hat{K}_\rho] &= -\eta_{\mu\rho}\,\hat{K}_\nu + \eta_{\nu\rho}\,\hat{K}_\mu\,, &
\\
 [\hat{M}_{\mu\nu},\,\hat{M}_{\rho\sigma}] &= -\eta_{\mu\rho}\,\hat{M}_{\nu\sigma}+\eta_{\mu\sigma}\,\hat{M}_{\nu\rho} + \eta_{\nu\rho}\,\hat{M}_{\mu\sigma}-\eta_{\nu\sigma}\,\hat{M}_{\mu\rho}\,.
\end{aligned}
\end{equation}

\section{Yang--Baxter deformations of the $\AdS{5} \times \rmS^5$ superstring}
\label{sec:YB-sigma}

Enlarging the reach of integrability techniques to other models brings many calculational advantages. A way of doing this is finding deformations of integrable models which retain the property of integrability. Being able to do so systematically instead of working case-by-case is a great advantage. The \ac{yb} deformation provides a systematic way of generating integrable deformations of the $\AdS{5} \times \rmS^5$ superstring. This section will give a comprehensive introduction to this topic.

\subsection{The action of the YB-deformed $\AdS{5} \times \rmS^5$ superstring}\label{subsec:YB-sigma-action}

The action of the \ac{yb}-deformed $\AdS{5} \times \rmS^5$ superstring is given by~\cite{Delduc:2013qra,Delduc:2014kha,Kawaguchi:2014qwa}
\begin{align}
 S_{\YB}=-\frac{T(1-c^2\eta^2)}{2}\int \rmd^2\sigma\,\Pg_{-}^{\WSa\WSb}\, \str\bigl[A_{\WSa}\, \hat{d}_-\circ\cO_-^{-1}(A_{\WSb})\bigr]\,, 
\label{eq:YBsM}
\end{align}
where $c^2$ is a real parameter, $\eta\in\mathbb{R}$ is a deformation parameter and $\hat{d}_{\pm}$ are the modified projection operators
\begin{align}
 \hat{d}_{\pm} &\equiv \mp P^{(1)}+2\hat{\eta}^{-2}\,P^{(2)}\pm P^{(3)} \,, & \hat{\eta} &= \sqrt{1+c^2\eta^2}\,.
\label{eq:YBdpm}
\end{align}
The linear operators $\cO_{\pm}$ are defined by
\begin{align}
 \cO_{\pm}&\equiv 1\pm\eta\, R_g\circ \hat{d}_\pm\,.
\label{eq:linearOpm-def}
\end{align}
For $\eta=0$ the deformed action (\ref{eq:YBsM}) reduces to the undeformed $\AdS{5}\times \rmS^5$ superstring sigma model action \eqref{AdS5S5-action}.

\paragraph{$R$-operator and classical $r$-matrix.}

A key ingredient of the \ac{yb} deformation is the $R$-operator
which is a skew-symmetric linear operator $R:\alg{g} \to \alg{g}$
and solves the \ac{cybe},
\begin{align}
\begin{split}
 \CYBE(X,Y) &\equiv [R(X),\,R(Y)] - R([R(X),\,Y]+[X,\,R(Y)])\\
&=-c^2\,[X,Y]\,,\qquad X,\,Y \in\alg{g}\,.
\label{eq:CYBE}
\end{split}
\end{align}
The dressed $R$-operator $R_g$ is defined by
\begin{align}
 R_g(X):= g^{-1}\, R(g\,X\,g^{-1})\,g ={\rm Ad}_g^{-1}\circ R \circ {\rm Ad}_g(X)\,,\qquad g\in SU(2,2|4)\,.
\end{align}
The operator $R_g$ is also a solution of the \ac{cybe} \eqref{eq:CYBE},
\begin{align}
 \CYBE_g(X,Y) \equiv [R_g(X),\,R_g(Y)] - R_g([R_g(X),\,Y]+[X,\,R_g(Y)])=-c^2\,[X,Y]\,,
\label{eq:CYBE-g}
\end{align}
if the linear operator $R$ satisfies the \ac{cybe}.
This is easily seen from the relation $\CYBE_g(X,Y)={\rm Ad}_g^{-1}\CYBE({\rm Ad}_g(X),{\rm Ad}_g(Y))$\,.

It is useful to rewrite the $R$-operator in tensorial notation.
Then, the $R$-operator can be expressed via a skew-symmetric classical $r$-matrix $r\in \mathfrak{g}\otimes \mathfrak{g}$\,.
Introducing the $r$-matrix
\begin{align}
 r&=\frac{1}{2}\,r^{ij}\,T_{i}\wedge T_{j}\,, & r^{ij}&=-r^{ji}\,, & T_{i}&\in \alg{g}\,,
\end{align}
the action of the $R$-operator can be defined as 
\begin{align}
 R(X) &= r^{ij}\, T_{i}\,\str(T_{j}\,X) \,, & X& \in\alg{g} \,.
\label{eq:R-operator}
\end{align}
This allows us to encode \ac{yb} deformations by classical $r$-matrices.

\paragraph{Classification of the CYBE.}

The \ac{cybe} can be of three types: 
\begin{enumerate}
	\item $c^2<0$\,,
	\item $c^2=0$\,,
	\item $c^2>0$\,.
\end{enumerate}
The \ac{cybe} with $c^2\neq 0$ is called the \emph{\ac{mcybe}}.
\ac{yb} deformations of principal chiral models with $c^2<0$ were originally developed by Klimcik~\cite{Klimcik:2002zj}, and the integrability of the deformed models was shown in~\cite{Klimcik:2008eq}.
These deformations were generalized to symmetric coset sigma models~\cite{Delduc:2013fga} and the AdS$_5\times$S$^5$ superstring~\cite{Delduc:2013qra,Delduc:2014kha}.
\ac{yb} deformations based on the \ac{mcybe} with $c^2<0$ are called $q$-deformations.\footnote{The $q$-deformed AdS$_5\times$S$^5$ background is also called the $\eta$-deformed AdS$_5\times$S$^5$ or the ABF background~\cite{Arutyunov:2013ega,Arutyunov:2015qva}. }
A typical solution of the \ac{mcybe} is the Drinfeld--Jimbo type $r$-matrix~\cite{Drinfeld:1985rx,Jimbo:1985zk},
\begin{align}
r_{\rm DJ}=c\,\sum_{1\leq i<j\leq 8}E_{ij}\wedge E_{ji}(-1)^{\bar{\imath}\bar{j}}\,,
\label{eq:DJ-r}
\end{align}
where $E_{ij}\,(i,j=1,\dots ,8)$ are the $\mathfrak{gl}(4|4)$ generators and the super skew-symmetric symbol is defined by
\begin{align}
E_{ij}\wedge E_{kl}\equiv E_{ij}\otimes E_{kl}-E_{kl}\otimes E_{ij}(-1)^{(\bar{\imath}+\bar{j})(\bar{k}+\bar{l})}\,.
\end{align}
Here we determined the parity of the indices as $\bar{\imath}=0$ for $i=1,\dots 4$ and $\bar{\imath}=1$ for $i=5,\dots 8$\,.
The associated $R$-operator acts on
\begin{align}
R(E_{ij})=
\begin{cases}
+c\,E_{ij}\qquad \text{if}\,\,\,i<j\\
0\qquad\quad\hspace{3mm}\text{if}\,\,\,i=j\\
-c\,E_{ij}\qquad \text{if}\,\,\,i>j
\end{cases}
\,.
\end{align}
The $r$-matrix (\ref{eq:DJ-r}) with $c^2<0$ was used for a $q$-deformation of the AdS$_5\times$S$^5$ superstring~\cite{Delduc:2013qra,Delduc:2014kha}\footnote{In~\cite{Hoare:2018ngg}, by using Drinfeld--Jimbo type r-matrices with different fermionic structures, other q-deformed
AdS$_2\times$S$^2\times T^6$ and AdS$_5\times$S$^5$ backgrounds were constructed and shown to be solutions of standard supergravity.}.
We often normalize the complex parameter $c$ as $c=i$\,.
The full explicit expression (\ref{eq:ABF-bg}) of the $q$-deformed AdS$_5\times$S$^5$ background is given in~\cite{Arutyunov:2015qva}.
Remarkably, this deformed background does not solve the standard supergravity equations but the \ac{gse}~\cite{Arutyunov:2015mqj}.
For the case $c^2>0$, the associated \ac{yb} deformations of the AdS$_5\times$S$^5$ superstring have been studied in~\cite{Hoare:2016ibq}.

\medskip

The second class, $c=0$, is frequently called the \emph{homogeneous} \ac{cybe}.
In terms of the $r$-matrix, the homogeneous \ac{cybe} \eqref{eq:CYBE} can be rewritten as
\begin{align}
 f_{l_1l_2}{}^i\,r^{jl_1}\,r^{kl_2} + f_{l_1l_2}{}^j\,r^{kl_1}\,r^{il_2} + f_{l_1l_2}{}^k\,r^{il_1}\,r^{jl_2} =0\,,
\label{eq:CYBE-r}
\end{align}
where $f_{ij}{}^k$ are the structure constants $[T_i,\,T_j]=f_{ij}{}^k\,T_k$ of $\mathfrak{g}$\,.
The homogeneous \ac{yb} deformations of principal sigma models and symmetric coset sigma models had been developed in~\cite{Matsumoto:2015jja}.
Moreover, it had been generalized to the AdS$_5\times$S$^5$ superstring case in~\cite{Kawaguchi:2014qwa}\footnote{The action of the homogeneous \ac{yb}-deformed $\AdS{5} \times \rmS^5$ superstring was constructed using the pure spinor formalism in~\cite{Benitez:2018xnh}.}.
A remarkable feature of this class is that we can consider partial deformations of a given background.
This is is due to the fact that the right-hand side of (\ref{eq:CYBE-r}) has no term proportional to $c^2$.
Thanks to this, we can find many nontrivial solutions of this equation.
In particular, as we will see later on, the associated deformations give deformed AdS$_5\times$S$^5$ backgrounds which solve not only the standard supergravity equations but also the \ac{gse}.

\subsection{Classical integrability}\label{sec:YBd_classInt}

In this subsection, we will show that also the deformed action (\ref{eq:YBsM}) admits a Lax pair.
Therefore, \ac{yb} deformations are \emph{integrable} deformations of the $\AdS{5} \times \rmS^5$ superstring.
To show this, we will explicitly give the Lax pair of the deformed system.

\paragraph{\ac{eom} and the flatness condition.}

To demonstrate the classical integrability of (\ref{eq:YBsM}),
we give the \ac{eom} of the deformed action (\ref{eq:YBsM}).
For this purpose, it is useful to introduce the deformed and the projected currents,
\begin{align}
J_{\alpha}&=\cO_-^{-1}A_{\alpha}\,, &
\tilde{J}_{\alpha}&=\cO_+^{-1}A_{\alpha}\,,\label{eq:deformed-c1}
\\
J_{(\pm)}^{\alpha}&=P_{\pm}^{\alpha\beta}J_{\beta}\,, &
\tilde{J}_{(\pm)}^{\alpha}&=P_{\pm}^{\alpha\beta}\tilde{J}_{\beta}\,.
\end{align}
Then, the \ac{eom} of the deformed action (\ref{eq:YBsM}) are given by
\begin{align}
\bar{\cE}=\cD_\alpha \hat{d}_-(J^\alpha_{(-)})+\cD_\alpha \hat{d}_+(\tilde{J}_{(+)}^\alpha)
+[\tilde{J}_{(+)\alpha}\,,\hat{d}_-(J_{(-)}^\alpha)]+[J_{(-)\alpha}\,,\hat{d}_+(\tilde{J}_{(+)}^\alpha)]=0\,.
\label{mEOM}
\end{align}
The flatness condition for the left-invariant current is 
\begin{align}
\bar{\cZ}&=\frac{1}{2}\epsilon^{\alpha\beta}(\cD_\alpha A_\beta-\cD_\beta A_\alpha+[A_\alpha\,,A_\beta])\nonumber \\
&=\cD_\alpha \tilde{J}_{(+)}^\alpha-\cD_\alpha J_{(-)}^\alpha+[J_{(-)\alpha}\,,\tilde{J}_{(+)}^\alpha]
-c^2\,\eta^2[\hat{d}_-(J_{(-)\alpha}), \hat{d}_+(\tilde{J}^{\alpha}_{(+)})]+\eta R_g(\bar{\cE})=0\,.
\label{mflat}
\end{align}
As in the undeformed case,
we decompose the \ac{eom} (\ref{mEOM}) and the flatness condition (\ref{mflat}) into the $\mathbb{Z}_4$-graded components.
The bosonic parts are
\begin{align}
\bar{\mathsf{B}}_1&:=\bar{\Z}^{(0)}=\cD_\alpha \tilde{J}_+^{\alpha(0)}-\cD_\alpha J_-^{\alpha(0)}
+[J_{-\alpha}^{(0)}\,,\tilde{J}_+^{\alpha(0)}]
+\left(\frac{1-c^2\eta^2}{1+c^2\eta^2}\right)^2[J_{-\alpha}^{(2)}\,,\tilde{J}_+^{\alpha(2)}]\no \\
&\qquad\qquad
+(1-c^2\eta^2)\left([J_{-\alpha}^{(1)}\,,\tilde{J}_+^{\alpha(3)}]+[J_{-\alpha}^{(3)}\,,\tilde{J}_+^{\alpha(1)}]\right)=0\,,\qquad \\ 
\bar{\mathsf{B}}_2&:=\frac{1}{4}(\bar{\cE}^{(2)}+2\,\bar{\Z}^{(2)})
=\cD_\alpha \tilde{J}_+^{\alpha(2)}+[J_{-\alpha}^{(0)}\,,\tilde{J}_+^{\alpha(2)}]
+(1+c^2\eta^2)[J_{-\alpha}^{(3)}\,,\tilde{J}_+^{\alpha(3)}]=0\,,\\
\bar{\mathsf{B}}_3&:=\frac{1}{4}(\bar{\cE}^{(2)}-2\,\bar{\Z}^{(2)})
=\cD_\alpha J_-^{\alpha(2)}-[J_{-\alpha}^{(2)}\,,\tilde{J}_+^{\alpha(0)}]-
(1+c^2\eta^2)[J_{-\alpha}^{(1)}\,,\tilde{J}_+^{\alpha(1)}]=0\,,
\label{bmEOM}
\end{align}
and the fermionic parts are given by
\begin{align}
\bar{\mathsf{F}}_1&:=\frac{1}{4}(3\bar{\Z}^{(1)}-\bar{\cE}^{(1)})=\cD_\alpha \tilde{J}_{(+)}^{\alpha(1)}-\cD_\alpha J_{(-)}^{\alpha(1)}+[J_{(-)\alpha}^{(0)}\,,\tilde{J}_{(+)}^{\alpha(1)}]+[J_{(-)\alpha}^{(1)}\,,\tilde{J}_{(+)}^{\alpha(0)}]\no\\
&\qquad\qquad\qquad\qquad+\frac{1-c^2\eta^2}{1+c^2\eta^2}[J_{(-)\alpha}^{(2)}\,,\tilde{J}_{(+)}^{\alpha(3)}]=0\,,\\
\bar{\mathsf{F}}_2&:=\frac{1}{4}(3\bar{\Z}^{(3)}+\bar{\cE}^{(3)})=\cD_\alpha \tilde{J}_{(+)}^{\alpha(3)}-\cD_\alpha J_{(-)}^{\alpha(3)}+[J_{(-)\alpha}^{(0)}\,,\tilde{J}_{(+)}^{\alpha(3)}]+[J_{(-)\alpha}^{(3)}\,,\tilde{J}_{(+)}^{\alpha(0)}]\no\\
&\qquad\qquad\qquad\qquad+\frac{1-c^2\eta^2}{1+c^2\eta^2}[J_{(-)\alpha}^{(1)}\,,\tilde{J}_{(+)}^{\alpha(2)}]=0\,,\\
\bar{\mathsf{F}}_3&:=\frac{1}{4}(\bar{\cE}^{(1)}+\bar{\Z}^{(1)})
=[J_{(-)\alpha}^{(3)}\,,\tilde{J}_{(+)}^{\alpha(2)}]=0\,,\\ 
\bar{\mathsf{F}}_4&:=\frac{1}{4}(-\bar{\cE}^{(3)}+\bar{\Z}^{(3)})
=[J_{(-)\alpha}^{(2)}\,,\tilde{J}_{(+)}^{\alpha(1)}]=0\,.
\label{fmEOM}
\end{align}

\paragraph{Construction of the Lax pair.}

Now, let us present the Lax pair of the \ac{yb}-deformed AdS$_5\times$S$^5$ superstring.
It is given by~\cite{Delduc:2013qra,Kawaguchi:2014qwa}
\begin{align}
\bar{\cL}_\alpha=\bar{L}_{(+)\alpha}+\bar{M}_{(-)\alpha}\,,
\label{eta-Lax}
\end{align}
where $\bar{L}^\alpha_{(+)}$ and $\bar{M}^\alpha_{(-)}$ are
\begin{align}
\bar{L}^\alpha_{(+)}=&\tilde{J}_{(+)}^{\alpha(0)}+u\sqrt{1-c^2\eta^2}\tilde{J}_{(+)}^{\alpha(1)}
+u^{-2}\frac{1-c^2\eta^2}{1+c^2 \eta^2}\tilde{J}_{(+)}^{\alpha(2)}+u^{-1}\sqrt{1-c^2\eta^2}\tilde{J}_{(+)}^{\alpha(3)}\no \\
\bar{M}^\alpha_{(-)}=&J_{(-)}^{\alpha(0)}+u \sqrt{1-c^2\eta^2}J_{(-)}^{\alpha(1)}+u^2\frac{1-c^2\eta^2}{1+c^2\eta^2}J_{(-)}^{\alpha(2)}+u^{-1}\sqrt{1-c^2\eta^2}J_{(-)}^{\alpha(3)}\,.
\end{align}
The expression has a similar form as in the undeformed case (\ref{uLax}).
We now evaluate the flatness condition of the Lax pair,
\begin{align}
\epsilon^{\alpha\beta}\left(\partial_{\alpha} \bar{\cL}_{\beta}-\partial_{\beta} \bar{\cL}_{\alpha}+[\bar{\cL}_{\alpha}\,,\bar{\cL}_{\beta}]\right)=0\,.
\label{YB-flat}
\end{align}
The left-hand side of this equation can be rewritten as
\begin{align}
{\rm LHS\,~of}\,~(\ref{YB-flat})=&u^0\,\bar{\mathsf{B}}_1+u^{-2}\,\frac{1-c^2\eta^2}{1+c^2\eta^2}\bar{\mathsf{B}}_2-u^2\,\frac{1-c^2\eta^2}{1+c^2\eta^2}\bar{\mathsf{B}}_3\no \\
&+\sqrt{1-c^2\eta^2}\left(u\,\bar{\mathsf{F}}_1+u^{-1}\bar{\mathsf{F}}_2
+u^{-3}\,\frac{1-c^2\eta^2}{1+c^2\eta^2}\bar{\mathsf{F}}_3+u^3\,\frac{1-c^2\eta^2}{1+c^2\eta^2}\bar{\mathsf{F}}_4\right) \,.
\end{align}
Therefore, the flatness condition (\ref{YB-flat}) is equivalent to the \ac{eom} (\ref{mEOM}) and the flatness condition (\ref{mflat})  on-shell.
We see that indeed, also
the deformed system (\ref{eq:YBsM}) is classically integrable.

\subsection{The $\kappa$-symmetry of the YB-deformed action}
\label{sec:kappa-YB}

The deformed action (\ref{eq:YBsM}) is also invariant under the $\kappa$-symmetry transformation\cite{Delduc:2013qra,Kawaguchi:2014qwa},
as we will show in the following.

The $\kappa$-symmetry transformation is given by~\cite{Delduc:2013qra,Kawaguchi:2014qwa}
\begin{align}
\mathcal{O}_-^{-1}g^{-1}\delta_{\kappa}g&=
P^{\alpha\beta}_-\{\gQ^1\kappa_{1\alpha},J_{-\beta}^{(2)}\}
+P^{\alpha\beta}_+\{\gQ^2\kappa_{2\alpha},J_{+\beta}^{(2)}\}
\,, \label{kappa1} \\
\delta_\kappa(\sqrt{-\ga} \ga^{\alpha\beta})&=
\frac{1+c^2 \eta^2}{4}\sqrt{-\ga}\,{\rm Str}\biggl[\Upsilon\Bigl([\gQ^1\kappa^\alpha_{1(+)},J_{+(+)}^{(1)\beta}]
+[\gQ^2\kappa^\alpha_{2(-)},J_{-(-)}^{(3)\beta}]\Bigr)+(\alpha\leftrightarrow \beta)\biggr]\,.
\label{kappa2}
\end{align}
It is easy to see that when we take $\eta=0$\,, this expression reduces to the undeformed transformation (\ref{ukappa1}), (\ref{ukappa2}).
As in the undeformed case,
we decompose the variation of the deformed action (\ref{eq:YBsM}) under the $\kappa$-symmetry transformation as
\begin{align}
\delta_\kappa S_{\YB}\equiv \delta_gS_{\YB}+\delta_\ga S_{\YB}\,,
\end{align}
where $\delta_g S_{\YB}$ and $\delta_\ga S_{\YB}$ are the variations with respect to the group element and the world-sheet metric, respectively.
Let us first consider $\delta_gS_{\YB}$\,.
By using (\ref{kappa1}), it is given by
\begin{align}
\delta_g S_{\YB}&=\frac{T(1-c^2\eta^2)}{2}\,\int \rmd^2\sigma\,
\sqrt{-\gamma}\,\str\left[\epsilon^{(1)}P_3\circ(1+\eta R_g)(\bar{\cE})
+\epsilon^{(3)}P_1\circ(1-\eta R_g)(\bar{\cE})\right]\no\\
&=-2T(1-c^2\eta^2)\,\int \rmd^2\sigma\,\sqrt{-\gamma}\,\str\left(\epsilon^{(1)}[J_{-\alpha}^{(2)}\,,\tilde{J}_+^{\alpha(1)}]+\epsilon^{(3)}[\tilde{J}_{+\alpha}^{(2)}\,,J_-^{\alpha(3)}]\right)\,,
\label{eq:kappa-g-vari}
\end{align}
where $\epsilon^{(1)}$ and $\epsilon^{(3)}$ are
\begin{align}
\epsilon^{(1)}&=(1+\eta R_g)P^{\alpha\beta}_-\{\gQ^1\kappa_{1\alpha},J_{-\beta}^{(2)}\}\,,&
\epsilon^{(3)}&=(1-\eta R_g)P^{\alpha\beta}_+\{\gQ^2\kappa_{2\alpha},J_{+\beta}^{(2)}\}\,.
\label{mkappaan}
\end{align}
In the second equation of (\ref{eq:kappa-g-vari}), we have used 
\begin{align}
P_1\circ(1-\eta R_g)(\bar{\cE})&=-4[\tilde{J}_{+\alpha}^{(2)}\,,J_-^{\alpha(3)}]-\bar{\cZ}^{(1)}\,,\\
P_3\circ(1+\eta R_g)(\bar{\cE})&=-4[J_{-\alpha}^{(2)}\,,\tilde{J}_+^{\alpha(1)}]+\bar{\cZ}^{(3)}
\end{align}
and ignored the flatness condition $\bar{\mathcal{Z}}$\,.
Each of the terms in (\ref{eq:kappa-g-vari}) can be rewritten as
\begin{align}
\begin{split}
\str\left(\epsilon^{(1)}[J_{-\alpha}^{(2)}\,,\tilde{J}_+^{\alpha(1)}]\right)
=\str\left(J_{-\alpha}^{(2)}J_{-\beta}^{(2)}[\tilde{J}_{+\alpha}^{(1)}\,,\gQ^1\kappa_+^{\beta}]\right)\,, \\
\str\left(\epsilon^{(3)}[\tilde{J}_{+\alpha}^{(2)}\,,J_-^{\alpha(3)}]\right)
=\str\left(\tilde{J}_{+\alpha}^{(2)}\tilde{J}_{+\beta}^{(2)}[J_{-\alpha}^{(3)}\,,\gQ^2\kappa^{\beta}_-]\right)\,.
\end{split}
\end{align}
By using equation (\ref{symformula}), the variation $\delta_g\,S_{\YB}$ becomes
\begin{align}
\delta_g S_{\YB}&=\frac{T(1-c^2\eta^2)}{4}\,\int \rmd^2\sigma\,
\sqrt{-\gamma}\,\biggl[\str\left(J^{(2)}_{\alpha-}J^{(2)}_{\beta-}\right)\str\left([\gQ^1\kappa_+^{\beta(1)}\,,\tilde{J}_+^{\alpha}]\right)\no \\
&\qquad\qquad\qquad\qquad+\str\left(\tilde{J}^{(2)}_{\alpha+}\tilde{J}^{(2)}_{\beta+}\right)
\str\left([\gQ^2\kappa_-^{\beta}\,,J_-^{\alpha(3)}]\right)\biggr]\,.
\label{gmkappa}
\end{align}

\medskip

Next, let us consider the variation $\delta_{\gamma}\,S_{\YB}$.
For this purpose, it is convenient to rewrite the deformed action (\ref{eq:YBsM}) in terms of the deformed current (\ref{eq:deformed-c1}) as
\begin{align}
S_{\YB}&=-\frac{\dlT}{2}\left(\frac{1-c^2\eta^2}{1+c^2\eta^2}\right)
\int \rmd^2\sigma\,\ga^{\WSa\WSb}\, \str\bigl[J_{\WSa}^{(2)}J_{\WSb}^{(2)}\bigr]\no\\
&\quad-\frac{\dlT(1-c^2\eta^2)}{2}\int \rmd^2\sigma\,\epsilon^{\WSa\WSb}\, 
\str\bigl[J_{\WSa}^{(1)}\,J_{\WSb}^{(3)}\bigr]\no\\
&\quad+\eta\frac{\dlT(1-c^2\eta^2)}{4}\int \rmd^2\sigma\,
\epsilon^{\WSa\WSb}\, \str\bigl[\hat{d}_-(J_{\WSa})\, R_{g}\circ \hat{d}_-(J_{\WSb})\bigr]\,.
\end{align}
Using relation (\ref{eq:AB-relation}), $\delta_{\gamma}\,S_{\YB}$ is given by
\begin{align}
\delta_\ga S_{\YB}&=-\frac{T(1-c^2\eta^2)}{4}\,\int \rmd^2\sigma\,\sqrt{-\gamma}\,
\str(J_\alpha^{(2)} J_\beta^{(2)})
\str\left[\Upsilon\left([\gQ^1\kappa_+^{\beta}\,,\tilde{J}_+^{(1)\alpha}]
+[\gQ^2\kappa_-^{\beta}\,,J_-^{(3)\alpha}]\right)\right]\no \\
&=-\frac{T(1-c^2\eta^2)}{4}\,\int \rmd^2\sigma\,\sqrt{-\gamma}\,
\biggl[\str\left(J^{(2)}_{\alpha-}J^{(2)}_{\beta-}\right)
\str\left([\gQ^1\kappa_+^{\beta}\,,\tilde{J}_+^{\alpha(1)}]\right)\no \\
&\qquad\qquad\qquad\qquad+\str\left(\tilde{J}^{(2)}_{\alpha+}\tilde{J}^{(2)}_{\beta+}\right)
\str\left([\gQ^2\kappa_-^{\beta}\,,J_-^{\alpha(3)}]\right)\biggr]\,.
\label{gamkappa}
\end{align}
This obviously cancels out the variation $\delta_g\,S_{\YB}$\,,
\begin{align}
\delta_\kappa S_{\YB}=(\delta_g+\delta_\ga)S_{\YB}=0\,.
\end{align}
As a result, the deformed action (\ref{eq:YBsM}) is $\kappa$-symmetric.

\medskip

Finally, let us comment on the implications of $\kappa$-invariance of the deformed \ac{gs} action.
As shown in~\cite{Wulff:2016tju}, $\kappa$-invariance ensures that all deformed backgrounds are solutions either to the standard supergravity equations or to the \acl{gse}~\cite{Arutyunov:2015mqj,Wulff:2016tju}.
In Sections~\ref{sec:ExampleYBAdS5} and~\ref{sec:YB-T-fold}, we will present \ac{yb}-deformed backgrounds which satisfy the \ac{gse}.

\subsection{YB-deformed backgrounds from the GS action}
\label{subsec:YBfromGS}

In the following, we will rewrite the \ac{yb}-deformed action in the form of the conventional \ac{gs} action.
In order to determine the deformed background, it is sufficient to expand the action up to quadratic order in the fermions,
\begin{align}
 S_{\YB}=S_{(0)}+S_{(2)}+\cO(\theta^4)\,.
\end{align}
The explicit expression of the $q$-deformed $\AdS{5}\times \rmS^5$ background was given in the pioneering work~\cite{Arutyunov:2015qva}.
It was subsequently generalized to the case of the homogeneous \ac{yb} deformations in~\cite{Kyono:2016jqy}.

In this subsection, we provide the general formula for homogeneous \ac{yb} deformed backgrounds.
For simplicity, we limit our analysis to the cases where the $r$-matrices are composed only of the bosonic generators of $\mathfrak{su}(2,2|4)$\,,
\begin{align}
r&=\frac{1}{2}r^{ij}T_{i}\wedge T_{j}\,, & T_{i}&\in \mathfrak{so}(2,4)\times \mathfrak{so}(6)\,,\label{eq:beta-r-matrix}
\end{align}
which is a solution of the homogeneous \ac{cybe}\footnote{The \ac{yb} sigma model action rewritten in the standard \ac{gs} form based on $\kappa$-symmetry was given in~\cite{Borsato:2016ose} to all orders in the fermionic variables. There, also the deformed background associated to a general $r$-matrix was determined.}. 

\subsubsection{Preliminaries.}

To be able to expand the action (\ref{eq:YBsM}) of the \ac{yb} sigma model,
we first need to introduce some notation.
Since the $r$-matrix consists of bosonic generators only,
the dressed $R$-operator $R_{g_{\bos}}$ acts on the generators as
\begin{equation}
\begin{aligned}
 R_{g_{\bos}}(\gP_{\Loa}) &=\lambda_{\Loa}{}^{\Lob}\,\gP_{\Lob}+\frac{1}{2}\,\lambda_{\Loa}{}^{\Lob\Loc}\,\gJ_{\Lob\Loc}\,, \\
 R_{g_{\bos}}(\gJ_{\Loa\Lob}) &=\lambda_{\Loa\Lob}{}^{\Loc}\,\gP_{\Loc}+\frac{1}{2}\,\lambda_{\Loa\Lob}{}^{\Loc\Lod}\,\gJ_{\Loc\Lod}\,, \\
 R_{g_{\bos}}(\gQ^I) &=0\,.
\label{eq:Rg-operation}
\end{aligned}
\end{equation}
The (dressed) $R$-operator is skew-symmetric,
\begin{align}
 \str\bigl[R_{g_{\bos}}(X)\,Y\bigr] = - \str\bigl[X\,R_{g_{\bos}}(Y)\bigr] \,.
\end{align}
If we take $X$ and $Y$ to be $\gP_{\Loa}$ or $\gJ_{\Loa\Lob}$\,, we obtain the relations
\begin{align}
 \lambda_{\Loa\Lob} \equiv \lambda_{\Loa}{}^{\Loc}\,\eta_{\Loc\Lob} &= -\lambda_{\Lob\Loa}\,, &
 \lambda_{\Loa\Lob}{}^{\Loc}&=-\frac{1}{2}\,\eta^{\Loc\Lod}\,R_{\Loa\Lob\Loe\Lof}\,\lambda_{\Lod}{}^{\Loe\Lof}\,, & 
 \lambda_{\Loa\Lob}{}^{\Loe\Lof}\,R_{\Loe\Lof\Loc\Lod} &= - \lambda_{\Loc\Lod}{}^{\Loe\Lof}\,R_{\Loe\Lof\Loa\Lob}\,,
\label{eq:lambda-properties}
\end{align}
where $R_{\Loa\Lob\Loc\Lod}$ is the Riemann tensor in the tangent space of the $\AdS{5} \times \rmS^5$ background.

Using the action (\ref{eq:Rg-operation}) for each generator of $R_{g_{\bos}}$ and the definitions (\ref{eq:linearOpm-def}) of $\cO_{\pm}$, the deformed currents $\cO_{\pm}^{-1}\,A$ can be expanded as
\begin{align}
 \cO_{\pm}^{-1}\,A&=\cO^{-1}_{\pm(0)}(A_{(0)})+\cO^{-1}_{\pm(0)}(A_{(1)})+\cO^{-1}_{\pm(1)}(A_{(0)})+\cO(\theta^2)
\nn\\
  &=e_{\pm}^{\Loa}\,\gP_{\Loa}-\frac{1}{2}\,W_{\pm}^{\Loa\Lob}\,\gJ_{\Loa\Lob} + \gQ^I\,D^{IJ}_{\pm}\theta_J +\cO(\theta^2)\,,
\label{eq:Jpm-expansion}
\end{align}
where we defined
\begin{align}
 &e_{\pm}^{\Loa}\equiv e^{\Lob}\,k_{\pm \Lob}{}^{\Loa}\,,\qquad 
 k_{\pm \Loa}{}^{\Lob} \equiv \bigl[(1\pm 2\,\eta\,\lambda)^{-1}\bigr]{}_{\Loa}{}^{\Lob}\,, \qquad
 W_{\pm}^{\Loa\Lob} \equiv \omega^{\Loa\Lob}\pm 2\,\eta\,e_{\pm}^{\Loc}\,\lambda_{\Loc}{}^{\Loa\Lob}\,, 
\label{eq:e-torsionful-spin-pm}
\\
 &D^{IJ}_{\pm}\equiv \delta^{IJ}\,D_{\pm} +\frac{\ii}{2}\,\epsilon^{IJ}\,e_{\pm}^{\Loa}\,\hat{\gamma}_{\Loa}\,, \qquad 
 D_{\pm}\equiv \rmd+\frac{1}{4}\,W_{\pm}^{\Loa\Lob}\, \gamma_{\Loa\Lob}\,. 
\end{align}
Here, $e_{\pm}^{\Loa}$ and $W_{\pm}^{\Loa\Lob}$ are the two vielbeins on the deformed background and the torsionful spin connections, respectively.
In fact, $e_{\pm}^{\Loa}$ satisfy
\begin{align}
g_{mn}'=\eta_{ab}e_{+}^{\Loa}e_{+}^{\Lob}=\eta_{ab}e_{-}^{\Loa}e_{-}^{\Lob}\,,
\end{align}
and describe the deformed metric $g_{mn}'$\,.
$W_{\pm}^{\Loa\Lob}$ are given by
\begin{align}
W_{\pm \Loa\Lob}=\omega_{[\mp] \Loa\Lob} \pm\frac{1}{2}\,e_{\mp}^{\Loc}\,H'_{\Loc\Loa\Lob}\,,
\end{align}
where $\omega_{[\pm]}$ are the spin connections (\ref{eq:spin-con}) associated to the vielbeins $e_{\pm}$ and
$H'_3$ is the $H$-flux on the deformed background.

\subsubsection{The NS--NS sector}
\label{sec:YB-NS-NS}

\paragraph{Metric and $B$-field.}

Let us first consider the metric and $B$-field of the \ac{yb}-deformed action
\begin{align}
 S_{(0)}=-\frac{\dlT}{2}\int \dd[2]{\sigma} \Pg_{-}^{\WSa\WSb} \STr[A_{\WSa (0)}\, d_-\circ\cO_{-(0)}^{-1}(A_{\WSb(0)}) ]\,. 
\end{align}
By using the leading term of the expansions (\ref{eq:A-ex}), \eqref{eq:Jpm-expansion} of $A$ and $J_{-}$ in fermions, the above action can be rewritten as
\begin{align}
 S_{(0)}
 =-\dlT\int \rmd^2\sigma\,\Pg_{-}^{\WSa\WSb}\, \eta_{\Loa\Lob}\,e_{\WSa}{}^{\Loa}\,e_{\WSb}{}^{\Loc}\,k_{-\Loc}{}^{\Lob}\,.
\label{eq:action-order0}
\end{align}
By comparing it with the canonical form (\ref{eq:GS-action-canonical-YBsec}) of the \ac{gs} action,
we can write down the expressions of the deformed metric and the $B$-field as
\begin{align}
 \CG'_{mn} &= e_{(m}{}^{\Loa}\,e_{n)}{}^{\Lob}\, k_{+\Loa\Lob} \,, &
 B'_{mn} &= e_{[m}{}^{\Loa}\,e_{n]}{}^{\Lob}\, k_{+\Loa\Lob} \,,
\label{eq:G-B-prime}
\end{align}
where we used $k_{+ab}=k_{+a}{}^{c}\eta_{cb}=k_{-ba}$.

\paragraph{Dilaton.}

Next we consider the \ac{yb}-deformed dilaton $\Phi'$\,.
The formula of the \ac{yb}-deformed dilaton $\Phi'$ had been proposed in~\cite{Kyono:2016jqy,Borsato:2016ose}:
\begin{align}
 \Exp{\Phi'} = (\det k_{+})^{\frac{1}{2}}=(\det k_{-})^{\frac{1}{2}} \,. 
\label{eq:dilaton-YB}
\end{align}
This expression is consistent with the \ac{eom} of supergravity in the string frame
and reproduces those of some well-known backgrounds (e.g. Lunin--Maldacena--Frolov~\cite{Lunin:2005jy,Frolov:2005dj} and Maldacena--Russo backgrounds~\cite{Maldacena:1999mh,Hashimoto:1999ut}).

\subsubsection{The R--R sector.}
\label{sec:YB-RR}

Next, we determine the R--R fields from the quadratic part $S_{(2)}$ of the \ac{yb}-deformed action.

As noted in~\cite{Arutyunov:2015qva,Kyono:2016jqy}, the deformed action naively does not have the canonical form of the \ac{gs} action (\ref{eq:GS-action-canonical-YBsec}), so we need to choose the diagonal gauge and perform a suitable redefinition of the bosonic fields $X^m$\,. 
Since the analysis is quite complicated, we only give an outline here (see~\cite{Sakamoto:2018krs} for the details of the computation).

The quadratic part of the deformed action $S_{(2)}$ can be decomposed into two parts,
\begin{align}
 S_{(2)} = S_{(2)}^{\rmc} + \delta S_{(2)}\,.
\end{align}
First, we focus only on the first part $S_{(2)}^{\rmc}$
since the second part $\delta S_{(2)}$ can be completely canceled by field redefinitions.
The explicit expression of $S_{(2)}^{\rmc}$ in terms of the $32\times 32$ gamma matrices is given by
\begin{align}
\begin{split}
 S^{\rmc}_{(2)}&=-\ii\,\dlT\int \rmd^2\sigma\,\biggl[
 \Pg_+^{\WSa\WSb}\,\brTheta_1\,e_{-\WSa}{}^{\Loa}\,\Gamma_{\Loa}\,D_{+\WSb}\Theta_1
 +\Pg_-^{\WSa\WSb}\,\brTheta_2\,e_{+\WSa}{}^{\Loa}\,\Gamma_{\Loa}D_{-\WSb}\Theta_2
\\
 &\qquad\qquad\qquad\qquad\quad -\frac{1}{8}\,\Pg_+^{\WSa\WSb}\,\brTheta_1\,e_{-\WSa}{}^{\Loa}\,\Gamma_{\Loa} \,\bisF_5\, e_{+\WSb}{}^{\Lob}\,\Gamma_{\Lob}\,\Theta_2 \biggr]\,,
\label{eq:quadratic-YBaction}
\end{split}
\end{align}
where $D_{\pm\WSa}\Theta_I\equiv \bigl(\partial_{\WSa}+\frac{1}{4}\,W_{\pm\WSa}{}^{\Loa\Lob}\, \Gamma_{\Loa\Lob}\bigr)\,\Theta_I$ and $\bisF_5$ is the undeformed R--R $5$-form field strength (\ref{eq:F5-bispinor}).
We see that the quadratic action (\ref{eq:quadratic-YBaction}) is slightly different from the canonical form of the \ac{gs} action.

\medskip

In order to rewrite the action (\ref{eq:quadratic-YBaction}) in the canonical form of the \ac{gs} action, we need to eliminate the vielbein $e_{+m}{}^{\Loa}$ by using the relations
\begin{align}
 e_{+m}{}^{\Loa}&=(\Lambda^{-1})^{\Loa}{}_{\Lob}\,e_{-m}{}^{\Lob}
 = \Lambda_{\Lob}{}^{\Loa}\,e_{-m}{}^{\Lob}\,, &
 \Lambda_{\Loa}{}^{\Lob} &\equiv (k_-^{-1})_{\Loa}{}^{\Loc}\, k_{+\Loc}{}^{\Lob}\,.
\label{eq:e--e+-relation}
\end{align}
As discussed in~\cite{Sakamoto:2018krs}, this procedure can be identified with the diagonal gauge fixing introduced in~\cite{Jeon:2011cn,Jeon:2012kd}.
Since the two vielbeins $e_{\pm m}{}^{\Loa}$ describe the same metric, the above relations (\ref{eq:e--e+-relation}) can be regarded as a local Lorentz transformation.
Indeed, it is easy to show that the matrix $ \Lambda$ is an element of the ten-dimensional Lorentz group $SO(1,9)$\,.
Furthermore, the matrix $ \Lambda$ satisfies the identity
\begin{align}
 \Omega^{-1} \,\Gamma_{\Loa}\,\Omega &= \Lambda_{\Loa}{}^{\Lob} \, \Gamma_{\Lob}\,, &
 \Omega &=(\det k_-)^{\frac{1}{2}}\,\text{\AE}\bigl(-\eta\,\lambda^{\Loa\Lob}\,\Gamma_{\Loa\Lob}\bigr) \,,
\label{eq:omega}
\end{align}
where $\Omega$ is a spinor representation of the local Lorentz transformation (\ref{eq:e--e+-relation}), and $\text{\AE}$ is an exponential-like function with the gamma matrices totally antisymmetrized~\cite{Hassan:1999mm},
\begin{align}
\text{\AE}\bigl(-\eta\,\lambda^{\Loa\Lob}\,\Gamma_{\Loa\Lob}\bigr)
=\sum_{p=0}^5 \frac{1}{2^pp!}(-2\eta\,\lambda_{\Loa_{1}\Loa_{2}})\cdots (-2\eta\,\lambda_{\Loa_{2p-1}\Loa_{2p}})\Gamma^{\Loa_1\cdots\Loa_{2p}}\,.
\end{align}
By performing the local Lorentz transformation (\ref{eq:e--e+-relation}) and using the identity (\ref{eq:omega}),
the action (\ref{eq:quadratic-YBaction}) becomes
\begin{align}
\begin{split}
 S^{\rmc}_{(2)}&=-\ii\,\dlT\int \rmd^2\sigma\,\biggl[
 \Pg_+^{\WSa\WSb}\,\brTheta_1\,e'_{\WSa}{}^{\Loa}\,\Gamma_{\Loa}\,D_{+\WSb}\Theta_1
 +\Pg_-^{\WSa\WSb}\,\brTheta_2\,\Omega^{-1}\,e'_{\WSa}{}^{\Loa}\,\Gamma_{\Loa}\,\Omega\,D_{-\WSb}\Theta_2
\\
 &\qquad\qquad\qquad\qquad\quad -\frac{1}{8}\,\Pg_+^{\WSa\WSb}\,\brTheta_1\,e'_{\WSa}{}^{\Loa}\,\Gamma_{\Loa} \,\bisF_5\,\Omega^{-1}\, e'_{\WSb}{}^{\Loc}\,\Gamma_{\Loc}\,\Omega\,\Theta_2 \biggr]\,,
\end{split}
\end{align}
where we redefined the deformed vielbein $e_{-\alpha}{}^a$ as $e'_{\alpha}{}^a$\,.
Next, we perform a redefinition of the fermionic variables $\Theta_I$\,,
\begin{align}
 \Theta'_1 &\equiv \Theta_1\,, &
 \Theta'_2 &\equiv \Omega \,\Theta_2 \,.
\label{eq:fermi-redef}
\end{align}
As the result of the redefinition, we obtain
\begin{align}
 S^{\rmc}_{(2)}&=-\dlT\int \rmd^2\sigma\,\biggl[
 \Pg_+^{\WSa\WSb}\,\ii\,\brTheta'_1 \,e'_{\WSa}{}^{\Loa}\,\Gamma_{\Loa}\,D'_{+\WSb} \Theta'_1 
 +\Pg_-^{\WSa\WSb}\,\ii\,\brTheta_2\,e'_{\WSa}{}^{\Loa}\,\Gamma_{\Loa}\,D'_{-\WSb} \Theta'_2 
\nn\\
 &\qquad\qquad\qquad\qquad\quad -\frac{1}{8}\,\Pg_+^{\WSa\WSb}\,\ii\,\brTheta_1\,e'_{\WSa}{}^{\Loa}\,\Gamma_{\Loa}\,\bisF_5\,\Omega^{-1}\, e'_{\WSb}{}^{\Lob}\,\Gamma_{\Lob}\,\Theta'_2 \biggr]\,,
\label{eq:ScYB2-2}
\end{align}
where the derivatives $D'_{\pm}$ are defined as
\begin{align}
\begin{split}
 D'_+ &\equiv D_+ = \rmd + \frac{1}{4}\,W_+^{\Loa\Lob}\,\Gamma_{\Loa\Lob}\,,
\\
 D'_- &\equiv \Omega\circ D_-\circ \Omega^{-1} = \rmd + \frac{1}{4}\,W_-^{\Loa\Lob}\,\Omega\,\Gamma_{\Loa\Lob}\,\Omega^{-1} + \Omega\,\rmd \Omega^{-1} 
\\
 &= \rmd + \frac{1}{4}\,\bigl[\Lambda^{\Loa}{}_{\Loc}\,\Lambda^{\Lob}{}_{\Lod}\,W_-^{\Loc\Lod} +(\Lambda\,\rmd\Lambda^{-1})^{\Loa\Lob}\bigr] \,\Gamma_{\Loa\Lob} \,. 
\end{split}
\end{align}
The spin connection $\omega'^{\Loa\Lob}$ associated with the deformed vielbein $e'^{\Loa}$ and the deformed $H$-flux $H'_{\Loa\Lob\Loc}$ satisfies
\begin{align}
\begin{split}
 \omega'^{\Loa\Lob}+\frac{1}{2}\,e'_{\Loc}\,H'^{\Loc\Loa\Lob} &=W_{+}^{\Loa\Lob}\,,
\\
 \omega'^{\Loa\Lob}-\frac{1}{2}\,e'_{\Loc}\,H'^{\Loc\Loa\Lob} &= \Lambda^{\Loa}{}_{\Loc}\,\Lambda^{\Lob}{}_{\Lod}\,W_-^{\Loc\Lod} +(\Lambda\,\rmd\Lambda^{-1})^{\Loa\Lob} \,.
\end{split}
\label{eq:torsionful-spin}
\end{align}
Therefore, $D'_{\pm}$ can be expressed as
\begin{align}
 D'_{\pm} = \rmd + \frac{1}{4}\,\Bigl(\omega'^{\Loa\Lob}\pm\frac{1}{2}\,e'_{\Loc}\,H'^{\Loc\Loa\Lob}\Bigr)\,\Gamma_{\Loa\Lob}\,.
\end{align}
In this way, the deformed action \eqref{eq:ScYB2-2} becomes the conventional \ac{gs} action at order $\cO(\theta^2)$
by identifying the deformed R--R field strengths as
\begin{align}
\begin{split}
 \bisF' &=\bisF_5\,\Omega^{-1}\,,\\
\bisF'&=\sum_{p=1,3,5,7,9}\frac{1}{p!}e^{\Phi'}\,\hat{F}'_{a_1\dots a_p}\Gamma^{a_1\dots a_p}\,. 
\label{eq:YBRR0}
\end{split}
\end{align}
Here the deformed dilaton $\Phi'$ is given by (\ref{eq:dilaton-YB}).
The transformation rule \eqref{eq:YBRR0} was originally given in~\cite{Borsato:2016ose}.
A different derivation based on the $\kappa$-symmetry variation is given in Appendix I of~\cite{Sakamoto:2018krs}.

\medskip

Finally, let us consider the remaining part $\delta S_{(2)}$\,.
It can be completely canceled by redefining the bosonic fields $X^m$~\cite{Arutyunov:2015qva,Kyono:2016jqy}\,,
\begin{align}
 X^m\ \to\ X^m + \frac{\eta}{4}\,\sigma_1^{IJ}\,e^{\Loc m}\,\lambda_{\Loc}{}^{\Loa\Lob}\,\brtheta_I\,\gamma_{\Loa\Lob}\,\theta_J + \cO(\theta^4)\,,
\label{eq:bosonic-shift}
\end{align}
as long as the $r$-matrix satisfies the homogeneous \ac{cybe}.
Indeed, this redefinition results in a shift $S_{(0)} \to S_{(0)} + \delta S_{(0)}$
and the sum of $\delta S_{(0)}$ and $\delta S_{(2)} $ is the quite simple expression
\begin{align}
 &\delta S_{(0)}+\delta S_{(2)} 
\nn\\
 &=\frac{\eta^2\,\dlT}{2}\int \rmd^2\sigma\,\, \Pg_-^{\WSa\WSb}\,\sigma_1^{IJ}\,
  \bigl[\CYBE^{(0)}_g\bigl(J_{+m}^{(2)},J_{-n}^{(2)}\bigr)\bigr]^{\Loa\Lob}\,\brtheta_I\, \gamma_{\Loa\Lob}\,\theta_J\,\partial_{\WSa}X^{m}\,\partial_{\WSb}X^n \,,
\label{eq:quadro-CYBE}
\end{align}
where $\CYBE^{(0)}_g (X,Y)$ represents the grade-$0$ component of $\CYBE_g (X,Y)$ defined in \eqref{eq:CYBE-g}.
This shows that $\delta S_{(2)}$ is completely canceled out by $\delta S_{(0)}$
when the $r$-matrix satisfies the homogeneous \ac{cybe}.

\subsection{Generalized supergravity}
\label{sec:GSE-YBsec}

In general, \ac{yb} deformations correspond to solutions both of the usual supergravity equations and of the \acf{gse}.
The \ac{gse} were originally proposed to support a $q$-deformed $\AdS5\times\rmS^5$ background as a solution~\cite{Arutyunov:2015mqj}.
It was shown subsequently that some homogeneous \ac{yb} deformed $\AdS5\times\rmS^5$ backgrounds are also solutions to the \ac{gse}~\cite{Orlando:2016qqu}.
In this subsection, we will give the explicit expression of the $q$-deformed $\AdS5\times\rmS^5$ background and then introduce the \ac{gse}.

\subsubsection{The $q$-deformed $\AdS{5} \times \rmS^5$ background.}

As explained in Subsection~\ref{subsec:YB-sigma-action}, the $q$-deformed background can be realized as a \ac{yb} deformation of the $\AdS5\times\rmS^5$ superstring~\cite{Delduc:2013qra} with a classical $r$-matrix of Drinfeld--Jimbo type 
satisfying the \ac{mcybe} (with $c=i$). 
The metric and $B$-field were originally derived in~\cite{Arutyunov:2013ega} 
and the full background including R-R fluxes and dilaton has been obtained by performing the
supercoset construction in~\cite{Arutyunov:2015qva}. It is given by 
\begin{align}
\begin{split}
\scalebox{0.85}{$\displaystyle\bd s^2$}&=
\scalebox{0.85}{$\displaystyle
\sqrt{1+\kappa^2}\biggl[-\frac{1+\rho^2}{1-\kappa^2\rho^2}\bd t^2
+\frac{\bd \rho^2}{(1-\kappa^2\rho^2)(1+\rho^2)}+\frac{\rho^2 (\bd\zeta^2+\cos^2\zeta\bd\psi_1^2)}{1+\kappa^2\rho^4\sin^2\zeta}
+\rho^2\sin^2\zeta\bd\psi_2^2 $}\\
&\quad\scalebox{0.85}{$\displaystyle
+\frac{1-r^2}{1+\kappa^2r^2}\bd \phi^2
+\frac{\bd r^2}{(1+\kappa^2r^2)(1-r^2)}+\frac{r^2 (\bd\zeta^2+\cos^2\xi\bd\phi_1^2)}{1+\kappa^2r^4\sin^2\xi}
+r^2\sin^2\xi\bd\phi_2^2$}\biggr]\,,\\
\scalebox{0.85}{$\displaystyle B_2$}
&=\scalebox{0.85}{$\displaystyle
\sqrt{1+\kappa^2}\biggl[\frac{\kappa \rho}{1-\kappa^2\rho^2}\bd t\wedge \bd \rho
-\frac{\kappa \rho^4 \sin\zeta \cos\zeta}{1+\kappa^2\rho^4\sin^2\zeta} \bd \zeta\wedge \bd \psi_1
 $}\\
&\quad\scalebox{0.85}{$\displaystyle
+\frac{\kappa r}{1+\kappa^2 r^2}\bd \phi\wedge \bd r
+\frac{\kappa r^4\sin\xi\cos\xi}{1+\kappa^2 r^4\sin^2\xi}\bd\xi\wedge \bd\phi_1
 $}\biggr]\,,\\
\scalebox{0.85}{$\displaystyle \hat{F}_1$}
&=\scalebox{0.85}{$\displaystyle
4\kappa^2\sqrt{1+\kappa^2}(\rho^4\sin^2\zeta\bd\psi_2
-r^4\sin^2\xi\bd\phi_2)
 $}\,,\\
\scalebox{0.85}{$\displaystyle \hat{F}_3$}
&=\scalebox{0.85}{$\displaystyle
-4\kappa (1+\kappa^2)\biggl[\frac{\rho}{1-\kappa^2\rho^2}
\bd t\wedge\bd\rho\wedge(\rho^2\sin^2\zeta\bd\psi_2-\kappa^2r^4\sin^2\xi\bd\phi_2)
 $}\\
&\quad\scalebox{0.85}{$\displaystyle
+\frac{r}{1+\kappa^2r^2}
\bd \phi\wedge\bd r\wedge(\kappa^2\rho^4\sin^2\zeta\bd\psi_2+r^2\sin^2\xi\bd\phi_2)
$}\\
&\quad\scalebox{0.85}{$\displaystyle
+\frac{\rho^4\sin\zeta\cos\zeta}{1+\kappa^2\rho^4\sin^2\zeta}
\bd\zeta\wedge\bd\psi_1\wedge (\bd\psi_2+\kappa^2 r^4 \sin^2\xi\bd\phi_2)
$}\\
&\quad\scalebox{0.85}{$\displaystyle
+\frac{r^4\sin\xi\cos\xi}{1+\kappa^2r^4\sin^2\xi}
\bd\xi\wedge\bd\phi_1\wedge (\kappa^2 \rho^4\sin^2\zeta\bd\psi_2+\bd\phi_2)
\biggr]
$}\,,\\
\scalebox{0.85}{$\displaystyle \hat{F}_5$}
&=\scalebox{0.85}{$\displaystyle
4(1+\kappa^2)^{3/2}\biggl[\frac{\rho^3\sin\zeta\cos\zeta}{(1-\kappa^2\rho^2)(1+\kappa^2\rho^4\sin^2\zeta)}
\bd t\wedge \bd \rho\wedge \bd\zeta\wedge\bd \psi_1\wedge (\bd\psi_2+\kappa^4\rho^2r^4\sin^2\xi\bd\phi_2)
$}\\
&\quad\scalebox{0.85}{$\displaystyle
+\frac{r^3\sin\xi\cos\xi}{(1+\kappa^2r^2)(1+\kappa^2r^4\sin^2\xi)}
\bd \phi\wedge \bd r\wedge \bd\xi\wedge\bd \phi_1\wedge (\kappa^4r^2\rho^4\sin^2\zeta\bd\psi_2-\bd\phi_2)
$}\\
&\quad\scalebox{0.85}{$\displaystyle
+\frac{\kappa^2\rho r}{(1-\kappa^2\rho^2)(1+\kappa^2r^2)}
\bd t\wedge\bd\rho\wedge\bd\phi\wedge\bd r\wedge (\rho^2\sin^2\zeta\bd \psi_2+r^2\sin^2\xi\bd \phi_2)
$}\\
&\quad\scalebox{0.85}{$\displaystyle
+\frac{\kappa^2\rho r^4\sin\xi\cos\xi}{(1-\kappa^2\rho^2)(1+\kappa^2r^4\sin^2\xi)}
\bd t\wedge\bd\rho\wedge \bd \xi\wedge\bd \phi_1 \wedge(\rho^2\sin^2\zeta\bd\psi_2+\bd \phi_2)
$}\\
&\quad\scalebox{0.85}{$\displaystyle
+\frac{\kappa^2r \rho^4\sin\zeta\cos\zeta}{(1+\kappa^2r^2)(1+\kappa^2\rho^4\sin^2\zeta)}
\bd \phi\wedge\bd r\wedge \bd \zeta\wedge\bd \psi_1 \wedge(\bd \psi_2-r^2\sin^2\xi\bd\phi_2)
$}\\
&\quad\scalebox{0.85}{$\displaystyle
+\frac{\kappa^2\rho^4r^4\sin\zeta\cos\zeta\sin\xi\cos\xi}{(1+\kappa^2\rho^4\sin^2\zeta)(1+\kappa^2r^4\sin^2\xi)}
\bd\zeta\wedge\bd\psi_1\wedge\bd\xi\wedge\bd\phi_1\wedge(\bd\psi_2-\bd\phi_2)\biggr]
$}\,,\\
\scalebox{0.85}{$\displaystyle \Phi$}
&=\scalebox{0.85}{$\displaystyle
\frac{1}{2}\log\left[\frac{1}{(1-\kappa^2\rho^2)(1+\kappa^2\rho^4\sin^2\zeta)(1+r^2\kappa^2)(1+\kappa^2r^4\sin^2\xi)}\right]
$}\,,
\label{eq:ABF-bg}
\end{split}
\end{align}
where we defined $\kappa=\frac{2\eta}{1-\eta^2}$\,.
Note here that we have kept total derivative terms of the $B$-field that are obtained after performing the supercoset construction (see for example footnote $19$ of~\cite{Arutyunov:2015qva})\footnote{Another expression for the dilaton $\Phi$\,, which is different from the one in~\cite{Arutyunov:2015mqj}, 
is obtained due to the existence of the total derivative terms of the $B$-field.}.
Remarkably, the deformed background is not a solution of type IIB supergravity~\cite{Arutyunov:2015qva} but of the \ac{gse}~\cite{Arutyunov:2015mqj} when we introduce the extra Killing vector
\begin{align}
I&=\frac{1}{\sqrt{1+\kappa^2}}\left(-4\kappa\partial_t
+2\kappa\partial_{\psi_1}
+4\kappa\partial_\phi
-2\kappa\partial_{\phi_1}\right)\,.\label{eq:ABF-bg-I}
\end{align}
We will next give the explicit expression for the generalized type IIB supergravity equations.

\subsubsection{Generalized supergravity equations.}

Our conventions for the type II \ac{gse}~\cite{Arutyunov:2015mqj,Wulff:2016tju,Sakatani:2016fvh,Baguet:2016prz,Sakamoto:2017wor} are as follows:
\begin{align}
\begin{split}
 R_{mn}- \frac{1}{4}\,H_{mpq}\,H_n{}^{pq} + 2\,\sfD_m \partial_n \Phi + \sfD_m U_n +\sfD_n U_m &= T_{mn} \,,
\\
 -\frac{1}{2}\,\sfD^k H_{kmn} + \partial_k\Phi\,H^k{}_{mn} + U^k\,H_{kmn} + \sfD_m I_n - \sfD_n I_m &= \cK_{mn} \,,
\\
 R  - \frac{1}{2}\,\abs{H_3}^2  + 4\,\sfD^m \partial_m \Phi - 4\,\abs{\partial \Phi}^2
  - 4\,\bigl(I^m I_m+U^m U_m + 2\,U^m\,\partial_m \Phi - \sfD_m U^m\bigr) &=0 \,,
\\
 \rmd *\hat{F}_n -H_3\wedge * \hat{F}_{n+2} -\iota_I B_2 \wedge * \hat{F}_n -\iota_I * \hat{F}_{n-2} &=0 \,,
\label{eq:GSEsecAdS}
\end{split}
\end{align}
where we have defined $\abs{\alpha_p}^2\equiv \frac{1}{p!}\alpha_{m_1\cdots m_p}\,\alpha^{m_1\cdots m_p}$\,.
$\sfD_m$ is the usual covariant derivative associated to $\CG_{mn}$\,,
and $T_{mn}\,, \cK_{mn}$ are defined by
\begin{align}
\begin{split}
 T_{mn} &\equiv \frac{1}{4}\Exp{2\Phi} \sum_p \biggl[ \frac{1}{(p-1)!}\, 
 \hat{F}_{(m}{}^{k_1\cdots k_{p-1}} \hat{F}_{n) k_1\cdots k_{p-1}} - \frac{1}{2}\, 
 \CG_{mn}\,\abs{\hat{F}_p}^2 \biggr] \,,
\\
 \cK_{mn}&\equiv \frac{1}{4}\Exp{2\Phi} \sum_p \frac{1}{(p-2)!}\, \hat{F}_{k_1\cdots k_{p-2}}\, 
 \hat{F}_{mn}{}^{k_1\cdots k_{p-2}}  \,. 
\end{split}
\end{align}
The relation between the R--R field strengths and potentials is given by (see~\cite{Sakamoto:2017wor} for details)
\begin{align}
 \hat{F}_n&= \rmd \hat{C}_{n-1} + H_3\wedge \hat{C}_{n-3} - \iota_I B_2 \wedge \hat{C}_{p-1} -\iota_I \hat{C}_{n+1}\,.
\end{align}
The Killing vector $I=I^m\,\partial_m$ is defined to satisfy
\begin{align}
 \Lie_I \CG_{mn} = 0\,, \qquad 
 \Lie_I B_2 + \rmd \bigl(U -\iota_I B_2\bigr) = 0\,,\qquad 
 \Lie_I \Phi =0 \,, \qquad 
 I^m\,U_m = 0\,. 
\end{align}
Here we have ignored the spacetime fermions. The full explicit expression of the type IIB \ac{gse} are given in~\cite{Wulff:2016tju}.

\medskip

We usually choose the particular gauge $U_m=I^nB_{nm}$\,(see~\cite{Arutyunov:2015mqj,Sakamoto:2017wor} for the details) in which the \ac{gse} (\ref{eq:GSEsecAdS}) become
\begin{align}
\begin{split}
 R_{mn}- \frac{1}{4}\,H_{mpq}\,H_n{}^{pq} + \sfD_m Z_n +\sfD_n Z_m &= T_{mn} \,,\\
 -\frac{1}{2}\,\sfD^k H_{kmn} + Z_{k}\,H^k{}_{mn} + \sfD_m I_n - \sfD_n I_m &= \cK_{mn} \,,\\
 R  - \frac{1}{2}\,\abs{H_3}^2 + 4\,\bigl( \sfD^m Z_{m} - I^m I_m-Z^m Z_m \bigr) &=0 \,,\\
 \rmd *\hat{F}_n -H_3\wedge * \hat{F}_{n+2} -\iota_I B_2 \wedge * \hat{F}_n -\iota_I * \hat{F}_{n-2} &=0 \,,
\label{eq:GSEsecAdS-gauge}
\end{split}
\end{align}
where we defined 
\begin{align}
Z_{m}\equiv\partial_{m} \Phi+I^n B_{nm}\,.\label{eq:Z-def}
\end{align}
In this gauge, we can show that the $q$-deformed $\AdS5\times \rmS^5$ background (\ref{eq:ABF-bg}) with the Killing vector (\ref{eq:ABF-bg-I}) solve the \ac{gse}.
When $I=0$, the \ac{gse} reduce to the usual supergravity \ac{eom}. 
Therefore, this deformation is characterized only by the Killing vector $I^m$\,. 
Note that due to the presence of this Killing vector, the solutions of the \ac{gse} are effectively nine dimensional.

\medskip

A remarkable feature of this theory is that the \ac{gse} can be reproduced from the requirement of $\kappa$-symmetry in the \ac{gs} formalism.
It has been known for a long time that the on-shell constraints of type II supergravity ensure $\kappa$-symmetry of the associated
\ac{gs} type string sigma model~\cite{Grisaru:1985fv,Bergshoeff:1985su}.
At the time it had been conjectured that the $\kappa$-symmetry requires the type II supergravity equations.
However, after about thirty years, Tseytlin and Wulff~\cite{Wulff:2016tju} solved this long-standing problem, showing that a general solution of the \(\kappa\)-symmetry constraint leads to solutions to the \ac{eom} of generalized supergravity.

\subsubsection{Weyl invariance of string theory on generalized supergravity backgrounds.}

Tseytlin--Wulff's result implies that, at the classical level, string theory is consistently defined on \emph{generalized} supergravity backgrounds. However, the quantum consistency of string theories defined on such backgrounds is not clear.
Indeed, the \ac{gse} (\ref{eq:GSEsecAdS}) were originally introduced as a scale-invariance condition for string theory.
The Weyl invariance of string theory on such backgrounds has been studied in~\cite{Sakamoto:2017wor,Fernandez-Melgarejo:2018wpg,Muck:2019pwj}.
We will briefly comment on the current status of this subject.

\medskip

For simplicity, we will consider the conventional (bosonic) string sigma model on a general background, 
\begin{align}
 S =-\frac{1}{4\pi\alpha'} \int \rmd^2\sigma \sqrt{-\gamma}\,\bigl(g_{mn}\,\gamma^{\alpha\beta} - B_{mn}\,\varepsilon^{\alpha\beta}\bigr)\, \partial_\alpha X^m\, \partial_\beta X^n \,,
\label{eq:string-action}
\end{align}
where $\varepsilon^{01}=1/\sqrt{-\gamma}$\,. 
The Weyl anomaly of this system takes the form
\begin{align}
 2\alpha'\,\langle T^\alpha{}_\alpha\rangle = \bigl(\beta^{g}_{mn}\,\gamma^{\alpha\beta} - \beta^{B}_{mn}\, \varepsilon^{\alpha\beta}\bigr)\, \partial_\alpha X^m\, \partial_\beta X^n \,.
\label{eq:Weyl-anomaly}
\end{align}
The $\beta$-functions at the one-loop level have been computed (for example in~\cite{Hull:1985rc}) and have the form
\begin{align}
 \beta^{\CG}_{mn} &= \alpha'\,\Bigl(R_{mn}-\frac{1}{4}\,H_{mpq}\,H_n{}^{pq}\Bigr) \,, &
 \beta^{B}_{mn} &= \alpha'\,\Bigl(- \frac{1}{2}\,\sfD^k H_{kmn}\Bigr) \,.
\end{align}
If the trace of the energy-momentum tensor (\ref{eq:Weyl-anomaly}) vanishes,
the system is Weyl invariant.

Quantum scale invariance is satisfied by requiring~\cite{Hull:1985rc}
\begin{align}
 \beta^{g}_{mn} &= - 2\,\alpha'\,D_{(m} Z_{n)}\,, & 
 \beta^{B}_{mn} &= - \alpha'\,\bigl(Z^k\,H_{kmn} + 2\,D_{[m} I_{n]}\bigr) \,,
\label{eq:scale-GSE}
\end{align}
where $I_m$ and $Z_m$ are certain vector fields in the target space. 
When the $\beta$-functions have the form \eqref{eq:scale-GSE}, the Weyl anomaly \eqref{eq:Weyl-anomaly} becomes
\begin{eqnarray}
 \langle T^\alpha{}_\alpha \rangle \! &=& \! -\cD_\alpha\bigl[(Z_m\,\gamma^{\alpha\beta} - I_m\,\varepsilon^{\alpha\beta})\,\partial_\beta X^m\bigr] 
 +Z^m\,\frac{2\pi\alpha'}{\sqrt{-\gamma}}\frac{\delta S}{\delta X^m} 
\nn\\
 \! &\overset{\text{e.o.m.}}{\sim}& \! -\cD_\alpha\bigl[(Z_m\,\gamma^{\alpha\beta} - I_m\,\varepsilon^{\alpha\beta})\,\partial_\beta X^m\bigr] \,. 
\label{eq:Weyl-general}
\end{eqnarray}
Note that this scale invariance condition takes the same form as the NS sector of the \ac{gse} (\ref{eq:GSEsecAdS}).
The full set of the \ac{gse} is obtained by generalizing the scale invariance condition to include the R-R fields~\cite{Arutyunov:2015mqj}.

\medskip

To obtain a consistent string theory at the quantum level, we need to cancel the Weyl anomaly (\ref{eq:Weyl-anomaly}).
It is well known that when $Z_m=\partial_m\Phi$ and $I^m=0$\,, we can cancel this by adding a counterterm, the so-called Fradkin--Tseytlin term~\cite{Fradkin:1984pq},
\begin{align}
 S_{\text{FT}} = \frac{1}{4\pi}\int \rmd^2\sigma \, \sqrt{-\gamma}\,R^{(2)}\,\Phi\,,
\label{eq:F-T}
\end{align}
to the original action \eqref{eq:string-action}. 
Compared to the sigma model action,
the counterterm (\ref{eq:F-T}) is of higher order in $\alpha'$ and should be regarded as a quantum correction. 

\medskip

In the more general \ac{gse} case, the situation is more subtle as the counterterm \eqref{eq:F-T} cannot cancel out the anomaly \eqref{eq:Weyl-general}. 
However, it is important to note that $I$ and $Z$ are arbitrary vectors in the scale invariance conditions (\ref{eq:scale-GSE}), while in the \ac{gse} case, $I$ is a Killing vector for \(g_{mn}\) and $Z$ is given by (\ref{eq:Z-def}).
In~\cite{Sakamoto:2017wor,Fernandez-Melgarejo:2018wpg}, by using this Killing property, a possible local and covariant counterterm for the bosonic string on generalized supergravity backgrounds was constructed as a generalization of the Fradkin--Tseytlin term (\ref{eq:F-T}) (see~\cite{Sakamoto:2017wor,Fernandez-Melgarejo:2018wpg} for details).
The Weyl invariance of the type I superstring in generalized supergravity backgrounds was discussed
in~\cite{Muck:2019pwj}.
A detailed study of the \ac{cft} picture is however still outstanding and in this sense, the full  consistency of string theory in generalized supergravity backgrounds is an open problem.

\subsection{The unimodularity condition}\label{subsec:unimod_cond}

As explained in the previous section, 
\ac{yb} deformed backgrounds can be solutions not only of the usual supergravity equations but also of the \ac{gse}.
Therefore, to obtain the usual supergravity solutions from \ac{yb} deformations,
we need to impose further constraints on the classical $r$-matrices.
This \emph{unimodularity} condition for the classical $r$-matrices is due to Borsato and Wulff~\cite{Borsato:2016ose}.
We will call an \(r\)-matrix unimodular if it satisfies the condition
\begin{align}
r^{ij}[T_i,T_j]&=0\,,& T_i&\in\mathfrak{su}(2,2|4)\,.
\label{unimodular1}
\end{align}

Let us briefly explain the origin of the name of the condition (\ref{unimodular1}).
For simplicity, we will consider the bosonic subalgebra $\mathfrak{so}(2,4)\oplus \mathfrak{so}(6)$ of $\mathfrak{su}(2,2|4)$\footnote{Recently~\cite{vanTongeren:2019dlq}, homogeneous \ac{yb} deformations associated with unimodular $r$-matrices including fermionic generators were considered, and the associated deformed backgrounds were constructed explicitly.}.
A constant solution of the homogeneous \ac{cybe} (\ref{eq:CYBE-r}) for a Lie algebra $\mathfrak{g}$ corresponds one-to-one to
a subalgebra $\mathfrak{f}\subset \mathfrak{g}$~\cite{Stolin1,Stolin2}.
Here, a constant solution means that $r^{ij}$ is a constant skew-symmetric matrix.
Furthermore, restricting the range of the indices of $r^{ij}$ to $i, j=1,\dots , {\rm dim}\,\mathfrak{f}$\,,
the matrices $r^{ij}$ are always invertible.
This implies that $\mathfrak{f}$ is always even dimensional.
Introduce the bi-linear map $\omega: \mathfrak{g}\times \mathfrak{g}\to \mathbb{R}$ defined by
\begin{align}
\omega(T_i, T_j):=(r^{-1})_{ij}\,.
\end{align}
The homogeneous \ac{cybe} implies that \(\omega\) is a $2$-cocycle 
\begin{align}
\begin{split}
\omega(x,y)&=-\omega(y,x)\,,\\
\omega([x,y],z)+\omega([x,y],z)+\omega([x,y],z)&=0\,,
\end{split}
\end{align}
where $x,y,z\in \mathfrak{f}$\,.
This means that $\mathfrak{f}$ is a quasi-Frobenius Lie algebra.
The $2$-cocycle condition can be rewritten as
\begin{align}
(r^{-1})_{i[j}f_{kl]}{}^{i}=0\,.
\label{2-cocycle-r}
\end{align}
By taking a contraction $r^{kl}$ with (\ref{2-cocycle-r}), we obtain
\begin{align}
0=3(r^{-1})_{i[j}f_{kl]}{}^{i}r^{kl}
=(r^{-1})_{ij}f_{kl}{}^{i}r^{kl}+2\,f_{ij}{}^{i}\,.
\end{align}
If the $r$-matrix satisfies the unimodularity condition, the equation becomes  
\begin{align}
f_{ij}{}^{i}=-\frac{1}{2}(r^{-1})_{ij}f_{kl}{}^{i}r^{kl}=0\,,
\label{eq:uni-r-structure}
\end{align}
then $\mathfrak{f}$ is also a unimodular Lie algebra.

\subsection{Classification of $r$-matrices}\label{subsec-class-r}

An $r$-matrix
\begin{align}
r =\frac{1}{2}\, r^{ij}\,T_i\wedge T_j\,,
\end{align}
is called \emph{Abelian} if it consists of a set of generators which commute with each other, $[T_i,\,T_j]=0$, otherwise it is called \emph{non-Abelian}. 
Most homogeneous \ac{yb} deformations studied in the literature are based on Abelian $r$-matrices, which are obviously unimodular.
Moreover, when $\mathfrak{g}$ is a compact Lie algebra (for example $\mathfrak{su}(N)$\,, $\mathfrak{so}(N)$),
all quasi-Frobenius Lie subalgebras $\mathfrak{f}$ are Abelian~\cite{Abelian-compact}.
Therefore, non-Abelian unimodular $r$-matrices only exist for non-compact Lie algebras $\mathfrak{g}$.

\paragraph{Non-Abelian unimodular $r$-matrices.}

We now discuss the classification of non-Abelian unimodular $r$-matrices. 
We define the rank of an $r$-matrix as 
\begin{equation}
	\mathrm{Rank}\ r^{ij} := {\rm dim}\,\mathfrak{f}\,. 
\end{equation}
Rank-2 unimodular $r$-matrices are by construction Abelian.
Non-Abelian unimodular $r$-matrices with rank four have been fully classified.

\paragraph{\bf Rank $4$.}
The rank-$4$ unimodular $r$-matrices for the bosonic isometries of $\AdS5$ have been classified in~\cite{Borsato:2016ose}. 
If we take a rank-$4$ $r$-matrix
\begin{align}
r&=T_1\wedge T_2+T_3\wedge T_4\,, & T_{1,\dots ,4}&\in \mathfrak{so}(2,4)\,,
\end{align}
we obtain the following four classes:
\begin{align}
	(i) && \mathfrak{h}_3&\oplus \mathbb{R} & [T_1, T_3]&=T_4 & &\\
	(ii) && \mathfrak{r}_{3,-1}&\oplus \mathbb{R} & [T_1,T_3]&=T_3\,, & [T_1,T_4]&=-T_4\\
	(iii) && \mathfrak{r}'_{3,0}&\oplus \mathbb{R} &[T_1,T_3]&=-T_4\,, &[T_1,T_4]&=T_3 \\
	(iv) && \mathfrak{n}_4& & [T_1,T_3]&=-T_2\,, &  [T_2,T_3]&=T_4.
\end{align}
Classes (i)--(iii) are called \emph{almost Abelian} $r$-matrices~\cite{vanTongeren:2016eeb} and cover most of the rank-4 examples studied in~\cite{Borsato:2016ose}. 
As argued in~\cite{Borsato:2016ose,vanTongeren:2016eeb}, this class of \ac{yb} deformations can be realized as a sequence of 
non-commuting TsT-transformations (see~\cite{Borsato:2016ose} for the explicit form for the rank-4 examples). 
They are obtained as combination of the usual TsT-transformations and appropriate diffeomorphisms.
The last class (iv) cannot be generated by performing non-commuting TsT-transformations.

\paragraph{\bf Rank $6$.}
There is to date no complete classification for rank $6$ unimodular $r$-matrices.
Some examples were given in~\cite{Borsato:2016ose}.

\paragraph{\bf Rank $8$.}
No rank-$8$ non-Abelian unimodular $r$-matrices exist.

\paragraph{Non-unimodular $r$-matrices.}

Finally, let us briefly comment on non-unimodular $r$-matrices.
The simplest example is the rank-$2$ $r$-matrix
\begin{align}
r&=\frac{1}{2}T_1\wedge T_2\,, & [T_1, T_2]&=T_2\,.
\end{align}
It is easy to see that the $r$-matrix does not satisfy the unimodularity condition (\ref{unimodular1}).
The $r$-matrix is also called a \emph{Jordanian} $r$-matrix.
A generalization of the rank-$2$ Jordanian $r$-matrix was given in~\cite{Tolstoy2004}.
\ac{yb} deformed backgrounds associated to non-unimodular $r$-matrices are solutions to the \ac{gse}~\cite{Borsato:2016ose,Orlando:2016qqu}.

\subsection{YB deformations as string duality transformations}\label{sec:stringDualityTrans}

In Subsection~\ref{subsec:YBfromGS}, we derived the general formulas (\ref{eq:G-B-prime}), (\ref{eq:dilaton-YB}), (\ref{eq:YBRR0}) for the homogeneous \ac{yb} deformed AdS$_5\times$S$^5$ backgrounds with an $r$-matrix composed of bosonic generators only.
An important observation of~\cite{Sakamoto:2017cpu} is that the homogeneous \ac{yb} deformation of the AdS$_5\times$S$^5$ background is nothing but the $\beta$-transformation which in turn is a $O(d,d)$ $T$-duality transformation\footnote{The deformed background can be reproduced from the requirement of invariance of non-zero Page forms and associated Page charges~\cite{Araujo:2017enj}.}~\cite{Sakamoto:2018krs}.
It follows that homogeneous \ac{yb}-deformations can be regarded as examples of string duality transformations.
In this section, we will briefly discuss this interpretation.%
\subsubsection{A brief review of DFT.}\label{sec:reviewDFT}

For this purpose, it is useful to utilize the manifestly T-duality covariant formulation of supergravity called \emph{\ac{dft}}~\cite{Hull:2009mi,Hohm:2010jy,Hohm:2010pp,Hull:2009zb,Siegel:1993bj,Siegel:1993th,Siegel:1993xq}.

\paragraph{DFT fields and their parametrization.}

In \ac{dft}, we consider a doubled spacetime with coordinates $(x^{M})=(x^m\,, \tilde{x}_m)$ $(M=1,\dots , 2d ;\,m=1,\dots ,d)$. %
Here  $x^m$ are the standard $d$-dimensional coordinates and $\tilde{x}^m$ are the dual coordinates.
The bosonic fields in \ac{dft} are the generalized metric $\mathcal{H}_{MN}$ $(M,N=0,\dots,2d)$, the \ac{dft} dilaton $d(x)$\,, and an $O(d, d)$ spinor of R-R fields $\hat{F}$ (see~\cite{Sakamoto:2018krs} for our conventions).

The generalized metric $\mathcal{H}_{MN}$ can be parameterized as
\begin{align}
\mathcal{H}\equiv (\mathcal{H}_{MN})=
\begin{pmatrix}
(g-Bg^{-1}B)_{mn}&B_{mk}g^{kn}\\
-g^{mk}B_{kn}& g^{mn}
\end{pmatrix}
\,,\label{eq:H-geometric}
\end{align}
in terms of the metric $g_{mn}$ and the Kalb--Ramond $B$-field $B_{mn}$.
The $2d$-dimensional indices $M, N, \cdots$ are raised and lowered with the $O(d, d)$ metric
\begin{align}
(\eta_{MN})&=
\begin{pmatrix}
~0~&\delta^n_{m}\\
\delta^{m}_{n}& ~0~
\end{pmatrix}
\,,  &
(\eta^{MN})&=
\begin{pmatrix}
~0~&\delta^m_{n}\\
\delta^{n}_{m}& ~0~
\end{pmatrix}
\,.
\end{align}
The familiar properties of the generalized metric,
\begin{align}
\mathcal{H}^{\rm T}&=\mathcal{H}\,, & \mathcal{H}^{\rm T} \eta \mathcal{H}&=\eta\,,
\end{align}
follow from the above parametrization, and imply that $\mathcal{H}$ is an element of $O(d,d)$\,.
As $\mathcal{H}\in O(d,d)$, the non-linear transformations of the T-duality group are covariantly realized as
\begin{align}
\mathcal{H} &\to h^{\rm T}\mathcal{H} h\,,&  h&\in O(d,d)\,.
\end{align}
The \ac{dft} dilaton $d(x)$ is related to the conventional dilaton $\Phi$ by
\begin{align}
e^{-2d(x)} =\sqrt{|g|} e^{-2\Phi}\,,
\end{align}
and is invariant under $O(d,d)$ duality transformations.

\medskip

For later discussions, it is convenient to introduce another parametrization of the generalized metric and the \ac{dft} dilaton,
\begin{align}
\begin{split}
\mathcal{H}&=(\mathcal{H}_{MN})=
\begin{pmatrix}
G_{mn}&G_{mk}\beta^{kn}\\
-\beta^{mk}G_{kn}& (G^{-1}-\beta G\beta)^{mn}
\end{pmatrix}
\,,\\
e^{-2d} &=\sqrt{|G|} e^{-2\tilde{\phi}}\,,\label{eq:H-non-geometric}
\end{split}
\end{align}
in terms of the dual fields $(G^{mn}, \beta^{mn}, \tilde{\phi})$~\cite{Duff:1989tf,Shapere:1988zv,Giveon:1988tt,Giveon:1994fu}.
This parametrization (\ref{eq:H-non-geometric}) is referred to as the \emph{non-geometric parametrization} of the generalized metric and the \ac{dft} dilaton.
If the matrix $E_{mn}\equiv g_{mn}+B_{mn}$ is invertible, the relation between the conventional fields $(g_{mn}, B_{mn}, \Phi)$ and the dual fields $(G^{mn}, \beta^{mn}, \tilde{\phi})$ is given by
\begin{align}
&E^{mn}\equiv (E^{-1})^{mn}=G^{mn}-\beta^{mn}\,,\label{eq:relation-open-closed}\\
&e^{-2d} =\sqrt{|g|} e^{-2\Phi}=\sqrt{|G|} e^{-2\tilde{\phi}}\label{eq:DFT-dilaton}
\,.
\end{align}
The dual metric $G_{mn}$ coincides with the open-string metric~\cite{Duff:1989tf,Seiberg:1999vs}, to be distinguished from the initial closed-string metric \(g_{mn}\).
As we will discuss in Section~\ref{sec:YB-T-fold}, the non-geometric parametrization of the generalized metric is useful to discuss the non-geometric aspects of a given background.

\paragraph{Section condition.}

For the consistency of \ac{dft}, we require that arbitrary fields or gauge parameters $A(x)$ and $B(x)$ satisfy the so-called \emph{section condition}~\cite{Siegel:1993th,Hull:2009mi,Hull:2009zb},
\begin{align}
\eta^{MN} \partial_{M}A(x) \partial_{N}B(x)=0\,,\\
\eta^{MN} \partial_{M} \partial_{N}A(x)=0\,,
\end{align}
where $\partial_{M}=(\partial_{m}, \tilde{\partial}^m)\equiv(\frac{\partial}{\partial x^m},\frac{\partial}{\partial \tilde{x}_m})\,.$
In general, under this condition, supergravity fields can depend on at most $d$ physical coordinates out of the $2d$ doubled coordinates $x^M$\,.

In the canonical solution, all fields and gauge parameters are independent of the dual coordinates; $\tilde{\partial}^m=0$ and \ac{dft} reduces to conventional supergravity.
If instead all fields depend on $(d-1)$ coordinates $x^i$ and only the \ac{dft} dilaton $d(x)$ has an additional linear dependence on a dual coordinate $\tilde{z}$, \ac{dft} reduces to generalized supergravity~\cite{Sakamoto:2017wor,Sakatani:2016fvh}.

\paragraph{$\beta$-transformation.}

A (local) $\beta$-transformation is a specific $\OO(d,d)$ transformation generated by
\begin{align}
\Exp{\bbeta} &= (\Exp{\bbeta}{}^{M}{}_{N})\equiv (h^{M}{}_{N})=
\begin{pmatrix}
~\delta^m_n ~& ~-\bmr^{mn}(x)~ \\
~ 0 ~& ~\delta_m^n ~
 \end{pmatrix}, & (\bmr^{mn}&=-\bmr^{nm})\,,
\label{eq:beta-rule-NS}
\end{align}
which induces a shift of the $\beta$-field or $E^{mn}$:
\begin{align}
\beta_{0}^{mn}(x) \to \beta^{mn}(x) = \beta_{0}^{mn}(x) - \bmr^{mn}(x) \,,
\label{eq:beta-rule-NS-sift}
\end{align}
where $\beta_0$ is the $\beta$-field on the original background.
The usual supergravity fields ($\CG_{mn}$, $B_{mn}$, $\Phi$, $\hat{F}$, $\hat{C}$) transform as
\begin{equation}
	\begin{aligned}
 \mathcal{H}'&=\Exp{\bbeta^{\rm T}}\,\mathcal{H}\Exp{\bbeta}\,, &
d'&=d\,,
\\
 \hat{F}'&=\Exp{-B_2'\wedge}\Exp{-\beta\vee}\Exp{B_2\wedge}\hat{F}\,,
 &
 \hat{C}'&=\Exp{-B_2'\wedge}\Exp{-\beta\vee}\Exp{B_2\wedge}\hat{C} \,,
\label{beta-formula-B}
\end{aligned}
\end{equation}
where $\hat{F}$\,, $\hat{C}$ are polyforms which are formal summations of all the original R--R field strengths and potentials,
\begin{equation}
	\begin{aligned}
\hat{F}&\equiv\sum_{p=1,3,5,7,9}\hat{F}_p\,, &
\hat{F}_p&\equiv\frac{1}{p!}\hat{F}_{m_1\cdots m_p}\dd x^{m_1}\wedge \cdots \wedge \rmd x^{m_p}\,,\\
\hat{C}&\equiv\sum_{p=0,2,4,6,8}\hat{C}_p\,, &
\hat{C}_p&\equiv\frac{1}{p!}\hat{C}_{m_1\cdots m_p}\dd x^{m_1}\wedge \cdots \wedge \rmd x^{m_p}\,.
\end{aligned}
\end{equation}
The operator $\beta\vee$ which acts on an arbitrary $p$-form $A_p$ is defined as
\begin{align}
\beta\vee A_p=\frac{1}{2}\beta^{mn}\iota_m\iota_n A_p\,,
\end{align}
where $\iota_m$ is the inner product along the $x^m$-direction.
Here it is also convenient to define the R-R fields $(F, A)$ and $(\check{F}, \check{C})$ as
\begin{align}
F&\equiv e^{B_2\wedge} \hat{F}\,, & F&\equiv e^{B_2\wedge} \hat{F}\,, & F&\equiv e^{B_2\wedge} \hat{F}\,, & F&\equiv e^{B_2\wedge} \hat{F}\,.
\label{eq:RR-defs}
\end{align}
In order to distinguish the three definitions of R-R fields, we call $(\hat{F}, \hat{C})$ $B$-untwisted R-R fields
while we call $(\check{F}, \check{C})$ $\beta$-untwisted R-R fields.
The $B$-untwisted fields are invariant under $B$-field gauge transformations while the $\beta$-untwisted
fields are invariant under local $\beta$-transformations.
$(F, A)$ will play an important role when we study the monodromy of T-folds in Section~\ref{sec:YB-T-fold}.

\medskip

Finally, we should stress that unlike the $B$-field gauge transformations, the local $\beta$-transformation is not a gauge transformation.
This fact implies that in general, the $\beta$-transformed background may not satisfy the (generalized) supergravity
equations (\ref{eq:GSEsecAdS}) even if the original background is a solution of the supergravity (or \ac{dft}).

\subsubsection{YB deformations and local $\beta$-transformations.}\label{subsubsec:YB-beta}

Now, let us explain the relation between local $\beta$-deformations and homogeneous \ac{yb} deformations. 
Since the original AdS$_5\times$S$^5$ background does not have a $B$-field,
the $\beta$-transformed background can be expressed as
\begin{equation}
	\begin{aligned}
 \CG_{mn}'+B_{mn}'&=\bigl[(G^{-1}-\beta)\bigr]^{-1}_{mn}\,,
& d'&=d\,,
\\
\hat{F}'&=\Exp{-B_2'\wedge}\Exp{-\beta\vee}\hat{F}_5\,,
&
 \hat{C}'&=\Exp{-B_2'\wedge}\Exp{-\beta\vee}\check{C}_4 \,,
\label{pre-beta-formula-noB}
\end{aligned}
\end{equation}
where $G_{mn}$ is the metric of the original AdS$_5\times$S$^5$ and
$\hat{F}_5$ is the undeformed R--R $5$-form field strength (\ref{eq:uF5})\,.
An important observation made in~\cite{Sakamoto:2017cpu} is that a \ac{yb} deformed background associated with the $r$-matrix (\ref{eq:beta-r-matrix}) can also be generated by
a local $\beta$-transformation with the $\beta$-field
\begin{align}
\beta^{mn}(x) =-\bmr^{mn}(x) = 2\,\eta\,r^{ij}\,\hat{T}^{m}_i(x)\,\hat{T}_j^{n}(x) \,,
\label{YB-beta}
\end{align}
where $\hat{T}^{m}_i(x)$ are Killing vector fields associated to the generators $T_i$ appearing in the $r$-matrix (\ref{eq:beta-r-matrix}).
This implies that the $\beta$-untwisted R-R fields $(\check{F}, \check{C})$ are invariant under homogeneous \ac{yb} deformations.

\medskip

In this review, we will only give a proof for the above relation in the NS sector (see~\cite{Sakamoto:2018krs} for the R-R sector). 
Since the original $\AdS{5}\times\rmS^5$ background does not include a $B$-field,
$E_{mn}$ is simply
\begin{align}
 E_{mn}&= \CG_{mn} = e_{m}{}^{\Loa}\,e_{n}{}^{\Lob}\,\eta_{\Loa\Lob}\,, & 
 E^{mn}&= \eta^{\Loa\Lob}\, e_{\Loa}{}^{m}\,e_{\Lob}{}^{n} \,. 
\end{align}
On the other hand, by using (\ref{eq:G-B-prime}), 
the inverse of $E_{mn}$ is deformed as
\begin{align}
 E'^{mn} \equiv \bigl[(\CG' + B')^{-1}\bigr]^{mn} = (k_+^{-1})^{\Loa\Lob}\, e_{\Loa}{}^{m}\,e_{\Lob}{}^{n} 
 = (\eta^{\Loa\Lob} + 2\,\eta\,\lambda^{\Loa\Lob})\, e_{\Loa}{}^{m}\,e_{\Lob}{}^{n} \,.
\end{align}
Therefore, the deformation can be summarized as
\begin{align}
 E^{mn} \ \to \ E'^{mn} = E^{mn} + 2\,\eta\,\lambda^{\Loa\Lob}\, e_{\Loa}{}^{m}\,e_{\Lob}{}^{n} \,.
\label{eq:EtodE}
\end{align}
By comparing this to the $\beta$-transformation rule \eqref{eq:beta-rule-NS} or (\ref{eq:beta-rule-NS-sift}),
the \ac{yb} deformation can be regarded as a local $\beta$-transformation with the parameter
\begin{align}
\bmr^{mn} = 2\,\eta\,\lambda^{\Loa\Lob}\, e_{\Loa}{}^{m}\,e_{\Lob}{}^{n} \,. 
\label{eq:bmr-YB}
\end{align}
Let us moreover rewrite $\bmr^{mn}$ in \eqref{eq:bmr-YB} using the $r$-matrix instead of $\lambda^{\Loa\Lob}$\,.
By using the $r$-matrix $r=\frac{1}{2}\,r^{ij}\,T_i\wedge T_j$\,,
$\lambda^{\Loa\Lob}$ can be expressed as
\begin{align}
\lambda^{\Loa\Lob} = \str\bigl[ R_{g_{\bos}}(\gP^{\Loa})\,\gP^{\Lob} \bigr]
 &= r^{ij}\, \str\bigl(g_{\bos}^{-1}\,T_{i}\,g_{\bos}\,\gP_{\Lob})\, \str(g_{\bos}^{-1}\,T_{j}\,g_{\bos}\,\gP_{\Loa}) \no\\
 &= -r^{ij}\,\bigl[\Ad_{g_{\bos}^{-1}}\bigr]_i{}^{\Loa}\,\bigl[\Ad_{g_{\bos}^{-1}}\bigr]_i{}^{\Lob}\,,
\end{align}
and (\ref{eq:bmr-YB}) becomes
\begin{align}
 \bmr^{mn} = -2\,\eta\,r^{ij}\,\bigl[\Ad_{g_{\bos}^{-1}}\bigr]_i{}^{\Loa}\,\bigl[\Ad_{g_{\bos}^{-1}}\bigr]_j{}^{\Lob}\, e_{\Loa}{}^{m}\,e_{\Lob}{}^{n} \,. 
\end{align}
By using the Killing vectors \eqref{eq:Killing-Formula}, we obtain the very simple expression
\begin{align}
 \bmr^{mn} = -2\,\eta\,r^{ij}\,\hat{T}_i^m\, \hat{T}_j^n \,.
\label{eq:bmr-YB2}
\end{align}
This implies that the $\beta$-field (\ref{eq:bmr-YB2}) is the bi-vector representation of the $r$-matrix characterizing a \ac{yb} deformation.
If we compute the dual fields $G_{mn}\,, \beta_{mn}$ in the deformed background from the relation (\ref{eq:relation-open-closed}), we obtain
\begin{align}
 \OG_{mn}&=\eta_{\Loa\Lob}\,e_m{}^{\Loa}\,e_n{}^{\Lob}\,, &
 \beta^{mn}&=2\,\eta\,r^{ij}\,\hat{T}_i^m\, \hat{T}_j^n\,.
\label{eq:beta}
\end{align}
The dual metric is invariant under the deformation $\OG_{0,mn}\to \OG_{0,mn} = \OG_{mn}$,
while the $\beta$-field, which is absent in the undeformed background, is shifted as $\beta_0^{mn}=0\ \to\ \beta'^{mn}= -\bmr^{mn}$\,.

\medskip

Next, let us compare (\ref{eq:dilaton-YB}) with the $\beta$-transformation law (\ref{beta-formula-B}) of the dilation.
The invariance of $\Exp{-2d}=\Exp{-2\Phi}\sqrt{-\CG}$ under $\beta$-deformations shows
\begin{align}
 \Exp{2\Phi'} = \frac{\sqrt{-\CG'}}{\sqrt{-\CG}} \Exp{-2\Phi} = \frac{\det (e_{\pm m}{}^{\Loa})}{\det (e_{m}{}^{\Loa})} \Exp{2\Phi}
 = (\det k_{\pm})\Exp{2\Phi} \,. 
\end{align}
Recalling $\Phi=0$ in the undeformed background, the transformation rule \eqref{eq:dilaton-YB} can be understood as a $\beta$-transformation. 
Therefore, the homogeneous \ac{yb} deformed NS--NS fields are precisely the $\beta$-transformed ones. 

\medskip

Finally, let us comment on the usefulness of the $\beta$-transformation rule (\ref{beta-formula-B}).
If the original backgrounds are not described by symmetric cosets or are supported by a non-trivial $H$-flux,
it is not straightforward to define \ac{yb} sigma models in general.
However, the $\beta$-transformation rule (\ref{beta-formula-B}) can be easily applied to almost any background.
More concretely, if a given background has a non-trivial isometry $G$, we can look for a skew-symmetric $r$-matrix that satisfies the homogeneous \ac{cybe} for the Lie algebra $\mathfrak{g}$ of $G$.
If such an $r$-matrix can be found, we can apply the associated $\beta$-transformation to a given background by using the $\beta$-field expressed in terms of the $r$-matrix as in (\ref{eq:beta}).
Indeed, we can consider deformations of Minkowski spacetime~\cite{Matsumoto:2015ypa,Borowiec:2015wua,Pachol:2015mfa,Kyono:2015zeu,Fernandez-Melgarejo:2017oyu} and AdS$_3\times$S$^3\times$T$^4$ supported by $H$-flux~\cite{Sakamoto:2018krs, Araujo:2018rho, Borsato:2018spz}, and show that the deformed backgrounds solve the (generalized) supergravity equations.
This shows that  $\beta$-transformations are a useful new tool to generate solutions to the (generalized) supergravity equations.

\paragraph{$R$-flux.}

When a $\beta$-field exists on a given background,
we can consider the associated tri-vector $R$ known as the \emph{non-geometric $R$-flux}.
This flux is defined as 
\begin{align}
 R\equiv [\beta,\,\beta]_S\,,
\label{Jacobi}
\end{align}
where $[\,,\,]_S$ denotes the Schouten bracket, which is defined for a $p$-vector and a $q$-vector as
\begin{align}
\begin{split}
 &[a_1\wedge \cdots \wedge a_p,\,b_1\wedge \cdots \wedge b_q]_{\rmS} \\ 
\equiv &\sum_{i,j}(-1)^{i+j} [a_i,\,b_j] \wedge a_1\wedge \cdots \check{a_i}\cdots \wedge a_p 
\wedge b_1\wedge \cdots \check{b_j}\cdots \wedge b_q \,,
\label{eq:Sch-bra}
\end{split}
\end{align}
where the czech $\check{a}_i$ denotes the omission of $a_i$\,.

The $\beta$-field on \ac{yb} deformed backgrounds takes the form
\begin{align}
 \beta^{mn} &= - \bmr^{mn} = 2\,\eta\,r^{ij}\,\hat{T}_i^m\, \hat{T}_j^n \,, &
 \beta &= \frac{1}{2}\,\beta^{mn}\,\partial_m\wedge\partial_n 
    = 2\,\eta\,\biggl(\frac{1}{2}\,r^{ij}\,\hat{T}_i \wedge \hat{T}_j\biggr) \,.
\end{align}
By using the Lie bracket for the Killing vector fields, $[\hat{T}_i,\,\hat{T}_j]=-f_{ij}{}^k\,\hat{T}_k$\,,
we obtain
\begin{align}
 R^{mnp} &=3\,\beta^{[m|q}\,\partial_q \beta^{|np]}\no\\
&= -8\,\eta^2\,\bigl(f_{l_1l_2}{}^i\,r^{jl_1}\,r^{kl_2} + f_{l_1l_2}{}^j\,r^{kl_1}\,r^{il_2} + f_{l_1l_2}{}^k\,r^{il_1}\,r^{jl_2}\bigr)\,\hat{T}_i^m\,\hat{T}_j^n\,\hat{T}_k^p = 0\,,
\end{align}
upon using the homogeneous \ac{cybe} \eqref{eq:CYBE-r}\cite{Sakamoto:2017cpu}.
This shows the absence of the $R$-flux in homogeneous \ac{yb}-deformed backgrounds.

\subsubsection{The divergence formula.}

The Killing vector $I$ in the \ac{gse} does not appear in the classical action of the string sigma model.
Therefore, we have to check whether a given background has a Killing vector that allows it to be a solution of the \ac{gse}.
Indeed, as discussed in~\cite{Sakamoto:2017wor,Fernandez-Melgarejo:2018wpg},
$I$ appears in the counterterm which is introduced to cancel the Weyl anomaly of the string sigma model defined on generalized supergravity backgrounds.

However, when we consider \ac{yb} deformations, we have a convenient formula to obtain the Killing vector $I^m$ for \ac{yb} deformed backgrounds.
As discovered in~\cite{Araujo:2017jkb}, the formula has a very simple form\footnote{A general formula for $I$ on the \ac{yb} deformed $\AdS5\times\rmS^5$ backgrounds
was originally obtained in~\cite{Borsato:2016ose}.}
\begin{align}
 I^m = \sfD_n \bmr^{nm} \,,
\label{div-formula}
\end{align}
where $\sfD_n$ is the covariant derivative associated to the original metric, and $\bmr^{mn}$ is given by (\ref{eq:bmr-YB2}).
If the $r$-matrix gives a non-zero $I$\,,
this implies the violation of the unimodularity condition (\ref{unimodular1}). 
To see this, we shall consider non-unimodular $r$-matrices satisfying
\begin{align}
r^{ij}\,[T_i,\,T_j] =r^{ij}\,f_{ij}{}^k\,T_k \neq 0 \,. 
\label{eq:cI-formula}
\end{align}
By using the concrete expression (\ref{eq:bmr-YB2}) for $\bmr^{mn}$,
the divergence formula (\ref{div-formula}) can be rewritten as
\begin{align}
\begin{split}
 &I^m =- \eta\,r^{ij}\,[\hat{T}_i,\,\hat{T}_j]^m = \eta\,r^{ij}\,f_{ij}{}^k\,\hat{T}_k^m
\,,
\end{split}
\label{eq:experimental}
\end{align}
where the Killing vectors $\hat{T}_i$ satisfy
\begin{align}
\,[\hat{T}_i,\,\hat{T}_j]^m = \Lie_{\hat{T}_i}\hat{T}_j^m= - f_{ij}{}^k\,\hat{T}_k^m\,.
\end{align}
The Killing vector $I^m$ represents the amount of the violation of the unimodularity condition.
In the next section, we will see that this formula works well for various non-unimodular $r$-matrices.

\section{Examples of homogeneous YB-deformed  $\AdS{5} \times \rmS^5$ backgrounds}
\label{sec:ExampleYBAdS5}

In this section, we will present a number of examples of homogeneous \ac{yb} deformed $\AdS{5} \times \rmS^5$ backgrounds.

\subsection{Abelian $r$-matrices}

First, let us consider homogeneous \ac{yb} deformations of the $\AdS{5} \times \rmS^5$ background associated to Abelian $r$-matrices.

\subsubsection{The Maldacena--Russo background.}

To demonstrate how to use formula \eqref{pre-beta-formula-noB},
let us consider the \ac{yb}-deformed $\AdS{5}\times S^5$ background associated 
to the classical $r$-matrix~\cite{Matsumoto:2014gwa}
\begin{align}
 r=\frac{1}{2}\,P_1\wedge P_2\,. \label{MR-r}
\end{align}
This $r$-matrix is Abelian and satisfies the homogeneous \ac{cybe} \eqref{eq:CYBE}. 
The associated \ac{yb} deformed background was derived in~\cite{Matsumoto:2014gwa,Kyono:2016jqy}.

The classical $r$-matrix \eqref{MR-r} leads to the associated $\beta$-field
\begin{align}
 \beta =\eta\,\hat{P}_1\wedge \hat{P}_2=\eta\,\partial_1\wedge \partial_2\,.
\end{align}
Then, the $\AdS{5}$ part of the $10\times 10$ matrix $(G^{-1}-\beta)$ is
\begin{align}
 \bigl(G^{-1}-\beta\bigr)^{mn}=
 \begin{pmatrix}
 z^2&0&0&0&0\\
 0&-z^2&0&0&0\\
 0&0&z^2&-\eta&0\\
 0&0&\eta&z^2&0\\
 0&0&0&0&z^2
 \end{pmatrix} \,,
\label{G-beta}
\end{align}
where we have ordered the coordinates as $(z\,,x^0\,,x^1\,,x^2\,,x^3)$\,.
By using the inverse of the matrix \eqref{G-beta} and formula \eqref{pre-beta-formula-noB},
we obtain the NS-NS fields of the \ac{yb}-deformed background,
\begin{align}
\begin{split}
 \rmd s^2&=\frac{\rmd z^2-(\rmd x^0)^2+(\rmd x^3)^2}{z^2}+\frac{z^2\,[(\rmd x^1)^2+(\rmd x^2)^2]}{z^4+\eta^2} +\rmd s_{\rm S^5}^2\,,
\\
 B_2&=\frac{\eta}{z^4+\eta^2}\,\rmd x^1\wedge \rmd x^2\,,\qquad
 \Phi=\frac{1}{2}\,\log\biggl[\frac{z^4}{z^4+\eta^2}\biggr]\,.
\label{MR-NS}
\end{split}
\end{align}
The next task is to derive the R-R fields of the deformed background.
Using the undeformed R-R $5$-form field strength \eqref{eq:uF5} of the $\AdS5\times\rmS^5$ background, we find that the deformed filed  $F=\Exp{-\beta\vee}\hat{F}_5$ is given by
\begin{align}
\begin{split}
 F&=\Exp{-\beta\vee}\hat{F}_5
 =4\,\bigl(\omega_{\AdS5}+\omega_{\rmS^5}\bigr)-4\,\beta\vee\omega_{\AdS5}
\\
 &=4\,\bigl(\omega_{\AdS5}+\omega_{\rmS^5}\bigr)-4\,\eta\,\frac{\rmd z\wedge \rmd x^0\wedge \rmd x^3}{z^5}\,,
\end{split}
\end{align}
which is a linear combination of the deformed R--R field strengths with different rank. 
Hence we can readily read off the following expressions: 
\begin{align}
 F_3&=-4\,\eta\,\frac{\rmd z\wedge \rmd x^0\wedge \rmd x^3}{z^5}\,, &
 F_5&=4\,\bigl(\omega_{\AdS5}+\omega_{\rmS^5}\bigr)\,.
\end{align}
The deformed R--R fields $\hat{F}'$ can be computed as
\begin{align}
 \hat{F}'&=\Exp{-B_2\wedge}F
\no\\
 &=-4\,\eta\,\frac{\rmd z \wedge \rmd x^0\wedge \rmd x^3}{z^5}
 +4\,\biggl(\frac{z^4}{z^4+\eta^2}\,\omega_{\AdS5}+\omega_{\rmS^5}\biggr)
 -4\,B_2\wedge\omega_{\rmS^5}
\end{align}
to obtain
\begin{align}
\begin{split}
 \hat{F}_1'&=0\,,\qquad \hat{F}_3'=-4\,\eta\,\frac{\rmd z \wedge \rmd x^0\wedge \rmd x^3}{z^5}\,,
\\
 \hat{F}_5'&=4\,\biggl(\frac{z^4}{z^4+\eta^2}\,\omega_{\AdS5}+\omega_{\rmS^5}\biggr)\,,
\\
 \hat{F}_7'&=-4\,B_2\wedge\omega_{\rmS^5}\,.
\label{MR-RR}
\end{split}
\end{align}
The full deformed background, given by \eqref{MR-NS} and \eqref{MR-RR}, is a solution of standard type IIB supergravity.
This is
the gravity dual of non-commutative gauge theory~\cite{Hashimoto:1999ut,Maldacena:1999mh}. 

\medskip 

This example shows how, instead of the cumbersome supercoset construction, we can derive the full expression of the \ac{yb}-deformed backgrounds using formula \eqref{pre-beta-formula-noB}, which only requires the knowledge of the classical $r$-matrix.

\subsubsection{Lunin--Maldacena--Frolov background.}

Next, we will consider $r$-matrix 
\begin{align}
r=\frac{1}{2}(\mu_3\,h_1\wedge h_2+\mu_1\,h_2\wedge h_3+\mu_2\,h_3\wedge h_1)\,,
\label{eq:abelian}
\end{align}
which is composed of the Cartan generators $h_1$\,, $h_2$\, and $h_3$ of $\mathfrak{su}(4)$\,.
Here, the $\mu_i$~($i=1,2,3$) are deformation parameters.
The metric and $B$--field were computed in~\cite{Matsumoto:2014nra} and
the full background was reproduced in~\cite{Kyono:2016jqy} by performing the supercoset construction.

The associated $\beta$-field is
\begin{align}
\beta=2\eta\,(\mu_3\,\partial_{\phi_1}\wedge \partial_{\phi_2}+\mu_1\,\partial_{\phi_2}\wedge \partial_{\phi_3}+\mu_2\,\partial_{\phi_3}\wedge \partial_{\phi_1})\,.
\end{align}
By using the formula (\ref{pre-beta-formula-noB}),
we obtain the deformed background
\begin{align}
\begin{split}
\rmd s^2 &= \rmd s^2_{\AdS5}+\sum_{i=1}^{3} \left( {d\rho_i}^2+G(\hat{\gamma}_i) {\rho_i}^2 {d\phi_i}^2 \right)
+G(\hat{\gamma}_i){\rho_1}^2 {\rho_2}^2 {\rho_3}^2 \left( \sum_{i=1}^{3} \hat{\gamma}_i d{\phi_i} \right)^2\,,\\
B_2&=G(\hat{\gamma}_i)\,(\hat{\gamma}_3\,{\rho_1}^2{\rho_2}^2 d\phi_1\wedge d\phi_2
+\hat{\gamma}_1\,{\rho_2}^2{\rho_3}^2 d\phi_2\wedge d\phi_3
+\hat{\gamma}_2\,{\rho_3}^2{\rho_1}^2 d\phi_3\wedge d\phi_1)\,,\\
\Phi&=\frac{1}{2}\log\,G(\hat{\gamma}_i)\,,\\
\hat{F}_3&=-4\,\sin^3\alpha\,\cos\,\alpha\,\sin\theta\,\cos\theta\left(\sum_{i=1}^3\hat{\gamma}_i\,d\phi_i\right)\wedge d \alpha \wedge d \theta\,,\\
\hat{F}_5&=4\,\left(\omega_{{\rm AdS}_5}+G(\hat{\gamma}_i)\,\omega_{{\rm S}^5}\right)\,,
\label{3p-Lunin-Maldacena}
\end{split}
\end{align}
where we defined new coordinates $\rho_i~(i=1,2,3)$ as
\begin{align}
\rho_1&= \sin r\cos  \zeta\,, & \rho_2&= \sin r\sin\zeta\,, & \rho_3 &=\cos r\,.
\end{align}
The deformation parameters $\hat{\gamma}_i$ are defined by 
\begin{align}
\hat{\gamma}_i = 8\,\eta\mu_i\,,
\end{align}
and the scalar function $G(\hat{\gamma}_i)$ is given by
\begin{align}
G^{-1}(\hat{\gamma}_i)& \equiv1+\sin^2 r(\hat{\gamma}_1^2\cos^2 r\sin^2\zeta+\hat{\gamma}_2^2\cos^2 r\cos^2\zeta
+\hat{\gamma}_3^2\sin^2 r\sin^2\zeta\cos^2\zeta)\,.
\end{align}
The background (\ref{3p-Lunin-Maldacena}) has originally been derived in~\cite{Frolov:2005dj}. Supersymmetry is completely broken.

\medskip 

This changes when all deformation parameters $\hat{\gamma}_i$ are set equal,
\begin{align}
\hat{\gamma}_1=\hat{\gamma}_2=\hat{\gamma}_3 \equiv \hat{\gamma}\,.
\end{align}
The background becomes
\begin{align}
\begin{split}
ds^2 &= \sum_{i=1}^{3} \left( {d\rho_i}^2+G {\rho_i}^2 {d\phi_i}^2 \right)
+G \hat{\gamma}^2{\rho_1}^2 {\rho_2}^2 {\rho_3}^2 \left( \sum_{i=1}^{3}d{\phi_i}\right)^2\,,\\
B_2&=G\,\hat{\gamma}({\rho_1}^2{\rho_2}^2 d\phi_1\wedge d\phi_2
+{\rho_2}^2{\rho_3}^2 d\phi_2\wedge d\phi_3+{\rho_3}^2{\rho_1}^2 d\phi_3\wedge d\phi_1)\,,\\
\Phi&=\frac{1}{2}\log\,G\,,\\
\hat{F}_3&=-4\hat{\gamma}\,\sin^3\alpha\,\cos\,\alpha\,\sin\theta\,\cos\theta\left(\sum_{i=1}^3\,d\phi_i\right)\wedge d \alpha \wedge d \theta\,,\\
\hat{F}_5&=4\,\left(\omega_{{\rm AdS}_5}+G\,\omega_{{\rm S}^5}\right)\,,
\label{Lunin-Maldacena}
\end{split}
\end{align}
where the scalar function $G$ is 
\begin{align}
G^{-1} \equiv 1+\frac{\hat{\gamma}^2}{4}(\sin^22r+\sin^4r\sin^22\zeta)\,.
\end{align}
In this special case the background preserves $8$ supercharges.
The gauge dual of the background (\ref{Lunin-Maldacena}) is known as $\beta$-deformed $\mathcal{N}$=4 \ac{sym}~\cite{Lunin:2005jy} which is an exactly marginal deformation of $\mathcal{N}$=4 \ac{sym}~\cite{Leigh:1995ep} preserving $\mathcal{N}=1$ supersymmetry.

\subsubsection{Schr\"odinger spacetime.}

Finally, we consider the Abelian $r$-matrix~\cite{Matsumoto:2015uja}
\begin{align}
r=\frac{1}{2}\,P_-\wedge(h_1+h_2+h_3)\,,\label{r-Sch}
\end{align}
where $P_-=(P_0-P_3)/\sqrt{2}$ is a light-cone transformation generator in $\mathfrak{so}(2,4)$\,.
Using the coordinate system given in (\ref{S1overCP2}), (\ref{S1overCP21}),
the $\beta$-field can be expressed as
\begin{align}
\beta=-\eta\,\partial_-\wedge \partial_{\chi}\,.
\end{align}
Via the formula (\ref{pre-beta-formula-noB}), the resulting deformed background is given by
\begin{align}
\begin{split}
\rmd s^2&=\frac{-2dx^+dx^-+(dx^1)^2+(dx^2)^2+dz^2}{z^2}-\eta^2\frac{(dx^+)^2}{z^4}
+ds_{\text{S}^5}^2\,, \\
B_2&=\frac{\eta}{z^2}\,dx^+\wedge (d\chi+\omega)\,,\qquad
\Phi=0\,,\qquad
\hat{F}_5=4\,\left(\omega_{{\rm AdS}_5}+\omega_{{\rm S}^5}\right)\,.
\end{split}
\end{align}
Here the metric of S$^5$ is described as an $S^1$-fibration over $\mathbb{C}\text{P}^2$
and its explicit form is
\begin{align} 
\rmd s^2_{\rm S^5}&=(d\chi+\omega)^2 +ds^2_{\rm \mathbb{C}P^2}\,, \nln 
\rmd s^2_{\rm \mathbb{C}P^2}&= d\mu^2+\sin^2\mu\,
\bigl(\Sigma_1^2+\Sigma_2^2+\cos^2\mu\,\Sigma_3^2\bigr)\,,
\label{S1overCP2}
\end{align}
where $\chi$ is the fiber coordinate and $\omega$ is a one-form potential of the K\"ahler form on $\mathbb{C}$P$^2$\,.
$\Sigma_i ~(i=1,2,3)$ and $\omega$ are defined by
\begin{align}
\begin{split}
\Sigma_1&\equiv \tfrac{1}{2}(\cos\psi\, d\theta +\sin\psi\sin\theta\, d\phi)\,, \\
\Sigma_2&\equiv \tfrac{1}{2}(\sin\psi\, d\theta -\cos\psi\sin\theta\, d\phi)\,, \\
\Sigma_3&\equiv \tfrac{1}{2}(d\psi +\cos\theta\, d\phi)\,, 
\qquad 
\omega \equiv \sin^2\mu\, \Sigma_3\,.
\label{S1overCP21}
\end{split}
\end{align}
Note that the ${\rm S}^5$ part of the metric, the R-R $5$-form field strength and the dilaton remain undeformed.

This deformed background is called \emph{Schr\"odinger spacetime} and was first introduced in~\cite{Israel:2003ry}.
It is the gravity dual of dipole CFTs~\cite{Herzog:2008wg,Maldacena:2008wh,Adams:2008wt}.
Its classical integrability was discussed from the perspective of T-duality~\cite{Orlando:2010ay} in~\cite{arXiv:1011.1771}.
The spectral problem was recently studied in~\cite{Guica:2017mtd} using integrability methods.

\subsection{Non-unimodular classical $r$-matrices}
\label{subsec:Non-unimodular-sol}

In this section, we consider \ac{yb} deformations associated to non-unimodular $r$-matrices.
As explained in the previous sections,
these deformed backgrounds are solutions to the \ac{gse}.
We also show that some of them reduce to the original $\AdS{5} \times S^5$ background
after performing a generalized TsT transformation.

\paragraph{\mbox{\boldmath $1.~r=P_1\wedge D$}.}

As a first example, let us consider the non-Abelian classical $r$-matrix
\begin{equation}
  \label{eq:space-r-matrix}
r=\frac{1}{2}P_1\wedge D\,.
\end{equation}
It is a solution of the homogeneous \ac{cybe} which was already used to study 
a \ac{yb} deformation of four-dimensional Minkowski spacetime~\cite{Matsumoto:2015ypa}. 

The corresponding $\beta$-field is
\begin{align}
\label{eq:nonunimod-beta}
\beta=\eta\,\partial_1 \wedge (t\,\partial_t+z\partial_z)\,,
\end{align}
where we have rewritten the four-dimensional Cartesian coordinates as
\begin{align}
  \label{eq:cartesian-coord-x0x2x3}
  x^0 &=t\sinh\phi\,,  & x^2 &=t\cosh\phi\cos\theta\,,  & x^3 &=t\cosh\phi\sin\theta\,.
\end{align}
Then, the deformed background is found to be\footnote{The metric and NS-NS two-form 
were computed in~\cite{vanTongeren:2015uha}.} 
\begin{equation}
  \begin{aligned}
    \dd{s}^2 &= \frac{z^2[\dd{t}^2+(\dd{x^1})^2+\dd{z}^2]+\eta^2(\dd{t}-t
      z^{-1}\dd{z})^2}{z^4+\eta^2(z^2+t^2)}
    +\frac{t^2(-\dd{\phi}^2+\cosh^2\phi \dd{\theta}^2)}{z^2}
    +\dd{s_{{\rm S}^5}^2}\,, \\
    B_2 &=-\eta\,\frac{t \dd{t} \wedge \dd{x^1} + z \dd{z}\wedge \dd{x^1}}{z^4+\eta^2(t^2+z^2)}\,, 
\qquad  \Phi = \frac{1}{2}\log
    \left[\frac{z^4}{z^4+\eta^2(t^2+z^2)}\right]\,,
\\
    \hat{F}_3 &= -\frac{4\eta\, t^2\cosh\phi}{z^4}\left[\dd{t} \wedge
      \dd{\theta} \wedge \dd{\phi}
      -\frac{t}{z}\dd{\theta}\wedge \dd{\phi} \wedge \dd{z}  \right]\,, \\
    \hat{F}_5 &=
    4\left[\frac{z^4}{z^4+\eta^2(t^2+z^2)}\omega_{\AdS5}+\omega_{{\rm
          S}^5}\right]\,,\qquad I=-\eta\,\partial_1\,.
  \end{aligned}
\label{space}
\end{equation}
Note here that the $\phi$ direction has time-like signature. 
These fields \emph{do not} satisfy the \ac{eom} of type IIB
supergravity, but solve the equations of generalized type IIB supergravity.
The \ac{gse} has the Killing vector $I=\eta\,\partial_1$ which satisfies the divergence formula
\begin{align}
 I^1 = \eta = -\sfD_m\beta^{1m} \,.
\end{align}
Let us now perform T-dualities on the deformed background
(\ref{space}). Following~\cite{Arutyunov:2015mqj}, the extra fields are traded for
a linear term in the dual dilaton. 
T-dualizing along the $x^1$ and $\phi_3$ directions, we find:
\begin{equation}
  \label{T-4.3}
  \begin{aligned}
   \dd{s}^2 ={}& z^2\,(\dd{x^1})^2+\frac{1}{z^2}\Bigl[ (\dd{t} + \eta t
    \dd{x^1})^2+(\dd{z}+\eta z \dd{x^1})^2
    - t^2\dd{\phi}^2+t^2\cosh^2\phi \dd{\theta}^2\Bigr] \\
    &+ \dd{r}^2 +\sin^2 r \dd{\xi}^2+\cos^2\xi \sin^2 r\,
    \dd{\phi_1}^2
    + \sin^2r\sin^2\xi \dd{\phi_2}^2 + \frac{ \dd{\phi_3}^2}{\cos^2r}\,,\\
    e^{\Phi}\hat{F}_5 ={}&
    \frac{4t^2\cosh\phi }{z^4\cos r} (\dd{t}+\eta t \dd{x^1})\wedge
    (\dd{z}+\eta z \dd{x^1})
    \wedge \dd{\theta}\wedge \dd{\phi}\wedge \dd{\phi_3}  \\
    & +
    2z\sin^3 r\sin2\xi \dd{x^1}\wedge \dd{r}\wedge \dd{\xi} \wedge
    \dd{\phi_1}\wedge \dd{\phi_2}
    \,, \\
    \Phi ={}& \eta x^1+\log\left[\frac{z}{\cos
        r}\right]\,.
  \end{aligned}
\end{equation}
Remarkably, this is a solution 
of the \emph{usual} type IIB supergravity equations rather than the generalized ones. 
Note that the dilaton has a linear dependence on \(x^1\).
This same strategy was used in~\cite{Hoare:2015wia} to show that 
the Hoare--Tseytlin solution is ``T-dual'' to the $\eta$-deformed background.

\medskip

The ``T-dualized'' background in \eqref{T-4.3} is a solution to the standard type
IIB equations and has a remarkable property: it is \emph{locally equivalent} to 
undeformed \(\AdS{5} \times S^5\). Let us first perform the following change of coordinates:
\begin{align}
  t &= \tilde{t}(1-\eta\,\tilde{x}^1)\,, & 
                                  z &= \tilde{z}(1-\eta\, \tilde{x}^1)\,, & 
                                       x^1 &= -\frac{1}{\eta}\log(1-\eta\,\tilde{x}^1)\,.
\end{align}
Note that the new coordinate system does {\it not} cover all
of spacetime: the new coordinate $\tilde{x}^1$ has to be restricted to the region $\tilde{x}^1<\eta^{-1}\,$.
The signature of $\eta$ was fixed when we chose the deformation. 
This change of coordinates achieves the following points:
\begin{itemize}
\item it diagonalizes the metric;
\item it absorbs the \(x^1\)-dependence of the dilaton into the
  \(\tilde z\) variable, such that \(\partial_1\) is now a symmetry of the
  full background.
\end{itemize}
Explicitly, we find
\begin{equation}
  \begin{aligned}
    \dd{s}^2 ={}&\tilde{z}^2\,(\dd{\tilde{x}^1})^2+\frac{1}{\tilde{z}^2}\Bigl[
    \dd{\rho}^2+\dd{\tilde{z}}^2
    -\rho^2\dd{\phi}^2+\rho^2\cosh^2\phi \dd{\theta}^2\Bigr] \\
    & +\dd{r}^2 +\sin^2 r \dd{\xi}^2+\cos^2\xi \sin^2 r\,
    \dd{\phi_1}^2
    + \sin^2r\sin^2\xi \dd{\phi_2}^2 + \frac{ \dd{\phi_3}^2}{\cos^2r}\,,\\
    e^{\Phi}\hat{F}_5 ={}&
    \frac{4\rho^2\cosh\phi }{\tilde{z}^4\cos r} \dd{\rho}\wedge
    \dd{\tilde{z}}
    \wedge \dd{\theta}\wedge \dd{\phi}\wedge \dd{\phi_3} \\
    & +
    2\tilde{z}\sin^3 r\sin2\xi \dd{\tilde{x}^1}\wedge \dd{r}\wedge
    \dd{\xi} \wedge \dd{\phi_1}\wedge \dd{\phi_2} \,,\\
    \Phi ={}& \log\left[\frac{\tilde{z}}{\cos r}\right]\,.
  \end{aligned}
\end{equation}
Now we can perform again the two standard T-dualities along \(\tilde{x}^1\) and
\(\phi_3\) to find, as mentioned above, the \emph{undeformed} \(\AdS{5} \times \rmS^5\) background\footnote{The
  usual Poincar\'e coordinates are found using the same change of
  coordinates as in \eqref{eq:cartesian-coord-x0x2x3}.}.

\medskip 

Let us summarize what we have done.
We have started with a \ac{yb} deformation of \(\AdS{5}\) described by the non-Abelian \(r\)-matrix \eqref{eq:space-r-matrix}. Using the formula (\ref{pre-beta-formula-noB}),
we have found the corresponding deformed background \eqref{space} which is a solution to the
generalized equations described in Section~\ref{sec:GSE-YBsec}.
Then we have ``T-dualized'' this background using the rules of~\cite{Arutyunov:2015mqj}
to find a new background \eqref{T-4.3} which solves the \emph{standard}
supergravity equations, but whose dilaton depends linearly on one of
the T-dual variables.
Finally, we have observed that after a change of variables,
this last background is locally equivalent to the T-dual of the undeformed \(\AdS{5} \times S^5\). 
This result implies that the \ac{yb} deformation with the classical $r$-matrix in \eqref{eq:space-r-matrix} 
can be interpreted as an integrable twist, just like in the case of Abelian classical $r$-matrices 
(see for example~\cite{Frolov:2005dj,Alday:2005ww,Vicedo:2015pna,Kameyama:2015ufa}).

\paragraph{\mbox{\boldmath $2.~r=(P_0-P_3)\wedge (D+M_{03})$}.}

Our next example is the classical \(r\)-matrix
\begin{equation}
  r = \frac{1}{2\sqrt{2}} (P_0-P_3)\wedge (D+M_{03})\,,
  \label{r-4.1}
\end{equation}
where \(M_{03}\) is the generator of the Lorentz rotation in the plane
\((x^0, x^3)\).
Then the $\beta$-field is
\begin{align}
\beta=\eta\,\partial_- \wedge (\rho\,\rmd \rho+z\,\rmd z)\,,
\end{align}
where the Cartesian coordinates of four-dimensional Minkowski
spacetime $x^{\mu}$ are
\begin{align}
  x^\pm &=\frac{1}{\sqrt{2}} (x^0\pm x^3)\,, & 
 x^1 &=\rho \cos\theta\,, & x^2 &=\rho \sin\theta\,. 
\end{align}
The divergence of the $\beta$-field is given by 
\begin{align}
 I^- = -2\eta = \sfD_m\beta^{-m} \,.
\end{align}
Performing the supercoset construction~\cite{Kyono:2016jqy}, we obtain the corresponding background:
\begin{equation}
  \label{KY-sol}
  \begin{aligned}
    \dd{s^2} &= \frac{-2\dd{x^+}\dd{x^-} + \dd{\rho^2} + \rho^2\dd{\theta^2} + \dd{z^2}}{z^2}
    -\eta^2\left[\frac{\rho^2}{z^6}+\frac{1}{z^4}\right](\dd{x^+})^2 + \dd{s_{S^5}^2}\,, \\
    B_2 &= -\eta\,\frac{\dd{x^+}\wedge (\rho \dd{\rho}+z\,\dd{z})}{z^4}\,, \\
    \hat{F}_3 &= 4\eta \left[\frac{\rho^2}{z^5} \dd{x^+} \wedge \dd{\theta} \wedge
      \dd{z}
      +\frac{\rho}{z^4} \dd{ x^+}\wedge \dd{\rho} \wedge \dd{\theta} \right]\,, \\
    \hat{F}_5 &= 4 (\omega_{\AdS5}+\omega_{{\rm S}^5}) \,, \\
    \Phi &= \Phi_0 \text{ (constant),} \qquad
 I=-2\eta\, \partial_-\,.
  \end{aligned}
\end{equation}
This background is a solution of the \ac{gse} characterized by the Killing vectors $I=-2\eta\, \partial_-$\,.

\medskip

Let us perform four ``T-dualities'' along the $x^+\,,x^-\,,\phi_1$ and $\phi_2$ directions\footnote{
To perform the T-dualities in the two light-like directions one can equivalently pass to 
Cartesian coordinates \((x^0, x^3)\), T-dualize in these and finally introduce light-like combinations 
for the T-dual variables.}. 
The resulting background is given by 
\begin{equation}
  \label{T-4.1}
  \begin{aligned}
    \dd{s}^2 ={}& -2z^2\dd{x^+}\dd{x^-}
    +\frac{(\dd{\rho}+\eta \rho\dd{x^-})^2+\rho^2\dd{\theta}^2+(\dd{z}+\eta z\dd{x^-})^2}{z^2} \\
     &+ \dd{r}^2 + \sin^2 r\, \dd{\xi}^2+\frac{ \dd{\phi_1}^2}{\cos^2\xi
      \sin^2 r}
    +\frac{ \dd{\phi_2}^2}{\sin^2r\sin^2\xi}+\cos^2r \dd{\phi_3}^2\,,   \\
    e^{\Phi}\hat{F}_5 ={}& \frac{4 i \rho}{z^3\sin\xi \cos\xi\sin^2 r}
    (\dd{\rho}+\eta \rho\dd{x^-})\wedge \dd{\theta} \wedge (\dd{z}+\eta z\dd{x^-})
    \wedge \dd{\phi_1}\wedge \dd{\phi_2}  \\
    & + 4i z^2\sin r\cos r \dd{x^+}\wedge \dd{x^-}\wedge \dd{r} \wedge \dd{\xi}\wedge \dd{\phi_3}\,, \\
    \Phi ={}& 2\eta x^-+\log \left[\frac{z^2}{\sin^2r
        \sin\xi\cos\xi}\right]\,,
  \end{aligned}
\end{equation}
where all other components are zero.

The ``T-dualized'' background in \eqref{T-4.1} is a solution to the standard type
IIB equations and is again \emph{locally} equivalent to 
undeformed \(\AdS{5} \times S^5\). Let us first change the coordinates as follows:
\begin{align}
  x^- &= -\frac{1}{2 \eta} \log( 1-2\eta\,\tilde{x}^-)\, , & \rho &= \tilde \rho \sqrt{1-2\eta\,\tilde{x}^-}  \, , z &= \tilde z \sqrt{1-2\eta\,\tilde{x}^-} \, .
\end{align}
Explicitly, we find
\begin{equation}
  \label{eq:light-like-IIB-solution}
  \begin{aligned}
    \dd{s}^2 ={}& -2\tilde{z}^2\dd{x^+} \dd{\tilde{x}^-} +\frac{\dd{\tilde{\rho}}^2
      +\tilde{\rho}^2\dd{\theta}^2+\dd{\tilde{z}}^2}{\tilde{z}^2} \\
    &+ \dd{r}^2 + \sin^2 r \dd{\xi}^2+\frac{ \dd{\phi_1}^2}{\cos^2\xi \sin^2 r}
    +\frac{ \dd{\phi_2}^2}{\sin^2r\sin^2\xi}+\cos^2r \dd{\phi_3}^2\,,   \\
    e^{\Phi}\hat{F}_5 ={}& \frac{4 i \tilde{\rho}}{\tilde{z}^3\sin\xi
      \cos\xi\sin^2 r}\,
    \dd{\tilde{\rho}}\wedge \dd{\theta}\wedge \dd{\tilde{z}}\wedge \dd{\phi_1}\wedge \dd{\phi_2} \\
    &+ 4i \tilde{z}^2\sin r\cos r\,
    \dd{x^+}\wedge \dd{\tilde{x}^-}\wedge \dd{r} \wedge \dd{\xi}\wedge \dd{\phi_3}\,,  \\
    \Phi ={}& \log \left[\frac{\tilde{z}^2}{\sin^2 r
        \sin\xi\cos\xi}\right]\,.
  \end{aligned}
\end{equation}
Now, rewriting the light-like coordinates in terms of the Cartesian
coordinates as
\begin{align}
  x^+ &\equiv \frac{1}{\sqrt{2}}(\tilde{x}^0 + \tilde{x}^3)\,, &
\tilde{x}^- &\equiv \frac{1}{\sqrt{2}}(\tilde{x}^0 - \tilde{x}^3) \,,
\end{align}
and performing four T-dualities along $\tilde{x}^0$, $\tilde{x}^3$, $\phi_1$ and $\phi_2$, 
we reproduce the \emph{undeformed} \(\AdS{5} \times S^5\) background. 

\paragraph{Mixing of Abelian and non-Abelian classical $r$-matrices.}

This example admits a generalization, obtained by mixing Abelian and non-Abelian classical $r$-matrices:
\begin{equation}
  r = \frac{1}{2\sqrt{2}}(P_0-P_3)\wedge \left[a_1 (D+M_{03})+a_2 M_{12}\right]\,. 
\label{4.7-r}
\end{equation}
When $a_2=0 $, the classical $r$-matrix reduces to the one
described above; when $a_1=0$, the $r$-matrix becomes Abelian 
and the associated background is the Hubeny--Rangamani--Ross solution
of~\cite{Hubeny:2005qu}, as shown in~\cite{Kyono:2016jqy}. 

In~\cite{Kyono:2016jqy} it was shown that with a supercoset construction, one
finds the following ten-dimensional background:
\begin{equation}
  \begin{aligned}
    \dd{s}^2 &=
    \frac{-2\dd{x^+}\dd{x^-}+\dd{\rho}^2+\rho^2\dd{\theta}^2+\dd{z}^2}{z^2}
    -\eta^2\left[(a^2_1+a^2_2)\frac{\rho^2}{z^6}+\frac{a_1^2}{z^4}\right](\dd{x^+})^2+\dd{s}_{\rm S^5}^2\,, \\
    B_2 &= -\frac{\eta}{z^4}\dd{x^+}\wedge \left[ a_1 ( \rho \dd{\rho} + z \dd{z} )
      - a_2 \rho^2 \dd{\theta} \right]\,, \\
    \hat{F}_3 &= \frac{4\eta\rho}{z^5}\dd{x^+}\wedge \left[ a_1 ( z \dd{\rho} - \rho \dd{z} ) \wedge \dd{\theta}
      + a_2 \dd{\rho}\wedge \dd{z} \right]\,, \\
    \hat{F}_5 &= 4 (\omega_{\AdS5} + \omega_{{\rm S}^5})\,, \\
    \Phi &= \Phi_0 \text{ (constant)}\,,\qquad
   I=-2\eta\,a_1\partial_-\,.
  \end{aligned}
\end{equation}
This background is still a solution of the \ac{gse} with the Killing vector $I=-2\eta\,a_1\partial_-$\,.
This background can be reproduced by using the formula (\ref{pre-beta-formula-noB}) with
\begin{align}
\beta=\eta\,\partial_- \wedge[a_1\,(\rho\,\rmd \rho+z\,\rmd z)-a_2\,\partial_{\theta}]\,.
\end{align}
The Killing vector $I$ also satisfies
\begin{align}
 I^- = -2\eta\,a_1 = \sfD_m\beta^{-m} \,.
\end{align}
In the special case $a_1=0$, the above background reduces to 
a solution of standard type IIB supergravity. 

\medskip

Let us next perform four ``T-dualities'' along the $x^+\,,x^-\,,\phi_1$ and $\phi_2$ directions.
Then we can obtain a solution of the usual type IIB supergravity:
\begin{equation}
  \label{T-4.7}
  \begin{aligned}
    \dd{s}^2 ={}&-2z^2\dd{x^+}\dd{x^-} +\frac{(\dd{\rho}+\eta a_1
      \rho\dd{x^-})^2+\rho^2(\dd{\theta}-\eta a_2\dd{x^-})^2
      +(\dd{z}+\eta a_1 z\dd{x^-})^2}{z^2} \\
    & + \dd{r}^2 + \sin^2 r \dd{\xi}^2+\frac{
      \dd{\phi_1}^2}{\cos^2\xi \sin^2 r}
    +\frac{ \dd{\phi_2}^2}{\sin^2r\sin^2\xi}+\cos^2r \dd{\phi_3}^2\,,  \\
    e^{\Phi}\hat{F}_5 ={}& \frac{4 i \rho}{z^3\sin\xi \cos\xi\sin^2 r}
    (\dd{\rho}+\eta a_1 \rho\dd{x^-})\wedge (\dd{\theta}-\eta
    a_2\dd{x^-})
    \wedge (\dd{z}+\eta a_1 z\dd{x^-})\wedge \dd{\phi_1}\wedge \dd{\phi_2}  \\
    & + 4i z^2\sin r\cos r \dd{x^+}\wedge \dd{x^-}\wedge \dd{r} \wedge \dd{\xi}\wedge \dd{\phi_3}\,,  \\
    \Phi ={}& 2\eta a_1x^-+\log\left[\frac{z^2}{\sin^2r
        \sin\xi\cos\xi}\right]\,, 
  \end{aligned}
\end{equation}
where all other components are zero.
It is easy to see that this is just a twist of the previous solution
(in \eqref{eq:light-like-IIB-solution}) and in fact there is a
change of variables 
\begin{align}
  \rho &= \tilde{\rho}\,{\rm e}^{-\eta a_1\, x^-}\,, &
                                                      z &= \tilde{z}\,{\rm e}^{-\eta a_1\, x^-}\,, &
\theta &= \tilde{\theta}-\eta\, a_2 x^-\,, &
x^- &= -\frac{1}{2\eta a_1}\log(1-2\eta a_1\, \tilde{x}^-)\,,
\end{align}
that maps this background to the same local form:
\begin{equation}
  \begin{aligned}
    \dd{s}^2 ={}& -2\tilde{z}^2\dd{x^+}d\tilde{x}^-+\frac{\dd{\tilde{\rho}}^2
      +\tilde{\rho}^2\dd{\tilde{\theta}}^2+\dd{\tilde{z}}^2}{\tilde{z}^2} \\
    & + \dd{r}^2 + \sin^2 r \dd{\xi}^2+\frac{
      \dd{\phi_1}^2}{\cos^2\xi \sin^2 r}
    +\frac{ \dd{\phi_2}^2}{\sin^2r\sin^2\xi}+\cos^2r \dd{\phi_3}^2\,,   \\
    e^{\Phi}\hat{F}_5 ={}& \frac{4 i \tilde{\rho}}{\tilde{z}^3\sin\xi
      \cos\xi\sin^2 r}\,
    \dd{\tilde{\rho}}\wedge \dd{\tilde{\theta}}\wedge \dd{\tilde{z}}\wedge \dd{\phi_1}\wedge \dd{\phi_2}  \\
    & + 4i \tilde{z}^2\sin r\cos r\,
    \dd{x^+}\wedge \dd{\tilde{x}^-}\wedge \dd{r} \wedge \dd{\xi}\wedge \dd{\phi_3}\,, \\
    \Phi ={}&\log\left[\frac{\tilde{z}^2}{\sin^2r
        \sin\xi\cos\xi}\right]\,,
  \end{aligned}
\end{equation}
which is a T-dual of the undeformed $\AdS{5}\times {\rm S}^5$ background.

\paragraph{\mbox{\boldmath $3.~r=(P_0-P_3)\wedge D$}.}

Our last example is the classical $r$-matrix
\begin{equation}
  r=\frac{1}{2\sqrt{2}} ( P_0 - P_3 ) \wedge D\,,
\end{equation}
which is another solution of the homogeneous \ac{cybe}. 
The associated $\beta$-field is
\begin{align}
\beta=\eta\,\partial_- \wedge (\rho\,\rmd \rho+z\,\rmd z+x^+\,\partial_{+})\,.
\end{align}
Here the following new coordinates have been introduced:
\begin{align}
x^0=\frac{x^++x^-}{\sqrt{2}}\,,\qquad
x^3=\frac{x^+-x^-}{\sqrt{2}}\,,\qquad
x^1=\rho\cos\theta\,,\qquad
x^2=\rho\sin\theta
\,.\label{lccoord}
\end{align}
Using the formula (\ref{pre-beta-formula-noB}), 
the associated background is found to be\footnote{The metric and NS-NS two-form were 
computed in~\cite{vanTongeren:2015uha}.} 
\begin{equation}
  \label{4.3}
  \begin{aligned}
    \dd{s}^2 ={}&\frac{1}{z^4-\eta^2(x^+)^2}\biggl[z^2(-2\dd{x^+}\dd{x^-}+\dd{z}^2)
    +2\eta^2z^{-2}x^+\rho \dd{x^+}\dd{\rho}-\eta^2 z^{-2}\rho^2(\dd{x^+})^2 \\
    &-\eta^2(\dd{x^+}-x^+z^{-1}\dd{z})^2\biggr]+\frac{\dd{\rho}^2+\rho^2\dd{\theta}^2}{z^2}
    +\dd{s_{{\rm S}^5}^2}\,, \\
    B_2 ={}&- \eta\,\frac{\dd{x^+} \wedge (z\dd{z}+\rho \dd{\rho}-x^+\dd{x^-})}{z^4-\eta^2(x^+)^2}\,, \\
    \hat{F}_3 ={}& 4\eta\,\frac{\rho}{z^4}\left[\frac{\rho}{z}\dd{x^+}\wedge \dd{\theta}\wedge \dd{z}
      + \dd{x^+}\wedge \dd{\rho}\wedge \dd{\theta}
      -\frac{ x^+}{z}\dd{\rho}\wedge \dd{\theta} \wedge \dd{z}\right]\,, \\
    \hat{F}_5 ={}& 4\left[\frac{z^4}{z^4-\eta^2(x^+)^2}\,\omega_{\AdS5} + \omega_{{\rm S}^5}\right]\,, \\
    \Phi ={}& \frac{1}{2}\log \left[\frac{z^4}{z^4-\eta^2(x^+)^2}\right]\,,\qquad
    I=-\eta\,\partial_-\,.
  \end{aligned}
\end{equation}
This background satisfies the \ac{gse} with the Killing vector $I=-\eta\,\partial_-$\,.
In particular, the divergence formula (\ref{div-formula}) works well.

\medskip

As of now, an appropriate T-dual frame in which this background is a solution to the standard type IIB equations with a linear dilaton has not been found.
However, the deformed background can be reproduced by a generalized diffeomorphism
which is a gauge symmetry of \ac{dft} (see (18) in~\cite{Sakamoto:2017cpu}).

\section{$T$-folds from YB deformations}
\label{sec:YB-T-fold}

In this chapter, we will concentrate on \ac{yb} deformations of Minkowski and 
$\AdS5\times \rmS^5$ backgrounds,
and show that the deformed backgrounds we consider here 
belong to a specific class of non-geometric backgrounds, called $T$-folds~\cite{Hull:2004in}. 
It is worth noting that these examples have the intriguing feature that 
also the R-R fields are twisted by the $T$-duality monodromy, as opposed to 
the well-known $T$-folds which include no R-R fields.

\subsection{A brief review of $T$-folds}
\label{sec:T-fold}

In this subsection, we briefly explain the notion of the $T$-fold.
A $T$-fold is a generalization of the usual notion of manifold. 
It locally looks like a Riemannian manifold, but its patches are glued together not just by diffeomorphisms but also by $T$-duality. 
T-folds play a significant role in the study of non-geometric fluxes 
beyond the effective supergravity description.
As illustrative examples, 
we revisit two well-known cases in the literature corresponding to 
a chain of duality transformations~\cite{Kachru:2002sk,Shelton:2005cf} 
and to a codimension-1 $5_2^2$-brane solution~\cite{Hassler:2013wsa,LozanoTellechea:2000mc,deBoer:2010ud}. 

\medskip 

Different string theories are related by discrete dualities. 
It is possible that via such a duality transformations, a flux configuration turns into a non-geometric flux, meaning it cannot be realized in terms of 
the usual fields in 10/11-dimensional supergravity. This suggests that 
we need to go beyond the usual geometric isometries to fully understand flux compactifications.

\medskip 

For the case of $T$-duality, one proposal to address this problem is the so-called doubled formalism. 
This construction is based on the generalization of a manifold in which all the local patches are geometric. 
However, the transition functions that are needed to glue these patches include not only the usual diffeomorphisms and gauge transformations, but also $T$-duality transformations.

\medskip 

$T$-fold backgrounds are formulated in an enlarged space with a $T^n\times \tilde T^n$ fibration. The tangent space is the doubled torus $T^n\times \tilde T^n$ and is described by a set of coordinates $Y^M=(y^m, \tilde y_m)$ which transform in the fundamental representation of $\OO(n,n)$. 
The physical internal space arises as a particular choice of a subspace of the double torus, $T^n_{\text{phys}}\subset T^n\times \tilde T^n$ (this is called a polarization). 
Then $T$-duality transformations $\OO(n,n;\mathbb{Z})$ act by changing the physical subspace $T^n_{\text{phys}}$ to a different subspace of the enlarged $T^n \times \tilde T^n$. 
For a geometric background, we have a spacetime which is a geometric bundle, $T^n_{\text{phys}}=T^n$\footnote{We can also have $T^n_{\text{phys}}=\tilde T^n$, which corresponds to a dual geometric description. 
}.
More general non-geometric backgrounds do not fit together to form a conventional manifold: despite being locally well-defined, they don't have a valid global geometric description. Instead, they are globally well-defined as $T$-folds.

\medskip 

This formulation is manifestly invariant under the $T$-duality group $O(n,n;\mathbb{Z})$, which is broken by the choice of polarization.
$T$-duality transformations allow to identify 
the backgrounds that belong to the same physical configuration or duality orbit and just differ 
by a choice of polarization\footnote{These orbits have been determined in terms of a classification of gauged supergravities in~\cite{Dibitetto:2012rk}.}.

\medskip 

Let us now review some examples of $T$-folds 
that have been studied in the literature.

\subsubsection{A toy example.}

We start by reviewing a toy model that involves several duality transformations 
of a given background. 
This example has been discussed in~\cite{Kachru:2002sk,Shelton:2005cf}. 
Before introducing the $T$-fold, we will discuss geometric cases 
such as the twisted torus and the torus with $H$-flux as a warm-up. 

\paragraph{Twisted torus.}

Let us consider the metric of a twisted torus,
\begin{align}
 \rmd s^2 
& = 
 \rmd x^2 + \rmd y^2 + (\rmd z-m\,x\,\rmd y)^2 
 \, ,
 & (m\in\mathbb{Z})
 \,.
\label{eq:twisted-torus-metric}
\end{align}
Note that this is not a supergravity solution for $m\neq 0$, 
but can serve to exemplify the non-geometric global property. 
As this background has isometries along the $y$ and $z$ directions, 
these directions can be compactified with certain boundary conditions. For example, let us take
\begin{align}
 (x,\,y,\,z) &\sim (x,\,y+1,\,z)\,, &
 (x,\,y,\,z) &\sim (x,\,y,\,z+1)\,. 
\end{align}
There is no isometry along the $x$ direction, but there is a Killing vector that can be thought of as a deformation with parameter \(m\):
\begin{align}
 k = \partial_x + m\,y\,\partial_z \,. 
\end{align}
Also this isometry direction can be compactified imposing
\begin{align}
 (x,\,y,\,z) \sim \Exp{k}(x,\,y,\,z)=(x+1,\,y,\,z+m\,y)\,.
\label{eq:twisted-torus-identification}
\end{align}
Under this identification, both the 1-form $e_z\equiv \rmd z-m\,x\,\rmd y$ and the metric \eqref{eq:twisted-torus-metric} are globally well-defined~\cite{Kachru:2002sk}. 

This background can be  regarded as a 2-torus $T^2_{y,z}$ fibered over a base $\rmS^1_x$\,. 
The metric of the 2-torus is
\begin{align}
 (\CG_{mn}) = 
 \begin{pmatrix}
 1 & -m\,x \\
 0 & 1
 \end{pmatrix}
 \begin{pmatrix}
 1 & 0 \\
 0 & 1 
 \end{pmatrix}
 \begin{pmatrix}
 1 & 0 \\
 -m\,x & 1 
 \end{pmatrix} \,. 
\end{align}
Then, as one moves around the base $\rmS_x^1$ , the metric is transformed 
by a $\GL(2)$ rotation. That is to say, for $x\to x+1$, the metric is given by
\begin{align}
 \CG_{mn}(x+1) &= \bigl[\Omega^{\rm T}\,\CG(x)\,\Omega\bigr]_{mn} \,,&
 \Omega^m{}_n &\equiv
 \begin{pmatrix}
 1 & 0 \\
 -m & 1 
 \end{pmatrix} \,. 
\end{align}
This monodromy twist can be compensated by a coordinate transformation
\begin{align}
 y &= y'\,, & z &= z'+m\,y'\,.
\end{align}
In other words, the metric is single-valued up to a diffeomorphism.
In this sense this background is geometric.

\paragraph{Torus with $H$-flux.}

When a $T$-duality is formally performed on the twisted torus \eqref{eq:twisted-torus-metric} 
along the $x$ direction, we obtain the background 
\begin{align}
 \rmd s^2 &= \rmd x^2+\rmd y^2+\rmd z^2\,,& B_2&=-m\,x\,\rmd y\wedge\rmd z\,,
\end{align}
equipped with the $H$-flux
\begin{align}
 H_3=\rmd B_2 = -m\,\rmd x\wedge\rmd y\wedge\rmd z\,.
\end{align}
If we consider the generalized metric \eqref{eq:H-geometric} on the doubled torus 
$(y,z,\tilde{y},\tilde{z})$ associated to this background, 
then we can easily identify the induced monodromy when $x \to x+1$. 
The monodromy matrix is given by
\begin{align}
 \cH_{MN}(x+1) = \bigl[\Omega^{\rm T}\,\cH(x)\,\Omega\bigr]_{MN}\,,\qquad 
 \Omega^M{}_N = \begin{pmatrix}
 \delta^m_n & 0 \\
 2\,m\,\delta_{[m}^y\,\delta_{n]}^z & \delta_m^n
 \end{pmatrix} \in \OO(2,2;\mathbb{Z})\,.
\end{align}
The induced monodromy can be compensated by a constant shift in the $B$-field,
\begin{align}
 B_{yz} ~~\to~~ B_{yz}-m. 
\end{align}
This shift transformation, which makes the background single-valued, 
is a gauge transformation of supergravity. 
In this sense, this background is also geometric.

\paragraph{$T$-fold.}

Finally, let us perform another $T$-duality transformation along the $y$-direction 
on the twisted torus \eqref{eq:twisted-torus-metric}. 
We obtain the background~\cite{Kachru:2002sk}
\begin{align}
 \rmd s^2 &= \rmd x^2 + \frac{\rmd y^2+\rmd z^2}{1+m^2\,x^2} \,, &
 B_2&= \frac{m\,x}{1+m^2\,x^2}\,\rmd y\wedge\rmd z\,. 
 \label{eq:H-Q-flux}
\end{align}
In this case, neither general coordinate transformations nor $B$-field gauge transformations 
are enough to remove the multi-valuedness of the background. 
This can also be seen by calculating the monodromy matrix. 
The associated generalized metric is given by
\begin{align}
 \cH(x)=
\begin{pmatrix}
 \delta_m^p & 0 \\
 -2\,m\,x\,\delta_y^{[m}\,\delta_z^{p]} & \delta^m_p
\end{pmatrix}
\begin{pmatrix}
 \delta_{pq} & 0 \\
 0 & \delta^{pq} 
\end{pmatrix}
\begin{pmatrix}
 \delta^q_n & 2\,m\,x\,\delta_y^{[q}\,\delta_z^{n]} \\
 0 & \delta_q^n
\end{pmatrix} \,. 
\label{eq:H-Q-flux-simple}
\end{align}
We find that, upon the transformation $x\to x+1$, the induced monodromy is 
\begin{align}
 \cH_{MN}(x+1) = \bigl[\Omega^{\rm T}\,\cH(x)\,\Omega\bigr]_{MN} \,,
\qquad \Omega^M{}_N \equiv \begin{pmatrix}
 \delta^m_n & 2\,m\,\delta_y^{[m}\,\delta_z^{n]} \\
 0 & \delta_m^n
\end{pmatrix}\in \OO(2,2;\mathbb{Z})\,. 
\label{eq:monodromy-simple}
\end{align}
The present $\OO(2,2;\mathbb{Z})$ monodromy matrix $\Omega$ takes an upper-triangular form 
i.e. it corresponds to a $\beta$-transformation which is not part of the gauge group of supergravity. 
Hence, to keep the background globally well defined, the transition functions that glue the local patches 
should be extended to the full set of $\OO(2,2;\mathbb{Z})$ transformations beyond 
general coordinate transformations and B-field gauge transformations. 
This is what happens in the $T$-fold case. 

\medskip

In summary, we conclude that a non-geometric background 
with a non-trivial $\OO(n,n;\mathbb{Z})$ monodromy transformation, 
such as a $\beta$-transformation, is a $T$-fold. 
The background \eqref{eq:H-Q-flux} is a simple example.

\medskip 

From the viewpoint of \ac{dft}, 
 $\beta$-transformations can be realized as gauge symmetries by choosing a suitable solution of the section condition. 
Indeed, the above $\OO(2,2;\mathbb{Z})$ monodromy matrix $\Omega$ can be canceled 
by a generalized coordinate transformation on the double torus coordinates $(y,z,\tilde y, \tilde z)$:
\begin{align}
 y&=y'+m\,\tilde{z}'\,,& z&=z'\,,& \tilde{y}&=\tilde{y}'\,,& \tilde{z}&=\tilde{z}' \,.
\label{eq:gen-coord-transf-simple}
\end{align}
In this sense, the twisted doubled torus is globally well-defined in \ac{dft}.

\medskip 

It is also possible to make the single-valuedness manifest by using the dual fields ($\OG_{mn}$\,, $\beta^{mn}$\,, $\tilde{\phi}$) defined by (\ref{eq:H-non-geometric}) or (\ref{eq:relation-open-closed}), (\ref{eq:DFT-dilaton}).
In the non-geometric parametrization (\ref{eq:H-non-geometric}),
the background \eqref{eq:H-Q-flux-simple} becomes
\begin{align}
 \rmd s_{\text{dual}}^2 &=\OG_{mn}\,\rmd x^m\,\rmd x^n 
= \rmd x^2+\rmd y^2+\rmd z^2\,,& \beta^{yz} &= m\,x \,,
\end{align}
and the $\OO(2,2;\mathbb{Z})$ monodromy matrix \eqref{eq:monodromy-simple} corresponds to 
a constant shift in the $\beta$-field, $\beta^{yz}\to \beta^{yz} + m$\,. 
We see that up to a constant $\beta$-shift, which is a gauge symmetry 
\eqref{eq:gen-coord-transf-simple} of \ac{dft}, the background becomes single-valued. 

\medskip 

We now define the non-geometric $Q$-flux~\cite{Grana:2008yw},
\begin{align}
 Q_p{}^{mn} \equiv \partial_p \beta^{mn} \,.
\end{align}
After the transformation $x \to x+1$, the induced monodromy on the $\beta$-field can be measured 
by an integral of the $Q$-flux,
\begin{align}
 \beta^{mn}(x+1)-\beta^{mn}(x) = \int_x^{x+1} \rmd x'^p\,\partial_p \beta^{mn}(x') 
= \int_x^{x+1} \rmd x'^p\,Q_p{}^{mn}(x') \,. 
\end{align}
This expression plays a central role in our argument. 

\medskip 

After this illustrative example we conclude that $Q$-flux backgrounds are globally well-defined 
as $T$-folds. In the next subsection, we give a codimension-1 example 
of the exotic $5_2^2$-brane using the above $Q$-flux. 

\subsubsection{The codimension-1 $5^2_2$-brane background.}

Our second example is a supergravity solution studied in~\cite{Hassler:2013wsa}. 
It is obtained by smearing the codimension-2 exotic $5^2_2$-brane solution 
~\cite{LozanoTellechea:2000mc,deBoer:2010ud}, which is related to the NS5-brane solution 
by two $T$-duality transformations. 
It is also referred to as a $Q$-brane, as it is a source of $Q$-flux as we will see in the following. 
The codimension-1 version of this solution is given by
\begin{align}
\begin{split}
 \rmd s^2 &= m\, x\,(\rmd x^2+\rmd y^2) + \frac{x\,(\rmd z^2+\rmd w^2)}{m\,(x^2+z^2)} 
+ \rmd s_{\mathbb{R}^6}^2 \,,
\\
 B_2&= \frac{x}{m\,(x^2+z^2)}\rmd z\wedge\rmd w \,,\qquad \Phi 
= \frac{1}{2}\,\log\biggl[\frac{x}{m\,(x^2+z^2)}\biggr] \,. 
\end{split}
\end{align}
With the non-geometric parametrization \eqref{eq:H-non-geometric}, 
this solution simplifies: 
\begin{align}
\begin{split}
 \rmd s_{\text{dual}}^2 &= m\, x\,(\rmd x^2+\rmd y^2) 
+ \frac{\rmd z^2+\rmd w^2}{m\,x} + \rmd s_{\mathbb{R}^6}^2 \,, 
\\
 \beta^{zw} &=m\,y \,,\qquad \tilde{\phi} = \frac{1}{2}\,\log \biggl[\frac{1}{m\,x}\biggr] \,. 
\end{split}
\end{align}
Assuming that the $y$ direction is compactified with $y\sim y+1$, 
the monodromy under $y\to y+1$ corresponds to a constant shift of the field $\beta$:
\begin{align}
 \beta^{zw}\to \beta^{zw}+m\,.
\end{align}
As the background is twisted by a $\beta$-shift, this  is an example of a $T$-fold with constant $Q$-flux
\begin{align}
 Q_y{}^{zw} = m \,. 
\end{align}
Finally, the monodromy matrix is given by
\begin{align}
 \cH_{MN}(y+1) &= \bigl[\Omega^{\rm T}\,\cH(y)\,\Omega\bigr]_{MN} \,, & 
 \Omega^M{}_N \equiv& \begin{pmatrix}
 \delta^m_n & 2\,m\,\delta_z^{[m}\,\delta_w^{n]} \\
 0 & \delta_m^n
\end{pmatrix}\in \OO(10,10;\mathbb{Z})\,. 
\end{align}
Employing the knowledge of $T$-folds introduced in this section, 
we will next elaborate on the non-geometric aspects of \ac{yb}-deformed backgrounds when seen as $T$-folds.

\subsection{Non-geometric aspects of YB deformations}
\label{sec:Non-geometric-YB}

Here we will show that various \ac{yb}-deformed backgrounds can be regarded as T-folds.

\subsubsection{$T$-duality monodromy of YB-deformed backgrounds.}
\label{sec:monodromy-YB}

As we explained in Section~\ref{subsubsec:YB-beta},
the homogeneous \ac{yb}-deformed background described by $(\cH',\,d',F')$ always has the structure
\begin{equation}
	\begin{aligned}
		\cH' &= \Exp{\bbeta^{\rm T}}\cH\Exp{\bbeta}\,, &
 d' &= d\,, &
F' &= \Exp{-\beta\vee} \check{F} \,,
\\
\Exp{\bbeta}  &=  \begin{pmatrix}
 \delta^m_n & \beta^{mn} \\
 0 & \delta_m^n
 \end{pmatrix} \,,&
 \beta^{mn} &= 2\,\eta\,r^{ij}\, \hat{T}_i^{m}\,\hat{T}_j^{n} \,,
	\end{aligned}
\end{equation}
where $(\cH,\,d,\,\check{F})$ represent the undeformed background and $F$ is defined in (\ref{eq:RR-defs}). 
In the following examples, the $B$-field vanishes in the undeformed background.
At this stage, we know only the local properties of the \ac{yb}-deformed background.

\medskip

In the examples considered in this section, the bi-vector $\beta^{mn}$ always has a linear coordinate dependence. 
Pick a frame in which  $\beta^{mn}$ depends linearly on a coordinate $y$,
\begin{align}
 \beta^{mn}=&\mathsf{r}^{mn}\,y + \mathsf{\bar{r}}^{mn} & (\mathsf{r}^{mn}:&\text{ constant},\quad \mathsf{\bar{r}}^{mn}:\text{ independent of $y$}) \,,
\end{align}
and the $\beta$-untwisted fields are independent of $y$. 
Then, from the Abelian property
\begin{align}
 \Exp{\bbeta_1+\bbeta_2}&=\Exp{\bbeta1}\Exp{\bbeta_2}=\Exp{\bbeta_2}\Exp{\bbeta_1}\,, &
 \Exp{-(\beta_1+\beta_2)\vee}&=\Exp{-\beta_1\vee}\Exp{-\beta_2\vee}=\Exp{-\beta_2\vee}\Exp{-\beta_1\vee}\,,
\end{align}
we obtain
\begin{equation}
  \begin{aligned}
    \cH_{MN}(y+a) &= \bigl[\Omega_a^{\rm T}\cH(y)\,\Omega_a\bigr]_{MN} \,, & 
    d(y+a) &= d(y) \,, &
    F(y+a) &= \Exp{-\omega_a\vee} F(y) \,,
    \\
    (\Omega_a)^M{}_N &\equiv\begin{pmatrix}
      \delta^m_n & a\,\mathsf{r}^{mn} \\
      0 & \delta_m^n
    \end{pmatrix}\,, & 
    (\omega_a)^{mn} &\equiv a\,\mathsf{r}^{mn} \,.
  \end{aligned}
\end{equation}
This is in general an element of $\OO(10,10;\,\mathbb{R})$.
If there is a value $a_0$ such that the matrix $\Omega_{a_0} \in \OO(10,10;\mathbb{Z})$), 
the background is consistent with an identification in the $y$ direction of the type $y\sim y + a_0$\,. 
This is because $\OO(10,10;\mathbb{Z})$ is a gauge symmetry of String Theory and the background can be identified up to a gauge transformation. 
In this example of a $T$-fold, the monodromy matrices for the generalized metric and R-R fields 
are $\Omega_{a_0}$ and $\Exp{-\omega_{a_0}\vee}$, respectively, 
while the dilaton $d$ is single-valued. 
Note that the R-R potential $A$ has the same monodromy as $F$\,. 

\subsubsection{YB-deformed Minkowski backgrounds.}
\label{sec:non-geometry-Minkowski}

In this subsection, we study \ac{yb}-deformations of Minkowski spacetime 
~\cite{Matsumoto:2015ypa,Borowiec:2015wua}. 
We begin with a simple example of an Abelian \ac{yb} deformation. 
Next we present two purely NS-NS solutions of the \ac{gse} and show that they are $T$-folds. 
These backgrounds have vanishing R-R fields and are the first examples of 
purely NS-NS solutions of the \ac{gse}. 

\paragraph{Abelian example.}

Let us consider the simple Abelian $r$-matrix~\cite{Matsumoto:2015ypa} 
\begin{align}
 r=-\frac{1}{2}\,P_1\wedge M_{23}\,. \label{Melvin-r}
\end{align}
The corresponding \ac{yb}-deformed background is given by 
\begin{align}
 \rmd s^2&= -(\rmd x^0)^2+\frac{(\rmd x^1)^2+\bigl[1+(\eta\, x^2)^2\bigr]\,(\rmd x^2)^2 + \bigl[1+(\eta\, x^3)^2\bigr]\,(\rmd x^3)^2 + 2\,\eta^2\,x^2\,x^3\,\rmd x^2\,\rmd x^3}{1+\eta^2 \, \bigl[(x^2)^2+(x^3)^2\bigr]} 
\nn\\
 & \qquad +\sum_{i=4}^9(\rmd x^i)^2\,,
\nn\\
 B_2 &= \frac{\eta\,\rmd x^1\wedge \bigl(x^2\,\rmd x^3 - x^3\, \rmd x^2\bigr)}{1+\eta^2\, \bigl[(x^2)^2+(x^3)^2\bigr]}\,,\qquad
 \Phi = \frac{1}{2} \log\biggl[\frac{1}{1+\eta^2\, \bigl[(x^2)^2+(x^3)^2\bigr]}\biggr]\,. 
 \label{Melvin}
\end{align}
These expressions appear complicated, but after moving to an appropriate polar coordinate system 
(see Section\,3.1 of~\cite{Matsumoto:2015ypa}), the background \eqref{Melvin} 
is found to be the dual to the well-known Melvin background~\cite{Tseytlin:1994ei,Gibbons:1987ps,Hashimoto:2004pb,Hellerman:2011mv,Hellerman:2012zf,Orlando:2013yea}. 
In~\cite{Matsumoto:2015ypa}, it was reproduced as a \ac{yb} deformation 
with the classical $r$-matrix \eqref{Melvin-r}. 
For later convenience, we will keep using the expression in \eqref{Melvin}. 

The dual parametrization of this background is given by 
\begin{align}
 \rmd s^2_{\text{dual}} = -(\rmd x^0)^2 + \sum_{i=1}^9(\rmd x^i)^2 \,, \quad
 \beta = \eta\,\bigl(x^2 \,\partial_1\wedge \partial_3 -x^3\,\partial_1\wedge \partial_2 \bigr)\,, \quad 
 \tilde{\phi}=0 \,. 
\end{align}
Hence, under a shift $x^2\to x^2+\eta^{-1}$\,, the background receives the $\beta$-transformation
\begin{align}
 \beta ~~\to~~ \beta + \partial_1\wedge \partial_3 \,. 
\end{align}
Therefore, if the $x^2$ direction is compactified with the period $\eta^{-1}$, 
then the monodromy matrix becomes
\begin{align}
 \cH_{MN}(x^2+\eta^{-1}) &= \bigl[\Omega^{\rm T}\cH(x^2)\,\Omega\bigr]_{MN} \,, &
 \Omega^M{}_N &\equiv \begin{pmatrix}
 \delta^m_n & 2\,\delta_1^{[m}\,\delta_3^{n]} \\
 0 & \delta_m^n \end{pmatrix} \in \OO(10,10;\mathbb{Z})\,. 
\end{align}
Thus this background turns out to be a $T$-fold. 

When the $x^3$ direction is also identified with the period $\eta^{-1}$, 
the corresponding monodromy matrix becomes
\begin{align}
 \cH_{MN}(x^3+\eta^{-1}) &= \bigl[\Omega^{\rm T}\cH(x^3)\,\Omega\bigr]_{MN} \,,& 
 \Omega^M{}_N &\equiv \begin{pmatrix}
 \delta^m_n & - 2\,\delta_1^{[m}\,\delta_2^{n]} \\
 0 & \delta_m^n \end{pmatrix} \in \OO(10,10;\mathbb{Z})\,. 
\end{align}

In terms of non-geometric fluxes, this background has a constant $Q$-flux. 
In the examples of $T$-folds given in Section~\ref{sec:T-fold}, 
a background with a constant $Q$-flux, $Q_p{}^{mn}$, 
is mapped to another background with a constant $H$-flux, $H_{pmn}$, 
under a double $T$-duality along the $x^m$ and $x^n$ directions. 
This is not possible in this case because \(\partial_2\) and \(\partial_3\) are not isometries and there is no T-dual frame in which the \(H\)-flux is constant.

\paragraph{Non-unimodular example 1: \ $r = \frac{1}{2} \, (P_0-P_1) \wedge M_{01}$.}
\label{sec:Minkowski-example1}

Let us consider the non-unimodular classical $r$-matrix~\cite{Fernandez-Melgarejo:2017oyu}
\begin{align}
 r = \frac{1}{2} \, (P_0-P_1) \wedge M_{01}\,.
\end{align}
The corresponding \ac{yb}-deformed background is given by
\begin{align}
\begin{split}
 \rmd s^2 &= \frac{-(\rmd x^0)^2+(\rmd x^1)^2}{1- \eta^2\,(x^0+x^1)^2} + \sum_{i=2}^9(\rmd x^i)^2\,, 
\\ 
 B_2 &= -\frac{\eta\, (x^0+x^1)}{1-\eta^2\,(x^0+x^1)^2}\, \rmd x^0 \wedge \rmd x^1\,, \quad 
 \Phi = \frac{1}{2}\log \biggl[\frac{1}{1-\eta^2\,(x^0+x^1)^2}\biggr]\,. 
\end{split}
\label{flatpen}
\end{align}
This background has a coordinate singularity at $x^0+x^1=\pm 1/\eta$ which is removed in the  dual parametrization \eqref{eq:H-non-geometric} where 
the dual fields are given by 
\begin{align}
 \rmd s^2_{\text{dual}} &= -(\rmd x^0)^2+\sum_{i=1}^9(\rmd x^i)^2 \,, &
 \beta &= \eta\,(x^0+x^1)\,\partial_0\wedge \partial_1 \,,\quad \tilde{\phi} = 0 \,,
\end{align}
and they are regular everywhere\footnote{A similar resolution of singularities in the dual parametrization has been used in~\cite{Malek:2012pw,Malek:2013sp} in the context of the exceptional field theory.}.

By introducing the Killing vector $I$ using the divergence formula \eqref{div-formula}, 
\begin{align}
 I = \tilde{\sfD}_{n} \beta^{mn}\,\partial_m = \partial_{n} \beta^{mn}\,\partial_m 
 = \eta\,(\partial_0 -\partial_1) \,,
\end{align}
the background \eqref{flatpen} with this $I$ solves the \ac{gse}. 
Here $\tilde{\sfD}_{n} $ is the covariant derivative associated to the original Minkowski spacetime.

\medskip

Since the $\beta$-field depends linearly on $x^1$, 
the background is twisted by the $\beta$-transformation as one moves along the $x^1$ direction. 
In particular, when the $x^1$ direction is identified with period $1/\eta$, 
this background becomes a $T$-fold with an $\OO(10,10;\mathbb{Z})$ monodromy,
\begin{align}
 \cH_{MN}(x^1+\eta^{-1}) &= \bigl[\Omega^{\rm T}\cH(x^1)\,\Omega\bigr]_{MN} \,, &
 \Omega^M{}_N &\equiv \begin{pmatrix} \delta^m_n & 2\,\delta_0^{[m}\,\delta_1^{n]} \\ 
 0 & \delta_m^n \end{pmatrix} \,. 
\end{align}
Note that an arbitrary solution of the \ac{gse} can be regarded as a solution of \ac{dft}~\cite{Sakamoto:2017wor}. 
Indeed, by introducing light-cone coordinates and a rescaled deformation parameter,
\begin{align}
 x^\pm \equiv \frac{x^0 \pm x^1}{\sqrt{2}} \,,\qquad \bar{\eta} = \sqrt{2}\,\eta \,,
\end{align}
the present \ac{yb}-deformed background can be regarded as the \ac{dft}-solution
\begin{align}
 \cH = \begin{pmatrix}
 0 & -1 & -\bar{\eta}\,x^+ & 0 \\
 -1 & 0 & 0 & \bar{\eta}\, x^+ \\
 -\bar{\eta}\, x^+ & 0 & 0 & (\bar{\eta}\,x^+)^2 -1 \\
 0 & \bar{\eta}\,x^+ & (\bar{\eta}\,x^+)^2 - 1 & 0 
\end{pmatrix}\,,\qquad d=\bar{\eta}\,\tilde{x}_-\,, 
\end{align}
where only the $(x^+,x^-,\tilde{x}_+,\tilde{x}_-)$-components of $\cH_{MN}$ are displayed. 
Note here that the dilaton has an explicit dual-coordinate dependence 
because we are now considering a non-standard solution of the section condition 
which makes this background a solution of the \ac{gse} rather than the usual supergravity. 

\medskip

Before performing the \ac{yb} deformation (\emph{i.e.} for~$\bar{\eta}=0$), 
there is a Killing vector $\chi \equiv\partial_+$\,, 
but the associated isometry is broken for non-zero $\bar{\eta}$\,. 
However, even after deforming the geometry, there exists a generalized Killing vector
\begin{align}
 \chi \equiv \Exp{\bar{\eta}\,\tilde{x}_-}\partial_+\,,
\end{align}
which turns into the original Killing vector in the undeformed limit, $\bar{\eta}\to 0$\,. 
Indeed, we can show that the generalized metric and the \ac{dft} dilaton are invariant under the generalized Lie derivative $\gLie_{\chi}$~\cite{Hull:2009zb,Siegel:1993th} associated to $\chi$,
\begin{align}
\gLie_{\chi}\cH_{MN}&=0\,,& \gLie_{\chi}e^{-2d}&=0\,.
\end{align}
Here, the generalized Lie derivative acts on $\cH_{MN}(x)$ and $d(x)$ as
\begin{align}
\gLie_{\chi}\cH_{MN}&=\chi^{K}\partial_{K} \mathcal{H}_{MN}+(\partial_{M}V^K-\partial^K V_{M})\mathcal{H}_{KN}
+(\partial_{N}V^K-\partial^{K}V_{N})\mathcal{H}_{MK}\,,\\
\gLie_{\chi}e^{-2d}&=\partial_{M}(e^{-2d}\chi^M)\,.
\end{align}
In order to make the generalized isometry manifest, 
let us consider a generalized coordinate transformation,
\begin{align}
 x'^+ &= \Exp{-\bar{\eta} \,\tilde{x}_-} x^+ \,,& \tilde{x}'_- 
 &= -\bar{\eta}^{-1}\Exp{-\bar{\eta}\,\tilde{x}_-}\,,& 
 x'^M &= x^M\quad (\text{others})\,. 
\end{align}
By employing Hohm and Zwiebach's finite transformation matrix~\cite{Hohm:2012gk},
\begin{align}
 \cF_M{}^N = \frac{1}{2}\,\Bigl(\frac{\partial x^K}{\partial x'^M}\frac{\partial x'_K}{\partial x_N}
        +\frac{\partial x'_M}{\partial x_{K}}\frac{\partial x^N}{\partial x'^K}\Bigr) \,,
\end{align}
the generalized Killing vector in the primed coordinates becomes constant, $\chi = \partial'_+$\,. 
We can also check that the generalized metric in the primed coordinate system 
is precisely the undeformed background. At least locally, 
the \ac{yb} deformation can be undone by the generalized coordinate 
transformation\footnote{In the study of \ac{yb} deformations of AdS$_5$\,, 
the similar phenomenon has already been observed in~\cite{Orlando:2016qqu}.}. 
This fact is consistent with the fact that \ac{yb} deformations can be realized 
as generalized diffeomorphisms~\cite{Sakamoto:2017cpu}. 

\paragraph*{Non-Riemannian background.} 
Since the above background has a linear coordinate dependence on $\tilde{x}_-$\,, 
let us rotate the solution to the canonical section 
(\emph{i.e.}~the section in which all of the fields are independent of the dual coordinates). 
By performing a $T$-duality along the $x^-$ direction, we obtain
\begin{align}
 \cH = \begin{pmatrix}
 0 & 0 & -\bar{\eta}\, x^+ & -1 \\
 0 & 0 & (\bar{\eta}\,x^+)^2-1 & \bar{\eta}\,x^+ \\
 -\bar{\eta}\,x^+ & (\bar{\eta}\,x^+)^2 -1 & 0 & 0 \\
 -1 & \bar{\eta}\,x^+ & 0 & 0 
\end{pmatrix}\,,\qquad 
 d=\bar{\eta}\,x^- \,.
\label{eq:non-Riemannian}
\end{align}
The resulting background is indeed a solution of \ac{dft} defined on the canonical section. 
However, this solution cannot be parameterized in terms of $(\CG_{mn},\,B_{mn})$ and 
is called a non-Riemannian background in the terminology of~\cite{Lee:2013hma}. 
This background does not even allow the dual parametrization \eqref{eq:H-non-geometric} in terms of $(\OG_{mn},\,\beta^{mn})$\footnote{For another example of non-Riemannian backgrounds, see~\cite{Lee:2013hma}. A classification of non-Riemannian backgrounds in \ac{dft} has been made in~\cite{Morand:2017fnv}. In the context of the exceptional field theory, non-Riemannian backgrounds have been found in~\cite{Malek:2013sp} even before~\cite{Lee:2013hma}. There, the type IV generalized metrics do not allow either the conventional nor the dual parametrization similar to our solution \eqref{eq:non-Riemannian}.}.

\paragraph{Non-unimodular example 2: \ $r=\frac{1}{2\sqrt{2}}\, \sum_{\mu=0}^4 \bigl(M_{0\mu}-M_{1\mu}\bigr) \wedge P^\mu$.}
\label{sec:Minkowski-example2}

Our next example is the classical $r$-matrix\cite{Borowiec:2015wua}
\begin{align}
 r=\frac{1}{2\sqrt{2}}\, \sum_{\mu=0}^4 \bigl(M_{0\mu}-M_{1\mu}\bigr) \wedge P^\mu\,. 
\end{align}
This classical $r$-matrix is a higher-dimensional generalization of the light-cone 
$\kappa$-Poincar\'e $r$-matrix in the four dimensional case. 
Using light-cone coordinates,
\begin{align}
 x^\pm \equiv \frac{x^0 \pm x^1}{\sqrt{2}} \,,
\end{align}
the corresponding \ac{yb}-deformed background becomes
\begin{align}
\begin{split}
 \rmd s^2&= \frac{-2\,\rmd x^+\, \rmd x^- - \eta^2\,\rmd x^+ \bigl[\sum_{i=2}^4(x^i)^2 \, \rmd x^+ - 2\,x^+ \sum_{i=2}^4x^i\,\rmd x^i \bigr]}{1-(\eta \, x^+)^2} + \sum_{i=2}^9(\rmd x^i)^2\,,
\\
 B_2&= \frac{\eta\,\rmd x^+\wedge \bigl(x^+\,\rmd x^- -\sum_{i=2}^4x^i\,\rmd x^i \bigr)}{1-(\eta \, x^+)^2} \,,\quad 
 \Phi = \frac{1}{2}\log\biggl[\frac{1}{1-(\eta \, x^+)^2}\biggr] \,.
\end{split}
\label{eq:penHvT}
\end{align}
In terms of the dual parametrization, this background becomes
\begin{align}
\begin{split}
 \rmd s^2_{\text{dual}} &= -2\,\rmd x^+\,\rmd x^- + \sum_{i=2}^9(\rmd x^i)^2 \,,\quad \tilde{\phi}=0 \,, 
\\
 \beta &= \eta\,\sum_{\mu=0}^4 \hat{M}_{-\mu} \wedge \hat{P}^\mu = \eta\, \partial_-\wedge \bigl(x^+\,\partial_+ + {\textstyle\sum}_{i=2}^4\, x^i\, \partial_i \bigr) \,.
\end{split}
\end{align}
Again, by introducing the Killing vector $I$ using the divergence formula \eqref{div-formula}, 
\begin{align}
 I = 4\,\eta\,\partial_- \,, 
\end{align}
the background \eqref{eq:penHvT} with this $I$ solves the \ac{gse}. 
This background can also be regarded as the following solution of \ac{dft}:
\begin{align}
\begin{split}
  \cH 
 &= {\tiny\begin{pmatrix}
 0 & -1 & 0 & 0 & 0 & -\eta\,x^+ & 0 & -\eta\,x^2 & -\eta\,x^3 & -\eta\,x^4 \\
 -1 & 0 & 0 & 0 & 0 & 0 & \eta\,x^+ & 0 & 0 & 0 \\
 0 & 0 & 1 & 0 & 0 & 0 & -\eta\,x^2 & 0 & 0 & 0 \\
 0 & 0 & 0 & 1 & 0 & 0 & -\eta\,x^3 & 0 & 0 & 0 \\
 0 & 0 & 0 & 0 & 1 & 0 & -\eta\,x^4 & 0 & 0 & 0 \\
 -\eta\,x^+ & 0 & 0 & 0 & 0 & 0 & (\eta\,x^+)^2 -1 & 0 & 0 & 0 \\
 0 & \eta\,x^+ & -\eta\,x^2 & -\eta\,x^3 & -\eta\,x^4 & (\eta\,x^+)^2 -1 
 & \eta^2\,\sum_{i=2}^4(x^i)^2 & \eta^2\, x^+\,x^2 & \eta^2 \, x^+\,x^3 & \eta^2\,x^+\,x^4 \\
 -\eta\,x^2 & 0 & 0 & 0 & 0 & 0 & \eta^2\, x^+\,x^2 & 1 & 0 & 0 \\
 -\eta\,x^3 & 0 & 0 & 0 & 0 & 0 & \eta^2\, x^+\,x^3 & 0 & 1 & 0 \\
 -\eta\,x^4 & 0 & 0 & 0 & 0 & 0 & \eta^2\, x^+\,x^4 & 0 & 0 & 1 
\end{pmatrix}} \,,
\\
 d &= 4\,\eta\,\tilde{x}_- \,, 
\end{split}
\end{align}
where only the $(x^+,\,x^-,\,x^2,\,x^3,\,x^4,\,\tilde{x}_+,\,\tilde{x}_-,\,\tilde{x}_2,\,
\tilde{x}_3,\,\tilde{x}_4)$-components of $\cH_{MN}$ are displayed. 

When one of the $(x^2,\,x^3,\,x^4)$-coordinates, say $x^2$, is compactified 
with the period $x^2 \sim x^2 + \eta^{-1}$, 
the monodromy matrix is given by 
\begin{align}
 \cH_{MN}(x^2 +\eta^{-1}) &= \bigl[\Omega^{\rm T}\cH(x^2)\,\Omega\bigr]_{MN} \,, & 
 \Omega^M{}_N &\equiv \begin{pmatrix} \delta^m_n & 2\,\delta_-^{[m}\,\delta_2^{n]} \\ 0 & \delta_m^n \end{pmatrix} \in \OO(10,10;\mathbb{Z})\,,
\end{align}
and in this sense the compactified background is a $T$-fold. 
In terms of the non-geometric $Q$-flux, 
this background has the following components: 
\begin{align}
 Q_+{}^{-+} = Q_2{}^{-2} = Q_3{}^{-3} = Q_4{}^{-4} = \eta \,.
\end{align}

\subsection{A non-geometric background from non-Abelian $T$-duality}
\label{sec:non-geometry-NATD}

Before considering \ac{yb}-deformations of $\AdS5\times\rmS^5$\,, let us consider another example with a pure NS-NS background, 
which was found in~\cite{Gasperini:1993nz} via a non-Abelian $T$-duality. 
It takes the form
\begin{align}
\begin{split}
 \rmd s^2&= -\rmd t^2 + \frac{(t^4+y^2)\,\rmd x^2-2\,x\,y\,\rmd x\,\rmd y+(t^4+x^2)\,\rmd y^2+t^4\,\rmd z^2}{t^2\,(t^4+x^2+y^2)} + \rmd s_{T^6}^2\,,
\\
 B_2 &= \frac{(x\,\rmd x+y\,\rmd y)\wedge \rmd z}{t^4+x^2+y^2}\,,\qquad 
 \Phi= \frac{1}{2} \log\biggl[\frac{1}{t^2\,(t^4+x^2+y^2)}\biggr] \,, 
\end{split}
\label{eq:bgNATD}
\end{align}
where $\rmd s_{T^6}^2$ is the flat metric of the 6-torus. 
In terms of the dual parametrization, this background takes 
a Friedmann--Robertson--Walker-type form,
\begin{align}
\begin{split}
 \rmd s^2_{\text{dual}}&= -\rmd t^2 + t^{-2}\,\bigl(\rmd x^2+ \rmd y^2 
 + \rmd z^2\bigr) + \rmd s_{T^6}^2\,,
\\
 \beta &= (x\,\partial_x+y\,\partial_y)\wedge \partial_z\,,\qquad 
 \tilde{\phi} = -\log t^3 \,. \label{3.69}
\end{split}
\end{align}
Note that this background cannot be represented by a coset or a Lie group itself. 
This is because the background \eqref{eq:bgNATD} contains a curvature singularity 
and is not homogeneous. 
Hence the background \eqref{eq:bgNATD} cannot be realized as a \ac{yb} deformation 
and is not included in the discussion of~\cite{Hoare:2016wsk,Borsato:2016pas,Hoare:2016wca}. 

The associated $Q$-flux is constant,
\begin{align}
 Q_y{}^{xy} = Q_z{}^{xz} = -1\,. 
\end{align}
If the $x$-direction is compactified as $x\sim x+1$, 
the background fields are twisted by an $\OO(10,10;\mathbb{Z})$ transformation:
\begin{align}
 \cH_{MN}(x+1) &= \bigl[\Omega^{\rm T}\cH(x)\,\Omega\bigr]_{MN} \,, & 
 \Omega^M{}_N &\equiv \begin{pmatrix} \delta^m_n & 2\,\delta_x^{[m} \,\delta_z^{n]} \\ 0 
 & \delta_m^n \end{pmatrix}\,, & 
 d(x+1)&=d(x) \,. 
\end{align}
Thus the background can be interpreted as a $T$-fold. 
If the $z$-direction is also compactified as $z\sim z+1$, another twist is realized as 
\begin{align}
 \cH_{MN}(y+1) &= \bigl[\Omega^{\rm T}\cH(y)\,\Omega\bigr]_{MN} \,, &
 \Omega^M{}_N &\equiv \begin{pmatrix} \delta^m_n & 2\,\delta_y^{[m} \,\delta_z^{n]} \\ 
 0 & \delta_m^n \end{pmatrix}\,, & 
 d(y+1)&=d(y) \,. 
\end{align}
As stated in~\cite{Gasperini:1993nz}, this background is not a solution of the usual supergravity. 
However, using again the divergence formula $I^m = \tilde{\sfD}_n\beta^{mn}$ and introducing 
the vector field
\begin{align}
 I=-2\,\partial_z \,,
\end{align}
we can see that the background \eqref{eq:bgNATD} together with the vector field $I$ 
satisfies the \ac{gse}. 
Thus, also this background can be regarded as a $T$-fold solution of \ac{dft}. 

\subsection{YB-deformed $\AdS5\times\rmS^5$ backgrounds}
\label{sec:non-geometry-AdS5xS5}

In this section, we will show that various \ac{yb} deformations of the $\AdS5\times\rmS^5$ background are $T$-folds. 
We consider here examples associated to the following five classical $r$-matrices:
\begin{enumerate}
\setlength{\leftskip}{0.7cm} 
\item \quad $r = \frac{1}{2\,\eta}\,\bigl[\eta_1\,(D+M_{+-})\wedge P_+ 
+ \eta_2\,M_{+2}\wedge P_3 \bigr]$\,, 

\item \quad $r = \frac{1}{2}\,P_0\wedge D$\,, 

\item \quad $r = \frac{1}{2}\,\bigl[P_0\wedge D + P^i \wedge (M_{0i}+ M_{1i})\bigr]$\,, 

\item \quad $r = \frac{1}{2\eta}\, P_- \wedge (\eta_1\,D-\eta_2\,M_{+-})$\,, 

\item \quad $r = \frac{1}{2}\, M_{-\mu}\wedge P^\mu$\,. 
\end{enumerate}
All the above $r$-matrices except the first are non-unimodular. 
Note here that the $\rmS^5$ part remains undeformed and only the $\AdS5$ part is deformed. 
As shown in Appendix A in~\cite{Fernandez-Melgarejo:2017oyu}
the second and third examples are reduced to the two examples discussed in the previous subsection taking a (modified) Penrose limit.

\subsubsection{Non-Abelian unimodular $r$-matrix.}

Let us consider the non-Abelian unimodular $r$-matrix 
(see $R_5$ in Tab.\,$1$ of~\cite{Borsato:2016ose}),
\begin{align}
 r = \frac{1}{2\,\eta}\,\bigl[\eta_1\,(D+M_{+-})\wedge P_+ + \eta_2\,M_{+2}\wedge P_3 \bigr] \,.
\end{align}
In light-cone coordinates\footnote{In the following, we use the light-cone convention $\varepsilon_{z+-23r\xi\phi_1\phi_2\phi_3} = + \sqrt{\abs{\CG}}$\,.},
\begin{align}
 x^\pm \equiv \frac{x^0 \pm x^1}{\sqrt{2}} \,. 
\end{align}
the corresponding \ac{yb}-deformed background is given by 
\begin{align}
\begin{split}
 \rmd s^2 &= 
  \frac{-2\,z^2\,\rmd x^+\,\rmd x^- + 4\,\eta_1^2\,z^{-1}\,x^-\,\rmd z\,\rmd x^-}{z^4- (2\,\eta_1\,x^-)^2}+\frac{z^2\,[(\rmd x^2)^2+(\rmd x^3)^2]}{z^4+(\eta_2\,x^-)^2} +\frac{\rmd z^2}{z^2} 
\\
&\quad - \frac{(\eta_1^2+\eta_2^2)\,(z\,x^2)^2 -2\,\eta_1\,\eta_2\,z^2\,x^2\,x^3 + \eta_1^2\,[z^4 + (z\,x^3)^2+ (\eta_2\,x^-)^2]}{[z^4-(2\,\eta_1\,x^-)^2]\,[z^4+(\eta_2\,x^-)^2]}\,(\rmd x^-)^2\\
 &\quad + \frac{2\,\{[x^2 (2\,\eta_1^2+\eta_2^2)-\eta_1\,\eta_2\,x^3]\,z^2\,x^-\,\rmd x^2+\eta_1\,(2\,\eta_1\,x^3-\eta_2\,x^2)\,\rmd x^3\}\,\rmd x^-}{[z^4-(2\,\eta_1\,x^-)^2]\,[z^4+(\eta_2\,x^-)^2]}+ \rmd s_{\rmS^5}^2\,,
\\
 B_2 &= -\biggl[\frac{\eta_1\,\{x^2\,[z^4+2\,(\eta_2\,x^-)^2]-2\,\eta_1\,\eta_2\,(x^-)^2\,x^3\}\,\rmd x^2}{[z^4- (2\,\eta_1\,x^-)^2]\,[z^4+(\eta_2\,x^-)^2]}
\\
&\quad+\frac{\{\eta_1\,z^4\,x^3-\eta_2\,x^2\,[z^4-2\,(\eta_1\,x^-)^2]\}\,\rmd x^3}{[z^4- (2\,\eta_1\,x^-)^2]\,[z^4+(\eta_2\,x^-)^2]}
+\frac{\eta_1\,(z\,\rmd z - 2\, x^-\,\rmd x^+)}{z^4-(2\,\eta_1\,x^-)^2}\biggr]\wedge \rmd x^-\\
 &\quad+\frac{\eta_2\,x^-\,\rmd x^2 \wedge \rmd x^3}{z^4+(\eta_2\,x^-)^2} \,,
\\
 \Phi &=\frac{1}{2} \log \biggl[\frac{z^8}{[z^4- (2\,\eta_1\,x^-)^2]\,[z^4+ (\eta_2\,x^-)^2]}\biggr]\,,
\\
 \hat{F}_1 &=\frac{4\,\eta_1\,\eta_2\,x^-\,(2\,x^-\,\rmd z -z\,\rmd x^-)}{z^5} \,,
\\
 \hat{F}_3 &= -B_2\wedge F_1 +\frac{4\,\eta_1}{z^5}\, \bigl(2\,x^-\,\rmd z - z\,\rmd x^-\bigr)\wedge \rmd x^2 \wedge \rmd x^3 
\\
 &\quad +\frac{4}{z^5}\, \rmd z \wedge \rmd x^- \wedge \bigl[\eta_1\,(x^3\,\rmd x^2- x^2\,\rmd x^3) + \eta_2\,(x^-\,\rmd x^+ - x^2\,\rmd x^2) \bigr] \,,
\\
 \hat{F}_5&=4\,\biggl[\frac{z^8}{[z^4-(2\,\eta_1\,x^-)^2]\,[z^4+ (\eta_2\,x^-)^2]}\,\omega_{\AdS5} + \omega_{\rmS^5}\biggr] \,,
\\
 \hat{F}_7&= -B_2\wedge F_5 \,,\qquad 
 \hat{F}_9= -\frac{1}{2}\,B_2\wedge F_7 \,. 
\end{split}
\end{align}
The expressions are greatly simplified in terms of dual fields:
\begin{align}
\begin{split}
 \rmd s_{\text{dual}}^2 &= \frac{ -2\,\rmd x^+\,\rmd x^-+ (\rmd x^2)^2
 +(\rmd x^3)^2+\rmd z^2}{z^2} + \rmd s_{\rmS^5}^2\,, \qquad 
 \tilde{\phi}=0\,, \\
 \beta &=\eta_1\,\bigl(2\,x^-\,\partial_-+x^2\,\partial_2+x^3\,\partial_3+z\,\partial_z\bigr)
 \wedge\partial_+ + \eta_2\,\bigl(x^2\,\partial_+ + x^-\,\partial_2\bigr)\wedge \partial_3 \,.
\end{split}
\end{align}
The R-R field strengths are given by 
\begin{align}
 \hat{F} &= \Exp{-B_2\wedge}F\,, & F &= \Exp{-\beta\vee}\check{F}\,, & 
 \check{F} &= 4\,\bigl(\omega_{\AdS5} + \omega_{\rmS^5}\bigr)\,.
\end{align}
We see that the $\beta$-untwisted R-R fields $\check{F}$ are invariant 
under the \ac{yb} deformation. 

The non-vanishing \(Q\)-flux components are
\begin{align}
 Q_z{}^{z+} &= \eta_1\,,& 
 Q_-{}^{-+} &= 2\,\eta_1\,,& 
 Q_2{}^{2+} &= \eta_1\,,& 
 Q_3{}^{3+} &= \eta_1\,,& 
 Q_2{}^{+3} &= \eta_2\,,& 
 Q_-{}^{23} &= \eta_2\,. 
\end{align}
This means that we can understand this background as a T-fold if we compactify for example the $x^3$ direction  with period 
$x^3\sim x^3+\eta_1^{-1}$.
The corresponding monodromy is
\begin{align}
 \cH_{MN}(x^3 +\eta_1^{-1}) &= \bigl[\Omega^{\rm T}\cH(x)\,\Omega\bigr]_{MN} \,, & 
 \Omega^M{}_N &\equiv \begin{pmatrix} \delta^m_n & 2\,\delta_3^{[m}\,\delta_+^{n]} \\ 0 & \delta_m^n \end{pmatrix} \in \OO(10,10;\mathbb{Z})\,. 
\end{align}
The R-R fields $F$ are also twisted by the same monodromy,
\begin{align}
 F(x^3+\eta_1^{-1}) &= \Exp{- \omega\vee}F(x^3)\,, & 
 \omega^{mn} &= 2\,\delta_3^{[m} \,\delta_+^{n]} \,. 
\end{align}
\subsubsection{$r=\frac{1}{2}\,P_0\wedge D$.}
\label{sec:AdS-P0-D}

Let us next consider the classical $r$-matrix\cite{vanTongeren:2015uha,Orlando:2016qqu}
\begin{align}
 r=\frac{1}{2}\,P_0\wedge D\,.
\end{align}
Since $[P_0, D] \neq 0$\,, this classical $r$-matrix does not satisfy the unimodularity condition. 
By introducing polar coordinates
\begin{align}
 x^1 &=\rho\sin\theta \cos\phi\,, & x^2 &=\rho\sin\theta\sin\phi\,, 
 & x^3 &=\rho\cos\theta\,,
\end{align}
the deformed background can be rewritten as~\cite{Orlando:2016qqu}\footnote{Only the metric 
and NS-NS two-form were computed in~\cite{vanTongeren:2015uha}.}
\begin{align}
\label{eq:AdS-P0-D}
\begin{split}
 \rmd s^2 &= \frac{z^2\,\bigl[ -(\rmd x^0)^2+\rmd\rho^2+\rmd z^2\bigr]-\eta^2\,(\rmd \rho -\rho\,z^{-1}\,\rmd z)^2}{z^4-\eta^2\,(z^2+\rho^2)}
    +\frac{\rho^2\,(\rmd \theta^2+\sin^2\theta\,\rmd \phi^2)}{z^2}\\
  &\quad  +\rmd s_{\rmS^5}^2 \,, 
\\
    B_2 &= -\eta\,\frac{\rmd x^0 \wedge ( \rho\,\rmd \rho+z\,\rmd z )}{z^4-\eta^2\,(z^2+\rho^2)}\,, \quad 
 \Phi = \frac{1}{2}\log \biggl[\frac{z^4}{z^4-\eta^2\,(z^2+\rho^2)}\biggr]\,,\quad 
 I = -\eta\,\partial_0 \,,
\\
 \hat{F}_1 &=0\,,\quad 
 \hat{F}_3 = \frac{4\,\eta\,\rho^2\sin\theta}{z^5}\, (z\,\rmd \rho - \rho \, \rmd z)\wedge \rmd \theta\wedge \rmd \phi \,,
\\
 \hat{F}_5 &= 4\,\biggl[\frac{z^4}{z^4-\eta^2\, (z^2+\rho^2)}\,\omega_{\AdS5} + \omega_{\rmS^5}\biggr]\,,
\\
 \hat{F}_7&=\frac{4\,\eta\,\rmd x^0 \wedge (\rho\,\rmd \rho+z\,\rmd z)}{z^4-\eta^2\, (z^2+\rho^2)}\wedge \omega_{\rmS^5}\,,\quad \hat{F}_9=0\,. 
\end{split}
\end{align}
This background is not a solution of the usual type IIB supergravity, 
but of the \ac{gse}~\cite{Arutyunov:2015mqj}. 
By setting $\eta=0$, this background reduces to the original $\AdS5\times \rmS^5$. 

\medskip

In the dual parametrization, the dual metric, the $\beta$-field and the dual dilaton 
are given by 
\begin{align}
\begin{split}
 \rmd s_{\text{dual}}^2 &= \frac{\rmd z^2 -(\rmd x^0)^2+(\rmd x^1)^2+(\rmd x^2)^2+(\rmd x^3)^2}{z^2} +\rmd s_{\rmS^5}^2 \,,\qquad \tilde{\phi}=0\,,
\\
 \beta &= \eta\,\hat{P}_0\wedge \hat{D}
=\eta\,\partial_0\wedge (  x^1\,\partial_1 + x^2\,\partial_2 + x^3\,\partial_3+z\,\partial_z) \no\\
 &= \eta\,\partial_0\wedge (\rho\,\partial_\rho+z\,\partial_z)\,. 
\end{split}
\end{align}
The Killing vector $I^m$ satisfies the divergence formula,
\begin{align}
 I^0 = -\eta = \sfD_{m}\beta^{0m} \,. 
\end{align}
The $Q$-flux has the following non-vanishing components:
\begin{align}
 Q_z{}^{0z} = Q_1{}^{01} = Q_2{}^{02} = Q_3{}^{03} = \eta\,. 
\end{align}
Thus, when at least one of the $(x^1,x^2,x^3)$ directions is compactified, 
the background can be interpreted as a $T$-fold. 
When for example the $x^1$ direction is compactified, the monodromy is given by 
\begin{align}
 \cH_{MN}(x^1+\eta^{-1}) &= \bigl[\Omega^{\rm T}\cH(x^1)\,\Omega\bigr]_{MN} \,, & 
 \Omega^M{}_N &\equiv \begin{pmatrix} \delta^m_n & 2\,\delta_0^{[m} \,\delta_1^{n]} \\ 0 & \delta_m^n \end{pmatrix}\,. 
\label{eq:monodromy-AdS-P0-D}
\end{align}
From \eqref{eq:AdS-P0-D}, the R-R potentials are found to be
\begin{align}
\begin{split}
 \hat{C}_0&=0\,,\quad 
 \hat{C}_2= \frac{\eta\,\rho^3\sin\theta}{z^4}\,\rmd \theta\wedge \rmd \phi\,, 
\\
 \hat{C}_4&= \frac{\rho^2\sin\theta}{z^4}\,\rmd x^0\wedge \rmd \rho\wedge \rmd \theta \wedge \rmd\phi + \omega_4 -B_2 \wedge \hat{C}_2 \,,
\\
 \hat{C}_6&= -B_2\wedge \omega_4 \,, \quad
 \hat{C}_8 = 0 \,,
\end{split}
\end{align}
where the $4$-form $\omega_{4}$ satisfies $\omega_{\rmS^5}=\rmd\omega_{4}$\,.
Via the $B$-twist, we obtain
\begin{equation}
\begin{aligned}
 F_1&=0\,, & 
 F_3 &= \frac{4\,\eta \,\rho^2 \sin \theta}{z^5}\, (\rho\,\rmd z- z\,\rmd \rho) \wedge \rmd \theta \wedge \rmd \phi\,,
\\
 F_5&=4\,\bigl(\omega_{\AdS5}+\omega_{\rmS^5} \bigr)\,, & 
F_7 &=0\,, & 
F_9 &=0\,,
\\
 A_0&=0\,, & 
A_2 &= \frac{\eta\,\rho^3 \sin \theta}{z^4}\,\rmd \theta \wedge\rmd \phi\,,
\\
 A_4&= \frac{\rho^2 \sin\theta}{z^4}\,\rmd x^0 \wedge \rmd \rho \wedge \rmd \theta \wedge \rmd \phi+\omega_4 \,, & 
A_6&=0\,, & 
A_8&=0\,. 
\end{aligned}
\end{equation}
We can further compute the $\beta$-untwisted fields,
\begin{equation}
\begin{aligned}
 \check{F}_1 &=0\,, & 
 \check{F}_3 &= 0\,, \quad
 \check{F}_5 =4\,\bigl(\omega_{\AdS5}+\omega_{\rmS^5}\bigr)\,,& 
\check{F}_7 &=0\,,\quad \check{F}_9 =0\,,
\\
 \check{C}_0 &=0\,,\quad \check{C}_2 = 0\,, &
 \check{C}_4 &= \frac{\rho^2 \sin\theta}{z^4}\,\rmd x^0 \wedge \rmd \rho \wedge \rmd \theta \wedge \rmd \phi +\omega_4 \,, & \check{C}_6&=0\,,& \check{C}_8&=0\,.
\end{aligned}
\end{equation}
As expected, the $\beta$-untwisted R-R fields are precisely the R-R fields of the undeformed background, and they are single-valued. 
In terms of the twisted R-R fields $(F,\,A)$, the R-R fields have the same monodromy as \eqref{eq:monodromy-AdS-P0-D},
\begin{align}
 A(x^1+\eta^{-1})= \Exp{- \omega \vee} A(x^1)\,,\quad 
F(x^1+\eta^{-1}) = \Exp{- \omega \vee}F(x^1)\,,\quad 
 \omega^{mn} = 2\,\delta_0^{[m} \,\delta_1^{n]} \,. 
\end{align}

\subsubsection{A scaling limit of the Drinfeld--Jimbo $r$-matrix.}
\label{sec:AdS-Drinfeld--Jimbo}

Let us now consider the classical $r$-matrix~\cite{Hoare:2016hwh,Orlando:2016qqu}
\begin{align}
 r = \frac{1}{2}\,\bigl[P_0\wedge D + P^i \wedge (M_{0i}+ M_{1i})\bigr] \,,
\end{align}
which can be obtained as a scaling limit of the classical $r$-matrix of 
Drinfeld--Jimbo type~\cite{Drinfeld:1985rx,Jimbo:1985zk}. 
Using polar coordinates $(\rho,\theta)$,
\begin{align}
 (\rmd x^2)^2 + (\rmd x^3)^2 = \rmd \rho^2+\rho^2\,\rmd \theta^2 \,,
\end{align}
the \ac{yb}-deformed background, which satisfies the \ac{gse}, is given by~\cite{Hoare:2016hwh,Orlando:2016qqu}
\begin{align}
\begin{split}
 \rmd s^2 &= \frac{-(\rmd x^0)^2+\rmd z^2}{z^2- \eta^2}
    +\frac{z^2\bigl[(\rmd x^1)^2+\rmd \rho^2\bigr]}{z^4+ \eta^2\,\rho^2}
    +\frac{\rho^2\,\rmd \theta^2}{z^2}+ \rmd s^2_{\rmS^5}\,, \\
 B_2 &= \eta\,\biggl[\frac{-\rmd x^0\wedge \rmd z }{z\,(z^2-\eta^2)} - \frac{\rho\,\rmd x^1 \wedge \rmd \rho}{z^4+ \eta^2\,\rho^2}\biggr]\,, \\
 \Phi &= \frac{1}{2}\log\biggl[\frac{z^6}{(z^2- \eta^2)(z^4+ \eta^2\,\rho^2)}\biggr]\,,\qquad 
 I = -\eta\,(4\,\partial_0 + 2\,\partial_1) \,, \\
 \hat{F}_1&= -\frac{4\,\eta^2\,\rho^2}{z^4}\, \rmd\theta\,, \\
 \hat{F}_3 &= 4\,\eta\,\rho\,\biggl(\frac{-\rho\, \rmd x^0\wedge \rmd z }{z\,(z^4-\eta^2\,z^2)} + \frac{\rmd x^1 \wedge \rmd \rho}{z^4+ \eta^2\,\rho^2}\biggr)\wedge \rmd \theta \,,\\
 \hat{F}_5&= 4\,\biggl[\frac{z^6}{(z^2- \eta^2) (z^4 + \eta^2\, \rho^2)}\,\omega_{\AdS5} +\omega_{\rmS^5}\biggr]\,,\\
 \hat{F}_7&=4\,\eta\,\biggl(\frac{\rmd x^0\wedge \rmd z }{z\,(z^2- \eta^2)}+\frac{\rho\,\rmd x^1 \wedge \rmd \rho}{z^4+ \eta^2\,\rho^2}\biggr) \wedge \omega_{\rmS^5} \,,\\
 \hat{F}_9&=\frac{4\,\eta^2\,\rho}{z\,(z^2- \eta^2) (z^4+ \eta^2\,\rho^2)}\,
\rmd x^0 \wedge \rmd x^1\wedge \rmd \rho \wedge\rmd z \wedge \omega_{\rmS^5}\,. 
\label{eq:HvT}
\end{split}
\end{align}
The R-R potentials are given by
\begin{align}
\begin{split}
 \hat{C}_0&= 0\,, \quad
 \hat{C}_2 = -\frac{\eta\,\rho^2}{z^4}\,\rmd x^0\wedge \rmd \theta\,,
\quad
 \hat{C}_4 = \frac{\rho}{z^4+ \eta^2\,\rho^2}\, \rmd x^0\wedge \rmd x^1\wedge \rmd\rho \wedge \rmd\theta + \omega_4 \,,
\\
 \hat{C}_6&= -B_2\wedge \omega_4 \,, \quad
 \hat{C}_8 = \frac{\eta^2\,\rho}{z\,(z^2-\eta^2)(z^4+\eta^2\,\rho^2)}\,\rmd x^0\wedge \rmd x^1\wedge \rmd\rho\wedge \rmd z\wedge \omega_4 \,.
\end{split}
\end{align}
The corresponding dual fields in the NS-NS sector are given by 
\begin{align}
\begin{split}
 \rmd s_{\text{dual}}^2 &= \frac{ - (\rmd x^0)^2+(\rmd x^1)^2+\rmd \rho^2+\rho^2\,\rmd \theta^2+\rmd{z}^2}{z^2} + \rmd s^2_{\rmS^5}\,,\qquad \tilde{\phi}=0\,, 
\\
 \beta &= \eta\,\bigl[\hat{P}_0\wedge \hat{D}+\hat{P}^i\wedge (M_{0i}+ M_{1i})\bigr]
 = \eta\,( - x^2\,\partial_1\wedge\partial_2 - x^3\,\partial_1\wedge\partial_3+z\,\partial_0\wedge \partial_z)
\nn\\
 &=\eta\,( -\rho\,\partial_1\wedge\partial_\rho+z\,\partial_0\wedge \partial_z) \,, 
\end{split}
\end{align}
and the Killing vector $I^m$ again satisfies the divergence formula,
\begin{align}
 I^0 &= -4\,\eta = \sfD_m\beta^{0m} \,,& I^1 &= -2\,\eta = \sfD_m\beta^{1m} \,.
\end{align}
Providing the $B$-twist to the R-R field strengths, we obtain
\begin{align}
\begin{split}
 F_1&=-\frac{4\,\eta^2 \rho^2}{z^4}\,\rmd \theta \,,\quad 
 F_3 = \frac{4\,\eta\,\rho}{z^5}\, \bigl(\rho\,\rmd z\wedge \rmd x^0 + z\,\rmd x^1\wedge \rmd\rho\bigr) \wedge \rmd \theta \,,
\\
 F_5&=4\,\bigl(\omega_{\AdS5}+\omega_{\rmS^5} \bigr)\,,
\quad F_7 =0\,,\quad F_9 =0\,,
\\
 A_0 &=0\,,\quad 
 A_2 = -\frac{\eta\,\rho^2}{z^4}\, \rmd x^0\wedge \rmd \theta \,,
\\
 A_4 &= \frac{\rho}{z^4} \, \rmd x^0 \wedge \rmd x^1 \wedge \rmd \rho \wedge \rmd \theta +\omega_4 \,, \quad
 A_6 =0\,,\quad 
 A_8 =0\,. 
\end{split}
\end{align}
Furthermore, the $\beta$-untwist leads to the following expressions: 
\begin{align}
\begin{split}
 \check{F}_1 &=0\,,\quad 
 \check{F}_3 = 0\,, \quad
 \check{F}_5 =4\,\bigl(\omega_{\AdS5}+\omega_{\rmS^5}\bigr)\,,\quad 
\check{F}_7 =0\,,\quad \check{F}_9 =0\,,
\\
 \check{C}_0 &=0\,,\quad \check{C}_2 = 0\,, \quad
 \check{C}_4 = \frac{\rho}{z^4} \, \rmd x^0 \wedge \rmd x^1 \wedge \rmd \rho \wedge \rmd \theta +\omega_4 \,,\quad \check{C}_6=0\,,\quad \check{C}_8=0\,. 
\end{split}
\end{align}
These are the same as the undeformed R-R potentials. 

The non-zero components of $Q$-flux are given by 
\begin{align}
 Q_z{}^{0z} = \eta\,,\qquad Q_2{}^{12} = -\eta\,,\qquad Q_3{}^{13} = -\eta\,. 
\end{align}
When the $x^2$-direction is compactified as $x^2 \sim x^2 +\eta^{-1}$, 
this background becomes a $T$-fold with monodromy
\begin{align}
\begin{split}
 \cH_{MN}(x^2+\eta^{-1}) &= \bigl[\Omega^{\rm T}\cH(x^2)\,\Omega\bigr]_{MN} \,,\qquad 
 \Omega^M{}_N \equiv \begin{pmatrix} \delta^m_n & -2\,\delta_1^{[m} \,\delta_2^{n]} \\ 0 & \delta_m^n \end{pmatrix}\,,
\\
 F(x^2+\eta^{-1}) &= \Exp{- \omega \vee}F(x^2)\,,\qquad 
 \omega^{mn} = -2\,\delta_1^{[m} \,\delta_2^{n]} \,. 
\end{split}
\end{align}

\subsubsection{$r=\frac{1}{2\,\eta}\,P_-\wedge (\eta_1\,D-\eta_2\,M_{+-})$.}

Let us next consider the non-unimodular $r$-matrix\footnote{This $r$-matrix includes the known examples studied in Section~4.3 ($\eta_1=-\eta_2=-\eta$) 
and 4.4 ($\eta_1=-\eta$, $\eta_2=0$) of~\cite{Orlando:2016qqu} as special cases. },
\begin{align}
 r=\frac{1}{2\,\eta}\,P_-\wedge (\eta_1\,D-\eta_2\,M_{+-})\,. 
\end{align}
Here we have introduced the light-cone coordinates and polar coordinates 
\begin{align}
 x^\pm &\equiv \frac{x^0 \pm x^1}{\sqrt{2}} \,, & 
 (\rmd x^2)^2 + (\rmd x^3)^2 =& \rmd \rho^2+\rho^2\,\rmd \theta^2 \,. 
\end{align}
The \ac{yb}-deformed background is given by
\begin{align}
\begin{split}
 \rmd s^2&= \frac{-2\, z^2\,\rmd x^+\,\rmd x^-}{z^4 - (\eta_1 + \eta_2)^2\,(x^+)^2} 
+\frac{ \rmd \rho^2 + \rho^2\,\rmd \theta^2+\rmd z^2 }{z^2}\\
 &\quad + \eta_1\,\rmd x^+\,\frac{2\,x^+\,(\eta_1 + \eta_2)\,(z\,\rmd z + \rho\,\rmd \rho) - \eta_1\,(z^2 + \rho^2)\,\rmd x^+}{z^2\,[z^4 - (\eta_1 + \eta_2)^2\,(x^+)^2]} + \rmd s^2_{\rmS^5}\,,\\
 B_2 &=-\frac{\bigl[\eta_1\,(  \rmd x^+\wedge (\rho\,\rmd \rho+z\,\rmd z)-x^+\,\rmd x^+\wedge \rmd x^- ) - \eta_2\,x^+\,\rmd x^+ \wedge \rmd x^-\bigr]}{z^4 - (\eta_1 + \eta_2)^2\,(x^+)^2}
   \,,\\
 \Phi &= \frac{1}{2} \log\biggl[\frac{z^4}{z^4 - (\eta_1 + \eta_2)^2\, (x^+)^2}\biggr]\,,\qquad 
 I = -(\eta_1 - \eta_2)\,\partial_-\,,\\
 \hat{F}_1&= 0\,,\\
 \hat{F}_3 &= -\frac{4\,\rho\,\bigl[\eta_1\,(\rmd x^+ \wedge (z\,\rmd \rho-\rho\, \rmd z)  - x^+\,\rmd z\wedge \rmd \rho)
                                    - \eta_2\,x^+\,\rmd z\wedge \rmd \rho\bigr]\wedge \rmd \theta}{z^5}\,,\\
 \hat{F}_5&= 4 \biggl[\frac{z^4}{z^4 - (\eta_1 + \eta_2)^2\,(x^+)^2}\,\omega_{\AdS5} + \omega_{\rmS^5}\biggr] \,,\\
 \hat{F}_7&= \frac{4\,\bigl[\eta_1\,(  \rmd x^+\wedge (\rho\,\rmd \rho+z\,\rmd z)-x^+\,\rmd x^+\wedge \rmd x^- ) - \eta_2\,x^+\,\rmd x^+ \wedge \rmd x^-\bigr]\wedge \omega_{\rmS^5}}{z^4 - (\eta_1+\eta_2)^2\,(x^+)^2} \,,\\
 \hat{F}_9&= 0\,. 
\end{split}
\end{align}
The R-R potentials are given by
\begin{align}
\begin{split}
 \hat{C}_0&= 0\,,\qquad
 \hat{C}_2 = \frac{\rho\,[\eta_1\,\rho\,\rmd x^+ - (\eta_1 + \eta_2)\,x^+\,\rmd \rho]\wedge \rmd \theta}{z^4}\,,
\\
 \hat{C}_4&= \frac{\rho\,\rmd x^+\wedge [z^3\,\rmd x^- - \eta_1\,(\eta_1 + \eta_2)\,x^+\,\rmd z]\wedge \rmd\rho \wedge \rmd\theta}{z^3\,[z^4 - (\eta_1 + \eta_2)^2\,(x^+)^2]} + \omega_4\,,
\\
 \hat{C}_6&= -B_2\wedge \omega_4\,,\qquad 
 \hat{C}_8 = 0\,.
\end{split}
\end{align}
The dual fields take the form
\begin{align}
\begin{split}
 \rmd s_{\text{dual}}^2&= \frac{ -2\,\rmd x^+\,\rmd x^- + \rmd \rho^2 + \rho^2\,\rmd \theta^2+\rmd z^2}{z^2} 
+ \rmd s^2_{\rmS^5} \,,\qquad 
 \tilde{\phi}=0 \,,
\\
 \beta &= \hat{P}_-\wedge (\eta_1\,\hat{D}+\eta_2\,\hat{M}_{+-}) 
= \eta_1 \, \partial_- \wedge ( x^+\,\partial_+ + \rho \,\partial_\rho+z\,\partial_z)
         +\eta_2 \,x^+\, \partial_- \wedge \partial_+ 
\\
 &= \eta_1 \, \partial_- \wedge ( x^+\,\partial_+ + x^2\,\partial_2 + x^3\,\partial_3+z\,\partial_z)
         +\eta_2 \,x^+\, \partial_- \wedge \partial_+ \,,
\end{split}
\end{align}
and the $Q$-flux has the following non-vanishing components:
\begin{align}
 Q_z{}^{-z} = Q_+{}^{-+} = Q_2{}^{-2} = Q_3{}^{-3} = \eta_1\,,\qquad 
 Q_+{}^{-+} = \eta_2 \,. 
\end{align}
In a similar manner as in the previous examples, by compactifying one of the $x^1$, $x^2$, and $x^3$ directions with a certain period, this background can also be regarded as a $T$-fold. 
If we make for example the identification $x^3\sim x^3 + \eta_1^{-1}$, 
the associated monodromy becomes
\begin{equation}
	\begin{aligned}
		\cH_{MN}(x^3+\eta_1^{-1}) &= \bigl[\Omega^{\rm T}\cH(x^3)\,\Omega\bigr]_{MN} \,,& 
 \Omega^M{}_N &\equiv \begin{pmatrix} \delta^m_n & 2\,\delta_-^{[m} \,\delta_3^{n]} \\ 0 & \delta_m^n \end{pmatrix}\,,
\\
 F(x^3+\eta_1^{-1}) &= \Exp{- \omega \vee}F(x^3)\,, & 
 \omega^{mn} &= 2\,\delta_-^{[m} \,\delta_3^{n]} \,. 
	\end{aligned}
\end{equation}

\paragraph{A solution of the generalized type IIA supergravity equations.}

In the background \eqref{eq:HvT}, by performing a $T$-duality along the $x^1$-direction (see~\cite{Sakamoto:2017wor} for the duality transformation rule), 
we obtain the following solution of the generalized type IIA \ac{eom}:
\begin{align}
\begin{split}
 \rmd s^2 &= \frac{ - (\rmd x^0)^2+\rmd z^2}{z^2 - \eta^2} + z^2\,(\rmd x^1)^2 
 + \frac{(\rmd \rho + \eta\, \rho\,\rmd x^1)^2 + \rho^2\,\rmd \theta^2}{z^2} + \rmd s_{\rmS^5} \,,\\
 B_2 &= -\frac{\eta\, \rmd x^0 \wedge \rmd z}{z\,(z^2 - \eta^2)}\,,\qquad 
 \Phi = -2\,\eta\,x^1 - \frac{1}{2} \log\Bigl(\frac{z^2 - \eta^2}{z^4}\Bigr)\,,\qquad I=-4\,\eta\,\partial_0\,,\\
 \hat{F}_2&= \frac{4\,\eta\Exp{2\,\eta\,x^1}\rho\,(\rmd\rho + \eta\,\rho\,\rmd x^1)\wedge \rmd \theta}{z^4} \,,\\
 \hat{F}_4&= \frac{4\Exp{2\,\eta\,x^1}\rho\,\rmd x^0 \wedge (\rmd \rho+\eta\,\rho\,\rmd x^1) \wedge \rmd \theta\wedge \rmd z}{z^3\,(z^2- \eta^2)} \,,\\
 \hat{F}_6&=-4 \Exp{2\,\eta\,x^1} \rmd x^1 \wedge \omega_{\rmS^5}\,,\qquad
 \hat{F}_8 = \frac{4\,\eta\Exp{2\,\eta\,x^1}\rmd x^0 \wedge \rmd x^1 \wedge \rmd z \wedge \omega_{\rmS^5}}{z\,(z^2 - \eta^2)} \,. 
\end{split}
\end{align}
Here the R-R potentials are given by 
\begin{equation}
	\begin{aligned}
		\hat{C}_1 &= 0 \,, &
 \hat{C}_3 &= \Exp{2\,\eta\,x^1} \frac{\rho\,\rmd x^0 \wedge (\rmd \rho + \eta\,\rho\, \rmd x^1)\wedge \rmd \theta}{z^4} \,,
\\
 \hat{C}_5 &= \Exp{2\,\eta\,x^1} \rmd x^1 \wedge \omega_4 \,, &
 \hat{C}_7 &= -\Exp{2\,\eta\,x^1} \frac{\eta\,\rmd z\wedge \rmd x^0 \wedge \rmd x^1 \wedge \omega_4}{z\,(z^2- \eta^2)} \,.
	\end{aligned}
\end{equation}
This background cannot be regarded as a $T$-fold, 
but it is the first example of a solution for the generalized type IIA supergravity equations. 

\subsubsection{$r = \frac{1}{2}\, M_{-\mu}\wedge P^\mu$.}

Our final example is associated to the $r$-matrix~\cite{Orlando:2016qqu}
\begin{align}
 r = \frac{1}{2}\, M_{-\mu}\wedge P^\mu\,.
\end{align}
This $r$-matrix is called light-like $\kappa$-Poincar\'e.
Again, by introducing the coordinates
\begin{align}
 x^\pm &\equiv \frac{x^0 \pm x^1}{\sqrt{2}} \,, & 
 (\rmd x^2)^2 + (\rmd x^3)^2 &= \rmd \rho^2+\rho^2\,\rmd \theta^2 \,,
\end{align}
the \ac{yb}-deformed background is given by (see Section~4.5 of~\cite{Orlando:2016qqu})\footnote{The metric and NS-NS two form were computed in~\cite{vanTongeren:2015uha}.}
\begin{align}
\begin{split}
 \rmd s^2&=\frac{z^2( -2\,\rmd x^+\,\rmd x^-+\rmd z^2) }{z^4 - (\eta\,x^+)^2}
- \eta^2 \,\frac{\rho^2 (\,\rmd x^+)^2 -2\,x^+ \rho\,\rmd x^+\,\rmd \rho+(x^+)^2\,\rmd z^2}{z^2(z^4 - (\eta\,x^+)^2)} \\
&\quad+ \frac{\rmd \rho^2 + \rho^2\,\rmd \theta^2}{z^2} 
 + \rmd s_{\rmS^5}^2 \,,\\
 B_2 &= \frac{\eta\,\rmd x^+\wedge (x^+\,\rmd x^- -\rho\,\rmd\rho)}{z^4-(\eta \,x^+)^2}\,,\qquad 
 \Phi =\frac{1}{2} \log\biggl[\frac{z^4}{z^4-(\eta\,x^+)^2}\biggr] \,,\qquad 
 I^- = 3\,\eta\,,\\
 \hat{F}_1&=0 \,,\quad
 \hat{F}_3= -\frac{4\,\eta\,\rho}{z^5}\,  \bigl(\rho\,\rmd x^+ - x^+\,\rmd \rho\bigr)\wedge \rmd\theta \wedge \rmd z \,,\\
 \hat{F}_5&= 4\,\biggl[\frac{z^4}{z^4-(\eta\,x^+)^2}\,\omega_{\AdS5}+\omega_{\rmS^5} \biggr] \,,\\
 \hat{F}_7&= - \frac{4\,\eta}{z^4-(\eta\,x^+)^2}\,\rmd x^+ \wedge \bigl(x^+\,\rmd x^- - \rho\,\rmd \rho\bigr)\wedge \omega_{\rmS^5} \,,\qquad
 \hat{F}_9= 0\,. 
\end{split}
\end{align}
The R-R potentials are found to be
\begin{align}
\begin{split}
 \hat{C}_0&=0\,,\quad 
 \hat{C}_2= \frac{\eta\,\rho}{z^4}\,\bigl(\rho\,\rmd x^+-x^+\,\rmd \rho\bigr)\wedge\rmd \theta \,,
\\
 \hat{C}_4& =\frac{\rho}{z^4-(\eta\,x^+)^2}\, \rmd x^+\wedge \rmd x^-\wedge \rmd\rho \wedge \rmd\theta + \omega_4\,, \quad 
 \hat{C}_6 = -B_2\wedge \omega_4 \,, \quad 
 \hat{C}_8=0 \,.
\end{split}
\end{align}
The corresponding dual fields are given by
\begin{align}
\begin{split}
 \rmd s_{\text{dual}}^2
 &= \frac{-2\rmd x^+\,\rmd x^- + \rmd \rho^2 + \rho^2\,\rmd \theta^2+\rmd z^2 }{z^2} + \rmd s^2_{\rmS^5} \,, \qquad \tilde{\phi}=0 \,,
\\
 \beta &= \eta\, \hat{M}_{-\mu}\wedge \hat{P}^\mu
= \eta\,\partial_-\wedge (x^+\,\partial_+ + \rho\,\partial_\rho) \no\\ 
 &= \eta\,\partial_-\wedge (x^+\,\partial_+ + x^2\,\partial_2 + x^3\,\partial_3) \,,
\end{split}
\end{align}
and it is easy to check that the divergence formula is satisfied,
\begin{align}
 I^- = 3\,\eta = \sfD_m \beta^{-m} \,.
\end{align}
We can calculate the other types of R-R fields:
\begin{align}
\begin{split}
 F_1&=0\,,\quad 
 F_3 =-\frac{4\,\eta\,\rho}{z^5}\, (\rho\,\rmd x^+ -x^+\,\rmd \rho)\wedge \rmd \theta\wedge \rmd z  \,,
\\
 F_5&=4\,\bigl(\omega_{\AdS5}+\omega_{\rmS^5} \bigr)\,,\quad 
F_7 =0\,,\quad F_9 =0\,,
\\
 A_0 &=0\,,\quad 
 A_2 = \frac{\eta\,\rho}{z^4}\,\bigl(\rho\,\rmd x^+ -x^+\,\rmd \rho\bigr)\wedge \rmd \theta \,,
\\
 A_4 &= \frac{\rho}{z^4} \, \rmd x^+ \wedge \rmd x^- \wedge \rmd \rho \wedge \rmd \theta +\omega_4 \,, \quad
 A_6 =0\,,\quad 
 A_8 =0\,,
\end{split}
\end{align}
and
\begin{align}
\begin{split}
 \check{F}_1 &=0\,,\quad 
 \check{F}_3 = 0\,, \quad
 \check{F}_5 =4\,\bigl(\omega_{\AdS5}+\omega_{\rmS^5}\bigr)\,,\quad 
\check{F}_7 =0\,,\quad \check{F}_9 =0\,,
\\
 \check{C}_0 &=0\,,\quad \check{C}_2 = 0\,, \quad
 \check{C}_4 =\frac{\rho}{z^4} \, \rmd x^+ \wedge \rmd x^-\wedge \rmd \rho \wedge \rmd \theta +\omega_4 \,,\quad 
\check{C}_6=0\,,\quad \check{C}_8=0\,.
\end{split}
\end{align}
The $\beta$-twisted fields are again invariant under the \ac{yb} deformation. 

The non-geometric $Q$-flux has the non-vanishing components
\begin{align}
 Q_+{}^{-+}=Q_2{}^{-2}=Q_3{}^{-3}= \eta \,,
\end{align}
and again by compactifying one of the $x^1$, $x^2$, and $x^3$ directions, 
this background becomes a $T$-fold. 
If we compactify the $x^3$-direction as $x^3\sim x^3 + \eta^{-1}$, 
the associated monodromy becomes
\begin{equation}
	\begin{aligned}
		\cH_{MN}(x^3+\eta^{-1}) &= \bigl[\Omega^{\rm T}\cH(x^3)\,\Omega\bigr]_{MN} \,, & 
 \Omega^M{}_N &\equiv \begin{pmatrix} \delta^m_n & 2\,\delta_-^{[m} \,\delta_3^{n]} \\ 0 & \delta_m^n \end{pmatrix}\,,
\\
 \sfF(x^3+\eta^{-1}) &= \Exp{- \omega \vee}\sfF(x^3)\,,& 
 \omega^{mn} &= 2\,\delta_-^{[m} \,\delta_3^{n]} \,. 
	\end{aligned}
\end{equation}

%% file: Killing.tex
\section{Killing spinors of the YB deformation}\label{sec:KillingYB}

We have discussed in the previous sections how to associate an integrable system to \(\AdS{5}\times S^5\) and how the deformations of the integrable system translate into deformations of the ten-dimensional background.
The starting system is remarkable for another reason, though: it has \(\mathcal{N} = 4\) supersymmetry (in four dimensions).
It is natural to wonder how this supersymmetric structure is affected by the integrable deformations discussed so far.
One first, crucial, observation is that, just like in the case of isometries, if \(T_i\) is a generator of the superalgebra \(\mathfrak{g}\), it is preserved by the \(R\)-matrix of Eq.~\eqref{eq:R-operator} if \(R\) is equivariant with respect to the (adjoint) action of \(T_i\) on \(\mathfrak{g}\), i.e.
\begin{align}
  \comm{T_{i}}{ R(X)} = R(\comm{T_{i}}{X}) \,,& \forall x \in \mathfrak{g}\,.
\end{align}

In the spirit of a geometrical interpretation, we want to describe the preserved supersymmetries in terms of Killing spinors of the deformed backgrounds.
Solving the Killing spinor equation is however in general a difficult task.
It is therefore much more convenient to have a frame-independent formalism allowing us to write an explicit formula for the spinors preserved by a given deformation.
In this section we will show how to write such a formula depending only on the non-commutativity parameter \(\Theta\)~\cite{Orlando:2018kms,Orlando:2018qaq}.
This is the same object defined via the \ac{sw} map in \eqref{eq:relation-open-closed}~\cite{Duff:1989tf,Seiberg:1999vs} and it coincides, up to a conventional sign, with the \(\beta\) field in \ac{dft}.
In our context it encodes all the information about the integrable deformation.

Such an explicit formalism is useful in a number of contexts.
The bilinear formalism of Killing spinors is for example useful for the classification of supergravity solutions.
It is necessary to solve the Killing spinor equations for supersymmetric localization calculations~\cite{Pestun:2016zxk}.
One needs to solve the Killing spinor equations in the construction of supersymmetric gauge theories realized on the D-branes embedded in deformed supergravity backgrounds (see e.g.~\cite{Lambert:2018cht,Choi:2017kxf,Choi:2018fqw}). 

Let us take the example of the $\Omega$-deformation of flat spacetime constructed in~\cite{Hellerman:2011mv}\,. From the perspective of integrable deformations, it corresponds to a TsT transformations, where one of the $U(1)$ isometries  acts freely.
Placing a probe brane in different configurations, one can construct either a non-Lorentz-invariant gauge theory a or massive gauge theory that can be studied from the viewpoint of string theory.
Having found these \(\Omega\)-deformed gauge theories as probe branes in a TsT-deformed spacetime, one may wonder if the near-horizon limit can be understood in terms of a similar deformation of the $\mathrm{AdS}_{5} \times S^{5}$ background.\footnote{See for example~\cite{Bobev:2019ylk} for related work, the Killing spinor analysis is also included.} 

\subsection{Killing spinors and T-dualities}

We have seen that \ac{yb} deformations are closely related to T-duality.
As a first step we will see how T-duality transforms the Killing spinors.

Given an isometry generated by a Killing vector \(\hat{T}_i\), a T-duality in this direction preserves only the Killing spinors that are covariantly constant with respect to the \ac{kl} derivative~\cite{Kosmann:1971sp,Bergshoeff:1994cb,Sfetsos:2010uq}
\begin{equation}
\mathcal{L}_{\hat{T}_{i}} \epsilon \equiv \hat{T}^{m}_{i}\nabla_{m}\epsilon + \frac{1}{4}(\nabla \hat{T}_{i})_{mn}\Gamma^{mn} \epsilon.
\end{equation}

This result can be extended to  generalized T-duality using the same argument  of~\cite{Kelekci:2014ima} for non-Abelian T-duality.
There always exists a frame where the isometry \(\hat{T}_i\) is represented as \(\del_z\).
Then the metric is parametrized by
\begin{equation}
ds^{2} = g_{\mu\nu}(x) \dd x^{\mu} \dd x^{\nu} + e^{2C(x)} \left(\dd z + A_{1}(x) \right)^{2},
\end{equation}
where $A_{1}$ is a one-form.
In this metric, the \ac{kl} derivative  along $\partial_{z}$ is
\begin{equation}
\mathcal{L}_{\partial_{z}}\epsilon = \partial_{z} \epsilon.
\end{equation}
Consider the generalized type IIA background:
\begin{equation}
\begin{aligned}
\label{eq:IIA-data}
ds^{2}_{\mathrm{IIA}} &= g_{\mu\nu} \dd x^{\mu} \dd x^{\nu} + e^{2C(x)}(\dd z + A_{1}(x))^{2}\,,\\
B_{2} &= B+ B_{1} \wedge \dd z\,,\\
\mathfrak{f}_{0} &= m\,,\\
\mathfrak{f}_{2} &= \mathfrak{g}_{2} + \mathfrak{g}_{1} \wedge (\dd z + A_{1}(x))\,,\\
\mathfrak{f}_{4} &= \mathfrak{g}_{4} + \mathfrak{g}_{3} \wedge (\dd z + A_{1}(x))\,,\\
\Phi &= - a z + \varphi + \frac{1}{2}C\,,\\
I &= \hat{a}\, \partial_{z},
\end{aligned}
\end{equation}
where $\hat{a}$ is a constant, and all the background fields $A_{1}$, $B_{1}$, $B$, $\mathfrak{g}_{k}$, $\varphi$, $C$ are functions of $x^{\mu}$ only.

The  dilaton has a linear dependence in $z$ and we cannot use the standard Buscher rule along $\partial_{z}$, but  we can still perform a generalized T-duality leading to the following generalized type IIB background\footnote{We will collectively denote the isometric coordinate and its T-dual by $z$. }: 
\begin{equation}
\begin{aligned}
\label{eq:IIB-data}
ds^{2}_{\mathrm{IIB}} &= g_{\mu\nu} \dd x^{\mu} \dd x^{\nu} + e^{-2C(x)}(\dd z + B_{1}(x))^{2}\,,\\
\wt{B}_{2} &= B + B_{1} \wedge \dd z\,,\\
\mathfrak{f}_{1} &= m\, (\dd z + B_{1}(x)) - \mathfrak{g}_{1}\,,\\
\mathfrak{f}_{3} &= \mathfrak{g}_{2} \wedge (\dd z + B_{1}(x)) - \mathfrak{g}_{3}\,,\\
\mathfrak{f}_{5} &= \left(1 + \star_{10} \right) \left(\mathfrak{g}_{4} \wedge (\dd z + B_{1}(x)) \right)\,,\\
\wt{\Phi} &= - \hat{a} z + \varphi - \frac{1}{2}C\,,\\
\wt{I} & = a\, \partial_{z}. 
\end{aligned}
\end{equation}
The background fluxes $\mathfrak{f}_{k}$ are usually rewritten in terms of generalized fluxes as
\begin{equation}
\begin{aligned}
\hat{\mathcal{F}}_{k} = 
\begin{cases}
e^{+C(x)/2}\ \mathfrak{f}_{k} & \mathrm{type\ IIA}\,,\\
e^{-C(x)/2}\ \mathfrak{f}_{k} & \mathrm{type\ IIB}\,.\\
\end{cases}
\end{aligned}
\end{equation}
Now we need to compare the supersymmetry variations in the two backgrounds above.
The supersymmetry variations of dilatini and gravitini for type IIA generalized supergravity are written as~\cite{Wulff:2016tju} 
\begin{equation}
\begin{aligned}
\delta \lambda_{\mathrm{IIA}} &= \frac{1}{2}\partial_{m}\Phi \Gamma^{m}\epsilon + \frac{1}{2}(I^{m}B_{mn} + I_{n} \sigma_{3}) \Gamma^{n} \epsilon - \frac{1}{8} H_{3} \sigma_{3}\epsilon + \frac{1}{8}\left[ 5 \hat{\mathcal{F}}_{0} \sigma_{1} + 3\hat{\mathcal{F}}_{2}(i\sigma_{2}) + \hat{\mathcal{F}}_{4} \sigma_{1} \right] \epsilon\,,\\
\delta \Psi_{ \mathrm{IIA} } &= \nabla_{m} \epsilon - \frac{1}{8} H_{3 mnp}\Gamma^{np} \sigma_{3} \epsilon + \frac{1}{8}\left[ \hat{\mathcal{F}}_{0}\sigma_{1} +  \hat{\mathcal{F}}_{2}(i\sigma_{2}) + \hat{\mathcal{F}}_{4}\sigma_{1} \right] \Gamma_{m} \epsilon\,,
\end{aligned}
\end{equation}
whereas for type IIB generalized supergravity
\begin{equation}
\begin{aligned}
\delta \lambda_{\mathrm{IIB}} &= \frac{1}{2}\partial_{m}\wt{\Phi} \Gamma^{m}\wt{\epsilon} + \frac{1}{2}(\wt{I}^{m}\wt{B}_{mn} + \wt{I}_{n} \sigma_{3}) \Gamma^{n} \wt{\epsilon} - \frac{1}{8} \wt{H}_{3} \sigma_{3} \wt{\epsilon} + \frac{1}{8}\left[ \mathcal{F}_{1} (i\sigma_{2}) + \frac{1}{2}\hat{\mathcal{F}}_{3} \sigma_{1} \right] \wt{\epsilon}\,,\\
\delta \Psi_{\mathrm{IIB}} &= \nabla_{m} \wt{\epsilon} - \frac{1}{8}\wt{H}_{3 mnp}\Gamma^{np} \sigma_{3} \wt{\epsilon} - \frac{1}{8}\left[ \hat{\mathcal{F}}_{1}(i\sigma_{2}) + \hat{\mathcal{F}}_{3}(i\sigma_{2}) + \frac{1}{2}\hat{\mathcal{F}}_{5}\sigma_{1} \right] \Gamma_{m} \wt{\epsilon}\,,
\end{aligned}
\end{equation}
where the background fluxes without explicit indices are contracted with curved Gamma matrices and divided by the symmetry factors as in \eqref{eq:bi-spinor}.
We write the Killing spinors as the doublet
\begin{equation}
  \epsilon = \begin{pmatrix}\epsilon_{+} \\ \epsilon_{-} \end{pmatrix}.
\end{equation}
Inserting the background data \eqref{eq:IIA-data} and \eqref{eq:IIB-data} into the  variations leads to the relations\footnote{The $+/-$ sign corresponds to $2/1$ in the conventions used so far.}
\begin{equation}
  \delta \Psi_{\mathrm{IIB} \mu +} = -\Gamma_{z} \delta \Psi_{\mathrm{IIA}\mu+},\qquad \delta \Psi_{\mathrm{IIB}\, \mu - } = \delta \Psi_{\mathrm{IIA}\mu -},
\end{equation}
provided that
\begin{equation}
  \partial_{z} \epsilon_{\pm} = 0 
\end{equation}
and that
\begin{align}
  \label{eq:Killing-transf}
  \wt{\epsilon}_{+} &= -\Gamma^{z} 
  \epsilon_{+} , & \wt{\epsilon}_{-} &= \epsilon_{-},
\end{align}
where $\Gamma_{z}$ and $\Gamma^{z}$ are flat Gamma matrices in  the initial background.

The conclusion is that a type IIA Killing spinor is mapped to a Killing spinor $\wt{\epsilon}$ in type IIB if and only if the Killing spinor vanishes under the \ac{kl} derivative along the isometry direction $z$. 
In the case of a TsT transformation the Killing spinor has to be independent of both isometry directions in order to be preserved (and mapped to the final background).

\subsection{Exponential factor from TsT transformation}

We start with a concrete example of a TsT transformation. %
The simplest non-trivial configuration is obtained as a deformation on the ten-dimensional flat spacetime of Lunin--Maldacena type.
The associated classical $r$-matrix is similar to the one in Eq.\eqref{eq:abelian}.
First we show how to construct the projector matrix using the Kosmann Lie derivative.
For flat space in polar coordinates,
\begin{equation}
  \dd s^{2} = -(\dd x^{0})^{2} + (\dd x^{1})^{2} + \sum_{i=1}^{3}\left( \dd \rho_{i}^{2} + \rho_{i}^{2} \dd \phi_{i}^{2}\right) + (\dd x^{8})^{2} + (\dd x^{9})^{2}\,.
\end{equation}
The Killing spinor equation is
\begin{equation}
  \nabla_{m}\epsilon = 0\,,
\end{equation}
which admits the solution 
\begin{equation}
  \epsilon = \prod_{i=1}^{3}\exp[\frac{\phi_{i}}{2}\Gamma_{\rho_{i}\phi_{i}}](\eta_{0} + i\chi_{0})\,,
\end{equation}
where $\eta_{0}, \chi_{0}$ are constant Majorana--Weyl spinors and all the Gamma matrices are flat.  To perform the deformation as in~\cite{Lunin:2005jy}, we introduce the following adapted angles:
\begin{equation}
  \phi_{1} = \psi - \varphi_{1}\,,\qquad \phi_{2} = \psi + \varphi_{1} + \varphi_{2} \,,\quad \mathrm{and}\quad  \phi_{3} = \psi - \varphi_{2}\,.
\end{equation}
We now consider the TsT transformation in the angles $(\varphi_{1}, \varphi_{2})$.
First, note that the Killing spinor is rewritten as
\begin{equation}
  \epsilon = e^{\frac{\psi}{2}(\Gamma_{\rho_{1}\phi_{1}} + \Gamma_{\rho_{2}\phi_{2}} + \Gamma_{\rho_{3}\phi_{3}})}e^{\frac{\varphi_{1}}{2}(\Gamma_{\rho_{2}\phi_{2}} - \Gamma_{\rho_{1}\phi_{1}})}e^{\frac{\varphi_{1}}{2}(\Gamma_{\rho_{2}\phi_{2}} - \Gamma_{\rho_{3}\phi_{3}})} \eta_{0}\,.
\end{equation}
When performing a T-duality along $\partial_{\varphi_{1}}$, in order to preserve supersymmetry we demand that the \ac{kl} derivative along $\partial_{\varphi_{1}}$ vanishes:
\begin{equation}
  \mathcal{L}_{\partial_{\varphi_{1}}}\epsilon = \partial_{\varphi_{1}} \epsilon = 0\,.
\end{equation}
This is equivalent to acting with the following projection on the constant spinors $\eta_{0}, \chi_{0}$:
\begin{equation}
  \label{eq:proj-1}
  \Pi^{\varphi_{1}} = \frac{1}{2}(1 - \Gamma_{\rho_{1}\phi_{1}\rho_{2}\phi_{2}})\,,
\end{equation}
which removes the $\varphi_{1}$-dependence from the Killing spinor. 
Now, we shift the angle $\varphi_{2}$ by $+ \eta \widetilde{\varphi}_{1}$, where $\eta$ denotes the deformation parameter and $\widetilde{\varphi}_{1}$ is the T-dual of $\varphi_{1}$, and we T-dualize on $\widetilde{\varphi}_{1}$.
Once more we demand the \ac{kl} derivative in the direction \(\del_{\wt{\varphi}_1}\) to vanish.
This is equivalent to asking for the derivative in the direction \(\del_{\varphi_2}\) to vanish and in turn
this is the same as inserting the projector 
\begin{equation}
  \label{eq:proj-2}
  \Pi^{\varphi_{2}} = \frac{1}{2}(1 - \Gamma_{\rho_{2}\phi_{2}\rho_{3}\phi_{3}})\,.
\end{equation}
As remarked earlier, the \ac{kl} derivative along $\partial_{\widetilde{\varphi}_{1}}$ acts equivalently to that along $\partial_{\varphi_{2}}$\,. In total, the two projectors~\eqref{eq:proj-1}, \eqref{eq:proj-2} preserve one quarter of the supersymmetry. 

Now we can determine the explicit form of the Killing spinors in the deformed background.
The TsT transformation of interest leads to
\begin{equation}
  \begin{aligned}
    \dd s^{2} &= -(\dd x^{0})^{2} + (\dd x^{1})^{2} + \sum_{i=1}^{3} \left(\dd \rho_{i}^{2}\right) + (\dd x^{8})^{2} + (\dd x^{9})^{2}\\
    &\quad + \frac{1}{\Delta^{2}}\left[ \rho_{1}^{2} (\dd \psi - \dd \varphi_{1})^{2} + \rho_{2}^{2} (\dd \psi + \dd \varphi_{1} + \dd \varphi_{2})^{2} + \rho_{3}^{2} (\dd \psi - \dd \varphi_{2})^{2} + 9 \lambda^{2} \rho_{1}^{2}\rho_{2}^{2}\rho_{3}^{2} \dd \psi^{2} \right]\,,\\
    e^{\Phi} &= \Delta^{2}\,,\\
    B_{2} &= \frac{\Delta^{2} - 1}{\eta \Delta^{2}} \dd \varphi_{1} \wedge \dd \varphi_{2} - \frac{\Delta^{2} - 1}{\eta \Delta^{2}} (\dd \varphi_{1} - \dd \varphi_{2}) \wedge \dd \psi + \frac{3\eta}{\Delta^{2}} (\rho_{1}^{2} \rho_{2} \dd \varphi_{1} - \rho_{2}^{2} \rho_{3}^{2} \dd \varphi_{2}) \wedge \dd \psi\,,\\
    \Delta^{2} &= 1 + \eta^{2} (\rho_{1}^{2} \rho_{2}^{2} + \rho_{2}^{2} \rho_{3}^{2} + \rho_{3}^{2} \rho_{1}^{2})\,,
  \end{aligned}
\end{equation}
and using the general formula in Eq.~(\ref{eq:general-Killing}) we obtain the following Killing spinors:
\begin{equation}
\label{eq:LM-spinor}
\widetilde{\epsilon}_{+} = \frac{1 + \eta \left( \rho_{1} \rho_{2} \Gamma_{\phi_{1} \phi_{2}} + \rho_{2} \rho_{3} \Gamma_{\phi_{2} \phi_{3}} +\rho_{3} \rho_{1} \Gamma_{\phi_{3} \phi_{1}}\right)}{1+ \eta^{2} (\rho_{1}^{2}\rho_{2}^{2}+\rho_{2}^{2}\rho_{3}^{2}+ \rho_{3}^{2}\rho_{1}^{2})} \epsilon_{+}\,,\qquad \widetilde{\epsilon}_{-} = \epsilon_{-}\,,
\end{equation}
where 
\begin{equation}
  \epsilon_{+} + i\epsilon_{-} = e^{\frac{\psi}{2} (\Gamma_{\rho_{1}\phi_{1}} +\Gamma_{\rho_{2}\phi_{2}} + \Gamma_{\rho_{3}\phi_{3}})} \Pi^{\varphi_{1}}\Pi^{\varphi_{2}} (\eta_{+} + i\eta_{-})\,.
\end{equation}

In order to generalize this result to any \ac{yb}-deformed background it is convenient to recast it in a form that depends explicitly on the parameters of the deformation encoded by the bivector \(\Theta\).
Applying the Seiberg--Witten map~\eqref{eq:relation-open-closed} to the background we find
\begin{equation}
  \label{eq:LM-bivector}
  \Theta = \eta (\partial_{\phi_{1}} \wedge \partial_{\phi_{2}} + \partial_{\phi_{2}} \wedge \partial_{\phi_{3}} + \partial_{\phi_{3}} \wedge \partial_{\phi_{1}})\,.
\end{equation}
The deformed Killing spinor $\widetilde{\epsilon}_{+}$ in \eqref{eq:LM-spinor} can be rewritten using this bi-vector as
\begin{equation}
\label{eq:general-Killingplus}
\widetilde{\epsilon}_{+} = \frac{1 + \frac{1}{2}\Theta^{mn}\Gamma_{mn}}{\sqrt{1 + \frac{1}{2}\Theta^{mn} \Theta_{mn}}} \epsilon_{+}= e^{\frac{1}{2} \omega(\Theta) \Theta^{mn} \Gamma_{mn}}\epsilon_{+}\,,
\end{equation}
where the Gamma matrices are curved in terms of the undeformed metric while the normalization factor $\omega(\Theta)$ satisfies
\begin{equation}
\tan \left(\omega(\Theta) \sqrt{\frac{1}{2}\Theta^{mn}\Theta_{mn}} \right) = \sqrt{\frac{1}{2}\Theta^{mn}\Theta_{mn}}\,. 
\end{equation}
This expression is reminiscent of a quantum R-matrix in the spinor representation.
This connection has not yet been explored in the literature.
In~\cite{Orlando:2018kms} other concrete examples of TsT deformations of flat space and $\mathrm{AdS}_{5}\times S^{5}$ background were studied and it was found that the form of the preserved Killing spinors in Eq.~\eqref{eq:general-Killingplus} is generic.

To sum up, we conjecture the following structure of \ac{yb} deformed Killing spinors expressed by the bi-vector $\Theta$ only:
\begin{equation}
\label{eq:general-Killing}
\widetilde{\epsilon}_{+} =  e^{\frac{1}{2} \omega(\Theta) \Theta^{mn} \Gamma_{mn}} \Pi^{\mathrm{TsT}}\epsilon_{+}\,,\qquad \widetilde{\epsilon}_{-} = \Pi^{\mathrm{TsT}}\epsilon_{-}\,,
\end{equation}
where $\epsilon_{\pm}$ are undeformed Killing spinors, and $\Pi^{\mathrm{TsT}}$ is the projector derived from the \ac{kl} derivatives.
What remains to be done in order to find a complete formula that only depends on \(\Theta\) alone is to relate the projector $\Pi^{\mathrm{TsT}}$ to the bi-vector.

\subsection{Supersymmetry projector formula}
We have seen that the projector matrix is needed for preserving the Killing spinors under TsT transformations. In this section, we show how to find such a projector for non-TsT \ac{yb} deformations, using only the bi-vector $\Theta$.

Let us assume that the initial undeformed background preserves some supersymmetry.
This means that
\begin{equation}
  \label{eq:gravitino-undeformed}
  \delta \Psi_{\mu} = \nabla_{\mu}\epsilon + \frac{1}{8} \mathcal{S}\Gamma_{\mu} \epsilon = 0\,,
\end{equation}
where $\mathcal{S}$ is the analog of the Ramond--Ramond flux bispinor in the initial background,
\begin{equation}
\mathcal{S} = - \hat{\mathcal{F}}_{1} \otimes (i\sigma_{2}) - \hat{\mathcal{F}}_{3} \otimes \sigma_{1} - \frac{1}{2} \hat{\mathcal{F}}_{5} \otimes (i\sigma_{2}) \,.
\end{equation}
As discussed before, for TsT transformations, the preserved Killing spinors have to be independent of both isometry directions.
So we ask the \ac{kl} derivatives of the Killing spinor to
vanish  along all the directions \(T_i\) that are included in the classical $r$-matrix:
\begin{equation}
\begin{aligned}
\label{eq:Kosmann-Killing}
\mathcal{L}_{\hat{T}_{i}}\epsilon = \hat{T}^{m}_{i}\nabla_{m} \epsilon + \frac{1}{4}(\nabla \hat{T}_{i})_{mn} \Gamma^{mn} \epsilon = 0\,.
\end{aligned}
\end{equation}
Multiplying \eqref{eq:gravitino-undeformed} by $\hat{T}^{m}_{i}$\,, we get
\begin{equation}
\label{eq:gravitino-undeformed2}
\hat{T}^{m}_{i} \delta \Psi_{m} = \hat{T}^{m}_{i}\nabla_{m} \epsilon + \frac{1}{8}\hat{T}^{m}_{i} \mathcal{S}\Gamma_{m} \epsilon = 0\,.
\end{equation}
Combining Eq.~\eqref{eq:Kosmann-Killing} and Eq.~\eqref{eq:gravitino-undeformed2}\,, we obtain
\begin{equation}
\left[ \frac{1}{8} \hat{T}^{m}_{i} \mathcal{S}\Gamma_{m} - \frac{1}{4} (\nabla \hat{T}_{i})_{mn} \Gamma^{mn}\right]\epsilon = 0 
\end{equation}
and multiplying by $r^{ij}\hat{T}_{j}^{l}\Gamma_{l}$\,, one finds
\begin{equation}
\label{eq:two-Killing-eq}
\left[ \frac{1}{8} r^{ij}\hat{T}_{j}^{n}\Gamma_{n}\hat{T}^{m}_{i}\mathcal{S}\Gamma_{m} - \frac{1}{4}r^{ij} \hat{T}_{j}^{l}(\nabla \hat{T}_{i})_{mn}\Gamma_{l}\Gamma^{mn}  \right] \epsilon = 0\,.
\end{equation}
In this step, there is one caveat.
If $\hat{T}_{j}$ is a light-like Killing vector, the matrix $r^{ij}\hat{T}_{j}^{\rho}\Gamma_{\rho}$ might not be invertible.
In this case, Eq.~\eqref{eq:two-Killing-eq} gives a necessary but not sufficient condition for the preservation of supersymmetry.

The above equation \eqref{eq:two-Killing-eq} can be further rewritten with respect to the bi-vector of the bi-Killing form
\begin{equation}
\Theta^{mn} = r^{ij}\hat{T}^{m}_{i}\hat{T}^{n}_{j}
\end{equation}
and its covariant derivative
\begin{equation}
\nabla_{m}\Theta^{nl} = 2r^{ij}(\nabla_{m}\hat{T}_{i}^{[n})\hat{T}^{l]}_{j}\,
\end{equation}
in the form
\begin{equation}
\label{eq:proj-formula}
\left[ \Theta^{mn}\Gamma_{m}\mathcal{S}\Gamma_{n} + \nabla_{m}\Theta^{nl}\Gamma^{m}{}_{nl} - 4 \nabla_{m}\Theta^{mn}\Gamma_{n} \right] \epsilon = 0\,.
\end{equation}
This result implies that given any classical $r$-matrix (or bi-vector $\Theta$), we can determine a projection matrix $\Pi^{\mathrm{TsT}}$ to preserve the Killing spinor via
\begin{equation}
\left[ \Theta^{mn}\Gamma_{m}\mathcal{S}\Gamma_{n} + \nabla_{m}\Theta^{nl}\Gamma^{m}{}_{nl} - 4 \nabla_{m}\Theta^{mn}\Gamma_{n} \right] \Pi^{\mathrm{TsT}} = 0\,. 
\end{equation}

In the case of the Lunin--Maldacena-like deformation of flat space in the previous section, due to the unimodularity of the classical $r$-matrix, only the second term in~\eqref{eq:proj-formula} contributes to the left-hand side.
It is evaluated as
\begin{equation}
\left[ \rho_{1}\Gamma_{\phi_{1}}( \Gamma_{\rho_{2}\phi_{2}} - \Gamma_{\rho_{3}\phi_{3}} ) - \rho_{2} \Gamma_{\phi_{2}} ( \Gamma_{\rho_{1}\phi_{1}} - \Gamma_{\rho_{3}\phi_{3}}) + \rho_{3}\Gamma_{\phi_{3}} ( \Gamma_{\rho_{1}\phi_{1}} - \Gamma_{\rho_{2}\phi_{2}})\right] \epsilon = 0\,.
\end{equation}
From this we deduce two independent conditions,
\begin{equation}
(\Gamma_{\rho_{1}\phi_{1}} - \Gamma_{\rho_{2}\phi_{2}}) \epsilon = (\Gamma_{\rho_{2}\phi_{2}} - \Gamma_{\rho_{3}\phi_{3}}) \epsilon = 0\,,
\end{equation}
which lead to the same projectors as \eqref{eq:proj-1} and \eqref{eq:proj-2}:
\begin{equation}
\Pi_{1} = \frac{1}{2}(1 - \Gamma_{\rho_{1}\phi_{1} \rho_{2}\phi_{2}})\,,\qquad \Pi_{2} = \frac{1}{2}(1 - \Gamma_{\rho_{2}\phi_{2} \rho_{3}\phi_{3}})\,.
\end{equation}

\subsection{Examples}

In this section we corroborate the conjecture of the validity the formula for the Killing spinor in Eq.~\eqref{eq:general-Killing} and with the projector in Eq.~\eqref{eq:proj-formula} also for generic \ac{yb} deformations not obtained as TsT transformations by reviewing two of the examples treated in~\cite{Orlando:2018qaq}. One is supersymmetric and the other non-supersymmetric.

Since all the deformations here act on the $\mathrm{AdS}_{5}$ part, we focus only on the Killing spinors derived from $\mathrm{AdS}_{5}$ spacetime. Given the Poincar\'e coordinate system for $\mathrm{AdS}_{5}$ space~\eqref{eq:AdS5S5-metric}\,, the Killing spinor of complex form can be readily solved: 
\begin{equation}
\epsilon_{\mathrm{AdS_{5}}} = \left[\left\{ \sqrt{z} + \frac{1}{\sqrt{z}}x^{\mu}\Gamma_{\mu}\Gamma_{z} \right\} \frac{1 - i\gamma \Gamma_{z}}{2} + \frac{1}{\sqrt{z}}\frac{1+i\gamma\Gamma_{z}}{2}\right] (\eta_{0} + i \chi_{0})\,,
\end{equation}
where $\epsilon_{0}, \chi_{0}$ are constant Majorana--Weyl spinors.  The matrix $\gamma$ is defined as $\gamma \equiv \Gamma_{56789}$, being the product of the flat $\Gamma$-matrices on $S^{5}$.

\subsubsection{$r = \frac{1}{2}\left[ P_{1} \wedge P_{3} + (P_{0} + P_{1} ) \wedge (M_{03} + M_{13}) \right]$.}

In this case the corresponding $\beta$-field is also divergence-free, but gives rise to a deformed background that cannot be obtained via TsT transformation. The full background is presented in~\cite{Borsato:2016ose}. The bi-vector is approximated as
\begin{equation}
\Theta = -2\eta \partial_{1} \wedge \partial_{3} + 2\eta (x^{0} - x^{1})(\partial_{0} + \partial_{1}) \wedge \partial_{3} + \mathcal{O}(\eta^{2})\,.
\end{equation}
The projector formula \eqref{eq:proj-formula} gives
\begin{equation}
\left[ \Gamma_{1} (1 + \gamma\Gamma_{z} \otimes (i\sigma_{2})) - (x^{0} - x^{1}) (1+ \gamma \Gamma_{01z} \otimes (i\sigma_{2}))( \Gamma_{0}+\Gamma_{1})\right] \epsilon = 0\,.
\end{equation}
It is not hard to see that non-zero solutions are constrained by two projectors
\begin{equation}
\Pi_{1} = \frac{1}{2}(1+ i\gamma\Gamma_{z}) \,,\qquad \Pi_{2} = \frac{1}{2}(1 + \Gamma_{01})\,,
\end{equation}
which implies that eight supercharges are preserved after the deformation.
Using the fact that
\begin{equation}
\begin{aligned}
\frac{1}{2}\Theta_{mn}\Theta^{mn} &= \frac{4\eta^{2}}{z^{4}}(1- 2(x^{0}-x^{1}))\,,\\
\frac{1}{2}\Theta_{mn}\Gamma^{mn} &= \frac{2\eta}{z^{2}}\left[ (x^{0} - x^{1})(\Gamma_{0} + \Gamma_{1}) - \Gamma_{1}  \right] \Gamma_{3}\,,
\end{aligned}
\end{equation}
we can write the Killing spinors at leading order:
\begin{equation}
\begin{aligned}
\widetilde{\epsilon}_{+} &= \left[ 1 + \frac{2\eta}{z^{2}}\left\{ (x^{0} - x^{1}) ( \Gamma_{0} + \Gamma_{1} ) - \Gamma_{1} \right\} \right] \Gamma_{3} \epsilon_{+}\,,\\
\widetilde{\epsilon}_{-} &= \epsilon_{-}\,,
\end{aligned}
\end{equation}
with
\begin{equation}
\epsilon_{+} + i \epsilon_{-} = \frac{1}{\sqrt{z}}\Pi_{1}\Pi_{2} \epsilon_{0}\,,\qquad \epsilon_{0}: const.
\end{equation}
We can verify this expression computing explicitly the supersymmetry variations.
It is enough to consider the first order in the deformation where the \(H\) and \(F_3\) fluxes appear.

First let us look at the gravitino variations.
Since the projector $\Pi_{2}$ acts on the $(x^{0}, x^{1})$-plane in Poincar\'e coordinates of the $\mathrm{AdS}_{5}$ space, we focus on the $x^{0}$ component.
At linear order we find
\begin{equation}
\begin{aligned}
\nabla_{0} &= \partial_{0} - \frac{1}{2z} \Gamma_{0z}\,,\\
\frac{1}{8}H_{0mn}\Gamma^{mn} &= -\frac{2\eta}{z^{3}}(x^{0} - x^{1})\gamma \Gamma_{012}\,,\\
\frac{1}{8}\mathcal{S}\Gamma_{0} &= - \frac{\eta}{z^{3}} \Gamma_{2z}\left[ (x^{0} - x^{1}) (1+ \Gamma_{01}) - 1\right] \otimes \sigma_{1} - \frac{1}{2z} \gamma\Gamma_{0} \otimes (i\sigma_{2}) \,.
\end{aligned}
\end{equation}
The gravitino variation becomes
\begin{equation}
\delta \Psi_{0+} = \frac{\eta}{z^{3}} \left[ 2(x^{0}-x{1}) \gamma\Gamma_{012} \widetilde{\epsilon}_{+} - \Gamma_{2z} \left\{ (x^{0} - x^{1}) (1+ \Gamma_{01}) +1\right\}\widetilde{\epsilon}_{-}\right] = 0\,.
\end{equation}
The dilatino variation is more involved.
The background fluxes contribute to the variation as
\begin{equation}
\begin{aligned}
\hat{H}_{3} &= -\frac{8}{z^{2}} \Gamma_{3z}\left[(x^{0} - x^{1})(\Gamma_{0} + \Gamma_{1}) - \Gamma_{1}\right] = -\frac{1}{2}\nabla_{m} \Theta^{np}\Gamma^{m}_{np}\,,\\
\Gamma_{m}\mathcal{S}\Gamma^{m} &= -\frac{32}{z^{2}}\Gamma_{2z}\left[ (x^{0} - x^{1})(\Gamma_{0} + \Gamma_{1}) - \Gamma_{0}\right] \otimes \sigma_{1} = 2 \Theta^{mn} \Gamma_{m} \mathcal{S}_{0} \Gamma_{n}\otimes \sigma_{3}\,,
\end{aligned}
\end{equation}
where $\mathcal{S}_{0}$ is the bi-spinor analog evaluated on the undeformed $\mathrm{AdS}_{5} \times S^{5}$ background. In total, we obtain
\begin{equation}
\delta \lambda = \frac{1}{8}\left[ \Theta^{mn} \Gamma_{m} \mathcal{S} \Gamma_{n} + \nabla_{n} \Theta^{np}\Gamma^{m}{}_{np} \right] \otimes \sigma_{3} \widetilde{\epsilon} = 0 \,.
\end{equation}
Remarkably, we end up with the projector formula \eqref{eq:proj-formula}.

\subsubsection{$r = \frac{1}{2}P_{1} \wedge D$.}

Finally let us comment on non-unimodular classical $r$-matrices which lead to a solution for the generalized supergravity \ac{eom}.
Consider the  background in \eqref{space}.
The corresponding $\beta$-field was computed in~\eqref{eq:nonunimod-beta}.
Using the projector formula, we find
\begin{equation}
\left[ \frac{1}{z} \Gamma_{1} + \frac{2}{z} \Gamma_{023}\otimes (i\sigma_{2}) -\frac{2}{z^{2}} \left( x^{1}\Gamma_{0} + x^{2} \Gamma_{2} + x^{3}\Gamma_{3} \right) \Gamma_{1z} (1- \Gamma_{0123} \otimes (i\sigma_{2})) \right] \epsilon = 0\,.
\end{equation}
Since the whole matrix acting on the spinor has a non-vanishing determinant, only $\epsilon=0$ solves the above equation.
All  supersymmetries are broken in this deformed background.  

\subsection{Comments.}

It would be interesting to derive the exponential factor \eqref{eq:general-Killingplus} as well as the projector formula \eqref{eq:proj-formula} in an alternative supergravity framework, such as the $\beta$-supergravity~\cite{Andriot:2013xca}.
The corresponding supersymmetry variations are given in~\cite{Andriot:2014qla} in absence of Ramond--Ramond fluxes.

We have restricted our attention to  Killing spinors in  type IIB supergravity. In~\cite{Orlando:2018kms}, the analysis was extended to the so-called M-theory TsT transformations using the M-theory T-duality~\cite{Sen:1995cf,Ganor:1996zk,Aharony:1996wp} on the $\mathrm{AdS}_{7} \times S^{4}$ background.
It might be interesting to look for a the general formula for TsT deformed Killing spinors  in terms of an antisymmetric tri-vector from the viewpoint of non-commutativity in M2-brane. To this end, the notion of the generalized Theta parameter in~\cite{Berman:2001rka} might be useful. For the tri-vector deformation of M-theory backgrounds, see~\cite{Bakhmatov:2019dow}.

%% file: appendix.tex
\section*{Appendix}

\section{Conventions}\label{app:conventions}

In this appendix, we collect our conventions.
\medskip

The antisymmeterization is defined as
\begin{align}
 A_{[m_1\cdots m_n]} \equiv \frac{1}{n!}\,\bigl(A_{m_1\cdots m_n} \pm \text{permutations}\bigr) \,.
\end{align}
For conventions of differential forms, we use
\begin{align}
\begin{split}
 &\varepsilon^{01}= \frac{1}{\sqrt{-\gga}}\,,\qquad 
 \varepsilon_{01}= - \sqrt{-\gga} \,, \qquad 
 \rmd^{2}\sigma = \rmd \tau \wedge\rmd \sigma \,,
\\
 &(*_{\gga} \alpha_q)_{\WSa_1\cdots\WSa_{p+1-q}} =\frac{1}{q!}\,\varepsilon^{\WSb_1\cdots\WSb_q}{}_{\WSa_1\cdots\WSa_{p+1-q}}\,\alpha_{\WSb_1\cdots\WSb_q} \,,
\\
 & *_{\gga} (\rmd \sigma^{\WSa_1}\wedge \cdots \wedge \rmd \sigma^{\WSa_q}) 
 = \frac{1}{(p+1-q)!}\,\varepsilon^{\WSa_1\cdots\WSa_q}{}_{\WSb_1\cdots\WSb_{p+1-q}}\,\rmd \sigma^{\WSb_1}\wedge \cdots \wedge \rmd \sigma^{\WSb_{p+1-q}} \,,
\end{split}
\end{align}
on the string worldsheet. In spacetime, we define
\begin{align}
\begin{split}
 &\varepsilon^{1\cdots D}=-\frac{1}{\sqrt{-\CG}}\,,\qquad 
 \varepsilon_{1\cdots D}= \sqrt{-\CG} \,, \qquad 
 \epsilon^{1\cdots D} = - 1\,,\qquad 
 \epsilon_{1\cdots D} = 1 \,, 
\\
 &(* \alpha_q)_{m_1\cdots m_{D-q}} =\frac{1}{q!}\,\varepsilon^{n_1\cdots n_q}{}_{m_1\cdots m_{D-q}}\,\alpha_{n_1\cdots n_q} \,,\qquad 
 \rmd^{D}x = \rmd x^1\wedge\cdots\wedge\rmd x^D \,,
\\
 &* (\rmd x^{m_1}\wedge \cdots \wedge \rmd x^{m_q}) = \frac{1}{(p+1-q)!}\,\varepsilon^{m_1\cdots m_q}{}_{n_1\cdots n_{p+1-q}}\,\rmd x^{n_1}\wedge \cdots \wedge \rmd x^{n_{p+1-q}} \,,
\\
 &(\iota_v \alpha_n) = \frac{1}{(n-1)!}\,v^n\,\alpha_{nm_1\cdots m_{n-1}}\,\rmd x^{m_1}\wedge\cdots\wedge \rmd x^{m_{n-1}}\,. 
\end{split}
\end{align}
The spin connection is defined as
\begin{align}
 \omega_m{}^{\Loa\Lob} \equiv 2\,e^{n[\Loa}\,\partial_{[m} e_{n]}{}^{\Lob]} - e^{p[\Loa}\,e^{\Lob]q}\,\partial_{[p} e_{q]}{}^{\Loc}\,e_{m\Loc} \,,
\label{eq:spin-con}
\end{align}
which satisfies
\begin{align}
 \rmd e^{\Loa} + \omega^{\Loa}{}_{\Lob}\wedge e^{\Lob} = 0\,, 
\end{align}
where $e^{\Loa}\equiv e_m{}^{\Loa}\,\rmd x^m$ and $\omega^{\Loa}{}_{\Lob}\equiv \omega_m{}^{\Loa}{}_{\Lob}\,\rmd x^m$\,. 
The Riemann curvature tensor is defined as
\begin{align}
 R^{\Loa}{}_{\Lob} \equiv \frac{1}{2}\,R^{\Loa}{}_{\Lob\Loc\Lod}\,e^{\Loc}\wedge e^{\Lod} \equiv \rmd \omega^{\Loa}{}_{\Lob} + \omega^{\Loa}{}_{\Loc}\wedge \omega^{\Loc}{}_{\Lob} \,,\qquad 
 R^{\Loa}{}_{\Lob\Loc\Lod} = e_m{}^{\Loa}\,e_{\Lob}{}^n\,e_{\Loc}{}^p\,e_{\Lod}{}^q\,R^m{}_{npq} \,. 
\end{align}

\section{$\alg{psu}(2,2|4)$ algebra}\label{app:psu-algebra}

In this appendix, we collect our conventions and useful formulas on the $\alg{psu}(2,2|4)$ algebra (see for example~\cite{Arutyunov:2009ga} for more details). 

\subsection{Matrix realization}

\subsubsection*{$8 \times 8$ supermatrix representation.}

The super Lie algebra $\alg{su}(2,2|4)$ can be realized by using $8 \times 8$ supermatrices $\cM$ satisfying $\str\cM =0$ and the reality condition
\begin{align}
 \cM^\dagger\, H+H\,\cM =0\,,\qquad 
 \cM = \begin{pmatrix} A & B \\ C & D \end{pmatrix} \,,
\label{eq:reality}
\end{align}
where $\str\cM\equiv \Tr A -\Tr D$ and the Hermitian matrix $H$ is defined as
\begin{align}
 H\equiv \begin{pmatrix} \Sigma & \bm{0_4} \\ \bm{0_4} & \bm{1_4} \end{pmatrix} \,,\qquad
 \Sigma \equiv \begin{pmatrix} \bm{0_2} & -\ii\,\sigma_3 \\ \ii\,\sigma_3 & \bm{0_2} \end{pmatrix}=\sigma_2\otimes\sigma_3 \,. 
\end{align}
A trivial element satisfying the above requirement is the $\alg{u}(1)$ generator
\begin{align}
 Z = \ii \begin{pmatrix} \bm{1_4} & \bm{0_4} \\ \bm{0_4} & \bm{1_4} \end{pmatrix} \,,
\label{eq:gZ-def}
\end{align}
and the $\mathfrak{psu}(2,2|4)$ is defined as the quotient $\mathfrak{su}(2,2|4)/\mathfrak{u}(1)$\,. 

\medskip

The $\mathfrak{psu}(2,2|4)$ has an automorphism $\Omega$ defined as
\begin{align}
 \Omega(\cM)=-\cK \,\cM^{\ST}\,\cK^{-1}\,,\qquad
 \cK = \begin{pmatrix} K & \bm{0_4} \\ \bm{0_4} & K \end{pmatrix} \,,
\end{align}
where $K$ is a $4\times 4$ matrix
\begin{align}
 K\equiv {\footnotesize\begin{pmatrix} 0 & -1 & 0 & 0 \\ 1 & 0 & 0 & 0 \\ 0 & 0 & 0 & -1 \\ 0 & 0 & 1 & 0 \end{pmatrix}}\,,\qquad 
 K^{-1} = - K \,,
\end{align}
and $\cM^{\ST}$ represents the supertranspose of $\cM$ defined as
\begin{align}
 \cM^{\ST} = \begin{pmatrix} A^\rmT & -C^{\rmT} \\ B^{\rmT} & D^{\rmT} \end{pmatrix} \,. 
\end{align}
By using the automorphism $\Omega$ (of order four), we decompose $\mathfrak{g}=\mathfrak{psu}(2,2|4)$ as
\begin{align}
 \mathfrak{g}=\mathfrak{g}^{(0)}\oplus\mathfrak{g}^{(1)}\oplus\mathfrak{g}^{(2)}\oplus\mathfrak{g}^{(3)}\,,
\end{align}
where $\Omega(\mathfrak{g}^{(k)})=\ii^k\,\alg{g}^{(k)}$ ($k=0,1,2,3$) and the projector to each vector space $\mathfrak{g}^{(k)}$ can be expressed as
\begin{align}
 P^{(k)}(\cM) \equiv \frac{1}{4}\,\bigl[\, \cM + \ii^{3k}\,\Omega(\cM)+\ii^{2k}\, \Omega^2(\cM) +\ii^k\,\Omega^3(\cM) \,\bigr]\,. 
\label{eq:P-i-projector}
\end{align}

\subsubsection*{Bosonic generators.}

The bosonic generators of $\alg{psu}(2,2|4)$ algebra, $\gP_{\Loa}$ and $\gJ_{\Loa\Lob}$, can be represented by the following $8\times 8$ supermatrices:
\begin{align}
\begin{split}
 &\{\gP_{\Loa}\}\equiv \{\gP_{\check{\Loa}}\,, \gP_{\hat{\Loa}}\}\,,\qquad 
 \{\gJ_{\Loa\Lob}\}\equiv \{\gJ_{\check{\Loa}\check{\Lob}}\,, \gJ_{\hat{\Loa}\hat{\Lob}}\}\,,
\\
 &\gP_{\check{\Loa}} = 
 \begin{pmatrix}
 \frac{1}{2}\,\bm{\gamma}_{\check{\Loa}} & \bm{0_4} \\ 
 \bm{0_4} & \bm{0_4} 
 \end{pmatrix} \,, \qquad 
 \gJ_{\check{\Loa}\check{\Lob}} = 
 \begin{pmatrix}
 -\frac{1}{2}\,\bm{\gamma}_{\check{\Loa}\check{\Lob}} & \bm{0_4} \\ 
 \bm{0_4} & \bm{0_4} \end{pmatrix}
 \qquad (\check{\Loa},\,\check{\Lob}=0,\dotsc,4)\,, 
\\
 &\gP_{\hat{\Loa}} = 
 \begin{pmatrix}
  \bm{0_4} & \bm{0_4} \\ 
  \bm{0_4} & -\frac{\ii}{2}\,\bm{\gamma}_{\hat{\Loa}} 
 \end{pmatrix} \,, \qquad 
 \gJ_{\hat{\Loa}\hat{\Lob}} = 
 \begin{pmatrix}
  \bm{0_4} & \bm{0_4} \\ 
  \bm{0_4} & -\frac{1}{2}\, \bm{\gamma}_{\hat{\Loa}\hat{\Lob}} 
 \end{pmatrix}
 \qquad (\hat{\Loa},\,\hat{\Lob}=5,\dotsc,9)\,, 
\end{split}
\label{eq:P-J-super}
\end{align}
where we defined $4\times 4$ matrices $\bm{\gamma}_{\check{\Loa}} \equiv (\bm{\gamma}_{\check{\Loa}\check{i}}{}^{\check{j}})$ $(\check{i},\check{j}=1,\dotsc,4)$ and $\bm{\gamma}_{\check{\Loa}} \equiv (\bm{\gamma}_{\check{\Loa}\check{i}}{}^{\check{j}})$ $(\hat{i},\hat{j}=1,\dotsc,4)$
\begin{align}
 &\{\bm{\gamma}_{\check{\Loa}}\} \equiv \bigl\{\brgamma_0\,,\brgamma_1\,,\brgamma_2\,,\brgamma_3\,,\brgamma_5\,\bigr\} \,, \qquad 
 \{\bm{\gamma}_{\hat{\Loa}}\} \equiv \bigl\{-\brgamma_4\,,-\brgamma_1\,,-\brgamma_2\,,-\brgamma_3\,,-\brgamma_5\,\bigr\} \,,
\\[1mm]
 \begin{split}
 &\brgamma_1=
 {\footnotesize\begin{pmatrix}
  0 & 0 & 0 & -1\\
  0 & 0 & 1 & 0\\
  0 & 1 & 0 & 0\\
  -1& 0 & 0 & 0 
 \end{pmatrix}} , \quad 
 \brgamma_2=
 {\footnotesize\begin{pmatrix}
 0 & 0 & 0 & \ii \\
 0 & 0 & \ii & 0 \\
 0 & -\ii& 0 & 0 \\
 -\ii& 0 & 0 & 0
 \end{pmatrix}}, \quad 
 \brgamma_3=
 {\footnotesize\begin{pmatrix}
 0 & 0 & 1 & 0 \\
 0 & 0 & 0 & 1 \\
 1 & 0 & 0 & 0 \\
 0 & 1 & 0 & 0 
 \end{pmatrix}},
\\
 &\brgamma_0= -\ii\,\brgamma_4=
 {\footnotesize\begin{pmatrix}
  0 & 0 & 1 & 0 \\
  0 & 0 & 0 &-1 \\
  -1& 0 & 0 & 0 \\
  0 & 1 & 0 & 0 
 \end{pmatrix}}, \quad 
 \brgamma_5=\ii\,\brgamma_1\brgamma_2\brgamma_3\brgamma_0=
 {\footnotesize\begin{pmatrix}
  1 & 0 & 0 & 0 \\
  0 & 1 & 0 & 0 \\
  0 & 0 &-1 & 0 \\
  0 & 0 & 0 &-1
\end{pmatrix}},
\end{split}
\end{align}
and their antisymmeterizations $\bm{\gamma}_{\check{\Loa}\check{\Lob}} \equiv \bm{\gamma}_{[\check{\Loa}}\,\bm{\gamma}_{\check{\Lob}]}$ and $\bm{\gamma}_{\hat{\Loa}\hat{\Lob}} \equiv \bm{\gamma}_{[\hat{\Loa}}\,\bm{\gamma}_{\hat{\Lob}]}$. 
Here, $\brgamma_{\mu}$ ($\mu=0,\dotsc,3$) and $(\bm{\gamma}_{\Loa})\equiv (\bm{\gamma}_{\check{\Loa}},\,\bm{\gamma}_{\hat{\Loa}})$ satisfy
\begin{align}
 \{\brgamma_{\mu}\,, \brgamma_{\nu}\} = 2\,\eta_{\mu\nu}\,,\qquad (\eta_{\mu\nu}) \equiv \diag(-1,1,1,1)\,,\qquad
 (\bm{\gamma}_{\Loa})^{\rmT}= K\,\bm{\gamma}_{\Loa}\,K^{-1}\,. 
\end{align}
The conformal basis, $\{P_{\mu},\, M_{\mu\nu},\,D,\,K_{\mu}\}$\,, of a bosonic subalgebra $\alg{su}(2,2)\cong \alg{so}(2,4)$ that corresponds to the AdS isometries, can be constructed from $\gP_{\check{\Loa}}$ and $\gJ_{\check{\Loa}\check{\Lob}}$ as
\begin{align}
 P_\mu \equiv \gP_\mu + \gJ_{\mu 4}\,,\qquad K_\mu \equiv \gP_\mu - \gJ_{\mu 4}\,,\qquad M_{\mu\nu}\equiv \gJ_{\mu\nu}\,,\qquad D\equiv \gP_4\,,
\end{align}
where $P_{\mu}$, $M_{\mu\nu}$, $D$, and $K_{\mu}$ represent the translation generators, the Lorentz generators, the dilatation generator, and the special conformal generators, respectively. 
On the other hand, a bosonic subalgebra $\alg{su}(4)\cong\alg{so}(6)$ that corresponds to the isometries of $\rmS^5$ are generated by $\gP_{\hat{\Loa}}$ and $\gJ_{\hat{\Loa}\hat{\Lob}}$\,. 
We choose the Cartan generators of $\alg{su}(4)$ as follows
\begin{align}
 h_1 \equiv \gJ_{57}\,,\qquad h_2 \equiv \gJ_{68}\,,\qquad h_3 \equiv \gP_9\,.
\end{align}
For later convenience, let us also define $16\times 16$ matrices $\gamma_{\Loa}$, $\hat{\gamma}_{\Loa}$, and $\gamma_{\Loa\Lob}$ as
\begin{align}
\begin{split}
 (\gamma_{\Loa})&\equiv(\gamma_{\check{\Loa}},\, \gamma_{\hat{\Loa}}) 
 =(\bm{\gamma}_{\check{\Loa}}\otimes \bm{1_4},\, \bm{1_4}\otimes \bm{\gamma}_{\hat{\Loa}})\,,
\\
 (\hat{\gamma}_{\Loa})&\equiv(\hat{\gamma}_{\check{\Loa}},\, \hat{\gamma}_{\hat{\Loa}}) 
 =(\bm{\gamma}_{\check{\Loa}}\otimes \bm{1_4},\, \bm{1_4}\otimes \ii\,\bm{\gamma}_{\hat{\Loa}})\,,
\\
 (\gamma_{\Loa\Lob})&\equiv (\gamma_{\check{\Loa}\check{\Lob}},\,\gamma_{\hat{\Loa}\hat{\Lob}})
 =(\bm{\gamma}_{\check{\Loa}\check{\Lob}}\otimes \bm{1_4}\,, \bm{1_4}\otimes\bm{\gamma}_{\hat{\Loa}\hat{\Lob}}) \,,
\end{split}
\label{eq:gamma-16}
\end{align}
which satisfy
\begin{align}
\begin{split}
 &(\gamma_{\check{\Loa}})^\dagger = \gamma_{\check{0}}\,\gamma_{\check{\Loa}}\,\gamma_{\check{0}} \,,\qquad 
 (\gamma_{\hat{\Loa}})^\dagger = -\gamma_{\check{0}}\,\gamma_{\hat{\Loa}}\,\gamma_{\check{0}} \,,\qquad 
 (\gamma_{\Loa})^\rmT = (K\otimes K)^{-1}\, \gamma_{\Loa}\,(K\otimes K) \,, 
\\
 &(\hat{\gamma}_{\check{\Loa}})^\dagger = \gamma_{\check{0}}\,\hat{\gamma}_{\check{\Loa}}\,\gamma_{\check{0}} \,,\qquad 
 (\hat{\gamma}_{\Loa})^\rmT = (K\otimes K)^{-1}\, \hat{\gamma}_{\Loa}\,(K\otimes K)\,,
\\
 &\{\gamma_{\Loa},\,\gamma_{\Lob}\}=2\,\eta_{\Loa\Lob}\,,\qquad
 \{\hat{\gamma}_{\check{\Loa}},\,\hat{\gamma}_{\check{\Lob}}\}=2\,\eta_{\check{\Loa}\check{\Lob}}\,,\qquad
 \{\hat{\gamma}_{\hat{\Loa}},\,\hat{\gamma}_{\hat{\Lob}}\}=-2\,\delta_{\hat{\Loa}\hat{\Lob}}\,.
\end{split}
\end{align}
We can easily see $\gamma_{\check{\Loa}\check{\Lob}}= \gamma_{[\check{\Loa}}\,\gamma_{\check{\Lob}]}$ and $\gamma_{\hat{\Loa}\hat{\Lob}}=\gamma_{[\hat{\Loa}}\, \gamma_{\hat{\Lob}]}$\,. 
If we also define $\hat{\gamma}_{\check{\Loa}\check{\Lob}}\equiv \hat{\gamma}_{[\check{\Loa}}\,\hat{\gamma}_{\check{\Lob}]}$ and $\hat{\gamma}_{\hat{\Loa}\hat{\Lob}}\equiv \hat{\gamma}_{[\hat{\Loa}}\,\hat{\gamma}_{\hat{\Lob}]}$\,, they satisfy
\begin{align}
 \hat{\gamma}_{\Loa\Lob} = -\frac{1}{2}\,R_{\Loa\Lob}{}^{\Loc\Lod}\, \gamma_{\Loc\Lod}\,,
\end{align}
where $R_{\Loa\Lob}{}^{\Loc\Lod}$ are the tangent components of the Riemann tensor in $\AdS{5}\times\rmS^5$, whose non-vanishing components are
\begin{align}
 R_{\check{\Loa}\check{\Lob}}{}^{\check{\Loc}\check{\Lod}} = -2\, \delta_{[\check{\Loa}}^{[\check{\Loc}}\,\delta_{\check{\Lob}]}^{\check{\Lod}]} \,,\qquad 
 R_{\hat{\Loa}\hat{\Lob}}{}^{\hat{\Loc}\hat{\Lod}} = 2\, \delta_{[\hat{\Loa}}^{[\hat{\Loc}}\,\delta_{\hat{\Lob}]}^{\hat{\Lod}]} \,.
\end{align}

\subsubsection*{Fermionic generators.}

The fermionic generators $(\gQ^I)^{\check{\SPa}\hat{\SPa}}$ $(\check{\SPa}, \hat{\SPa}=1,\dotsc, 4)$ are given by
\begin{align}
 (\gQ^1)^{\check{\SPa}\hat{\SPa}} &=
 \begin{pmatrix}
  \bm{0_4} & \ii\,\delta^{\check{\SPa}}_{\check{i}}\,K^{\hat{j}\hat{\SPa}} \\
  -\delta_{\hat{i}}^{\hat{\SPa}}\, K^{\check{\SPa}\check{j}} & \bm{0_4}
 \end{pmatrix} ,
\qquad 
 (\gQ^2)^{\check{\SPa}\hat{\SPa}} =
 \begin{pmatrix}
  \bm{0_4} & - \delta^{\check{\SPa}}_{\check{i}} \,K^{\hat{j}\hat{\SPa}} \\
  \ii\,\delta_{\hat{i}}^{\hat{\SPa}}\, K^{\check{\SPa}\check{j}}& \bm{0_4}
 \end{pmatrix}\,. 
\label{eq:Q-matrix}
\end{align}
As discussed in \cite{Arutyunov:2015qva}, these matrices do not satisfy the reality condition \eqref{eq:reality} but rather their redefinitions $\mathcal{Q}^I$ do. 
The choice, $\gQ^I$ or $\mathcal{Q}^I$, is a matter of convention, and we here employ $\gQ^I$ by following \cite{Arutyunov:2015qva}. 
We also introduce Grassmann-odd coordinates $\theta_I\equiv (\theta_{\check{\SPa}\hat{\SPa}})_I$ which are 16-component Majorana--Weyl spinors satisfying
\begin{align}
 (\gQ^I\,\theta_I)^{\dagger}\, H+H\,(\gQ^I\,\theta_I) = 0 \,. 
\label{eq:Q-theta-reality}
\end{align}
Since the matrices $\gQ^I$ satisfy
\begin{align}
\begin{split}
 &(\gQ^I)^{\dagger}_{\check{\SPa}\hat{\SPa}}
 =-\ii \,K^{-1}_{\check{\SPa}\check{\SPb}}\,(\gQ^I)^{\check{\SPb}\hat{\SPb}}\,K^{-1}_{\hat{\SPb}\hat{\SPa}} \,,
\\
 &H\,(\gQ^I)^{\check{\SPa}\hat{\SPa}}\,H^{-1}=\ii\,(\bm{\gamma}^0)_{\check{\SPb}}{}^{\check{\SPa}}\,(\gQ^I)^{\check{\SPb}\hat{\SPa}}\,,
\end{split}
\end{align}
the condition \eqref{eq:Q-theta-reality} is equivalent to the Majorana condition
\begin{align}
 \brtheta_I \equiv\theta^\dagger_I\, \gamma^0 = \theta_I^{\rmT}\,(K\otimes K)\,,
\end{align}
or more explicitly,
\begin{align}
 \brtheta_I^{\check{\SPa}\hat{\SPa}} = \theta_{I\check{\SPb}\hat{\SPb}}\, K^{\check{\SPb}\check{\SPa}}\,K^{\hat{\SPb}\hat{\SPa}}\,. 
\end{align}

\subsubsection*{Commutation relations.}

The generators of $\alg{su}(2,2|4)$ algebra, $\gP_{\Loa}$, $\gJ_{\Loa\Lob}$, $\gQ^I$, and $Z$ satisfy the following commutation relations:
\begin{align}
\begin{split}
 [\gP_{\Loa},\,\gP_{\Lob}] &= \frac{1}{2}\,R_{\Loa\Lob}{}^{\Loc\Lod}\,\gJ_{\Loc\Lod}\,, 
\qquad 
 [\gJ_{\Loa\Lob},\,\gP_{\Loc}] = \eta_{\Loc\Loa}\,\gP_{\Lob} - \eta_{\Loc\Lob}\,\gP_{\Loa} \,, 
\\
 [\gJ_{\Loa\Lob},\,\gJ_{\Loc\Lod}] 
 &= \eta_{\Loa\Loc}\,\gJ_{\Lob\Lod}-\eta_{\Loa\Lod}\,\gJ_{\Lob\Loc}-\eta_{\Lob\Loc}\,\gJ_{\Loa\Lod}+\eta_{\Lob\Lod}\,\gJ_{\Loa\Loc} \,, 
\\
 [\gQ^I\,\theta_I,\,\gP_{\Loa}] &= \frac{\ii}{2}\,\epsilon^{IJ}\,\gQ^J\,\hat{\gamma}_{\Loa}\,\theta_I\,, \qquad 
 [\gQ^I\,\theta_I,\,\gJ_{\Loa\Lob}] = \frac{1}{2}\,\delta^{IJ}\,\gQ^I\,\gamma_{\Loa\Lob}\,\theta_J\,,
\\
 [\gQ^I\,\theta_I,\,\gQ^J\,\psi_J] 
 &= -\ii\, \delta^{IJ}\, \brtheta_I\,\hat{\gamma}^{\Loa}\,\psi_J\, \gP_{\Loa} 
   - \frac{1}{4}\,\epsilon^{IJ}\, \brtheta_I\,\gamma^{\Loa\Lob}\,\psi_J\,R_{\Loa\Lob}{}^{\Loc\Lod}\, \gJ_{\Loc\Lod} 
   - \frac{1}{2}\, \delta^{IJ}\, \brtheta_I\,\psi_J\,Z \,,
\end{split}
\label{eq:su(2,2|4)}
\end{align}
and the $\alg{psu}(2,2|4)$ algebra is obtained by dropping the last term proportional to $Z$\,. 

On the other hand, the bosonic generators $\{P_{\mu},\, M_{\mu\nu},\,D,\,K_{\mu}\}$ satisfy the $\alg{so}(2,4)$ algebra,
\begin{align}
\begin{split}
 [P_\mu,\, K_\nu]&= 2\,\bigl(\eta_{\mu\nu}\, D - M_{\mu\nu}\bigr)\,,\quad 
 [D,\, P_{\mu}]= P_\mu\,,\quad [D,\,K_\mu]= -K_\mu\,,
\\
 [M_{\mu\nu},\, P_\rho] &= \eta_{\mu\rho}\, P_\nu-\eta_{\nu\rho}\, P_\mu \,,\quad 
 [M_{\mu\nu},\, K_\rho] = \eta_{\mu\rho}\,K_\nu-\eta_{\nu\rho}\,K_\mu\,, 
\\
 [M_{\mu\nu},\,M_{\rho\sigma}]&= \eta_{\mu\rho}\,M_{\nu\sigma}-\eta_{\mu\sigma}\,M_{\nu\rho} - \eta_{\nu\rho}\,M_{\mu\sigma}+\eta_{\nu\sigma}\,M_{\mu\rho}\,.
\end{split}
\label{eq:so(2-4)-algebra}
\end{align}

\subsubsection*{Supertrace and Projections.}

For generators of the $\alg{psu}(2,2|4)$ algebra, the supertrace becomes
\begin{align}
\begin{split}
 &\str(\gP_{\Loa}\,\gP_{\Lob})=\eta_{\Loa\Lob}\,,\qquad
 \str(\gJ_{\Loa\Lob}\,\gJ_{\Loc\Lod}) =R_{\Loa\Lob\Loc\Lod}\,,
\\
 &\str(\gQ^I\theta_I\,\gQ^J\lambda_J) =-2\,\epsilon^{IJ}\,\brtheta_I\,\lambda_J\,, 
\end{split}
\end{align}
where $R_{\Loa\Lob\Loc\Lod}\equiv R_{\Loa\Lob}{}^{\Loe\Lof}\,\eta_{\Loe\Loc}\,\eta_{\Lod\Lof}$ and
\begin{align}
 \eta_{\Loa\Lob} \equiv \begin{pmatrix} \eta_{\check{\Loa}\check{\Lob}} & 0 \\ 0 & \eta_{\hat{\Loa}\hat{\Lob}} \end{pmatrix}\,, \quad 
 \eta_{\check{\Loa}\check{\Lob}} \equiv \diag (-1,1,1,1,1)\,,\quad 
 \eta_{\hat{\Loa}\hat{\Lob}} \equiv \diag (1,1,1,1,1)\,. 
\end{align}
Each $\mathbb{Z}_4$-component $\alg{g}^{(i)}$ is spanned by the following generators:
\begin{align}
 \alg{g}^{(0)}\! = \Span_{\mathbb{R}}\{\gJ_{\Loa\Lob}\}\,,\quad 
 \alg{g}^{(1)}\! = \Span_{\mathbb{R}}\{\gQ^1\}\,,\quad 
 \alg{g}^{(2)}\! = \Span_{\mathbb{R}}\{\gP_{\Loa}\}\,,\quad 
 \alg{g}^{(3)}\! = \Span_{\mathbb{R}}\{\gQ^2\}\,. 
\end{align}
Then, from the definition of $d_{\pm}$ \eqref{eq:dpm},
\begin{align}
 d_{\pm} \equiv \mp P^{(1)}+2\,P^{(2)}\pm P^{(3)}\,.
\end{align}
we obtain
\begin{align}
 d_\pm(\gP_{\Loa}) = 2\, \gP_{\Loa}\,,\qquad d_\pm(\gJ_{\Loa\Lob}) =0\,,\qquad 
 d_\pm(\gQ^I) = \mp \sigma_3^{IJ}\,\gQ^J \,. 
\end{align}

\subsection{Connection to ten-dimensional quantities}

By using the $16\times 16$ matrices $\gamma_{\Loa}$ defined in \eqref{eq:gamma-16}, the $32\times 32$ gamma matrices $(\Gamma_{\Loa})^{\SPa}{}_{\SPb}$ are realized as
\begin{align}
 (\Gamma_{\Loa}) \equiv \bigl(\Gamma_{\check{\Loa}},\, \Gamma_{\hat{\Loa}}\bigr) 
 \equiv \bigl(\sigma_1\otimes \gamma_{\check{\Loa}},\, \sigma_2\otimes \gamma_{\hat{\Loa}}\bigr)\,. 
\end{align}
We can also realize the charge conjugation matrix as
\begin{align}
 C= \ii\,\sigma_2\otimes K \otimes K \,. 
\end{align}
The $32$-component Majorana--Weyl fermions $\Theta_I$ expressed as
\begin{align}
 \Theta_I = \begin{pmatrix} 1 \\ 0\end{pmatrix}\otimes \theta_I \,,
\label{eq:Theta-theta}
\end{align}
which satisfies the chiral conditions
\begin{align}
 \Gamma^{11}\,\Theta_I = \Theta_I\,. 
\end{align}
The Majorana condition is given by
\begin{align}
 \brTheta_I = \Theta_{I}^\rmT\,C = \begin{pmatrix} 0 & 1 \end{pmatrix}\otimes \brtheta_I \,.
\label{eq:brTheta-brtheta}
\end{align}
This decomposition leads to the following relations between $32$- and $8$-component fermions:
\begin{align}
 &\brtheta_I \hat{\gamma}_{\Loa} \theta_J = \brTheta_I \Gamma_{\Loa} \Theta_J \,,
\label{eq:lift-32-AdS5-1}
\\
 &\brtheta_I\,\hat{\gamma}_{\Loa}\,\hat{\gamma}_{\Lob}\,\theta_J 
 = -\ii\,\brTheta_I\,\Gamma_{\Loa}\, \Gamma_{01234}\,\Gamma_{\Lob}\,\Theta_J 
 = \ii\,\brTheta_I\,\Gamma_{\Loa}\, \Gamma_{56789}\,\Gamma_{\Lob}\,\Theta_J \,,
\\
&\ii\,\sigma_1\otimes \bm{1_4}\otimes \bm{1_4} = \Gamma_{01234}\,,\qquad 
 \sigma_2\otimes \bm{1_4}\otimes \bm{1_4} = \Gamma_{56789} \,,
\end{align}
The second relation plays an important role for a supercoset construction of the $\AdS{5} \times \rmS^5$ background since the R--R bispinor in the $\AdS{5}\times \rmS^5$ background takes the form
\begin{align}
 \bisF_5 
 =\frac{1}{5!}\,\bisF_{\Loa_1\cdots \Loa_5}\,\Gamma^{\Loa_1\cdots \Loa_5}
 =4\,(\Gamma^{01234}+\Gamma^{56789})\,.
\end{align}
Indeed, we obtain
\begin{align}
 \brtheta_I\,\hat{\gamma}_{\Loa}\,\hat{\gamma}_{\Lob}\theta_J
 =\frac{\ii}{8}\,\brTheta_I\,\Gamma_{\Loa}\,\bisF_5\,\Gamma_{\Lob}\,\Theta_J\,. 
\label{eq:lift-32-AdS5-2}
\end{align}
We can also show the following relations:\footnote{Recall that $\gamma_{\Loa\Lob}$ has only the components $(\gamma_{\Loa\Lob})=(\gamma_{\check{\Loa}\check{\Lob}},\,\gamma_{\hat{\Loa}\hat{\Lob}})$.}
\begin{align}
 &\brtheta_I\,\hat{\gamma}_{\Loa}\, \gamma_{\Lob\Loc}\,\theta_J = \brTheta_I\,\Gamma_{\Loa}\, \Gamma_{\Lob\Loc} \,\Theta_J \,, 
\label{eq:lift-32-AdS5-3}
\\
 &\brtheta_I\, \gamma_{\Loa\Lob} \,\theta_J
 = -\ii\,\brTheta_I\,\Gamma_{01234}\, \Gamma_{\Loa\Lob} \,\Theta
 = -\ii\,\brTheta_I\,\Gamma_{56789}\, \Gamma_{\Loa\Lob} \,\Theta_J \,, 
\\
 & \brtheta_I\, \gamma_{\Loa\Lob}\,\gamma_{\Loc\Lod}\,\theta_J
 = -\ii\,\brTheta_I\,\Gamma_{01234}\, \Gamma_{\Loa\Lob}\,\Gamma_{\Loc\Lod} \,\Theta_J 
 = -\ii\,\brTheta_I\,\Gamma_{56789}\, \Gamma_{\Loa\Lob}\,\Gamma_{\Loc\Lod} \,\Theta_J \,. 
\end{align}